\documentclass[journal,twocolumn,romanappendices]{IEEEtran-v17}

\pdfoutput=1

\usepackage{xspace}
\usepackage{url}
\usepackage[cmex10]{amsmath}
\usepackage{bbm}
\usepackage{graphicx}
\usepackage{tikz}
\usepackage{pgfplots}

\usepackage{ifthen}

\newboolean{conf}
\setboolean{conf}{false} 
\usepackage{fancyref} 
\usepackage{amsmath}
\usepackage{amssymb}
\usepackage{dsfont}
\usepackage{footnote}
\usepackage[nosort]{cite}

\pgfplotsset{width=13cm, height=9.5cm,  compat=1.11}
\pgfplotsset{major grid style={
color=black,
line width=0.1mm,
loosely dotted,
},
}
\pgfplotsset{
tick style={
color = black
},
}
 \pgfplotsset{every x tick label/.append style={font=\footnotesize}}
 \pgfplotsset{every y tick label/.append style={font=\footnotesize}}

\usepackage{color}
\definecolor{links}{rgb}{0.7,0,0}   
\definecolor{urls}{rgb}{0,0,0.8}    
\definecolor{cites}{rgb}{0,0,0.8}   

\usepackage[colorlinks,hyperindex,linkcolor=links,citecolor=cites,urlcolor=urls]{hyperref}

\usepackage{bm}
\usepackage{upgreek}
\usepackage{amssymb}
\usepackage{amsfonts}
\usepackage{mathrsfs}
\usepackage{xspace}

\makeatletter
\def\amsbb{\use@mathgroup \M@U \symAMSb}
\makeatother

\newcommand{\lefto}{\mathopen{}\left}
\newcommand{\safemath}[2]{\newcommand{#1}{\ensuremath{#2}\xspace}}
\safemath{\opE}{\amsbb{E}}
\newcommand{\Ex}[2]{\ensuremath{\amsbb{E}_{#1}\mathopen{}\left[#2\right]}} 	
\safemath{\prob}{\amsbb{P}}
\safemath{\bigO}{\mathcal{O}}
\safemath{\littleo}{\mathit{o}}

\safemath{\extendreal}{\overline{\realset}}

\newcommand{\tp}[1]{\ensuremath{#1^{\mathrm{T}}}} 		
\newcommand{\herm}[1]{\ensuremath{#1^{\mathrm{H}}}} 	

\newcommand{\fnorm}[1]{\ensuremath{\left\|#1\right\|_{\mathsf{F}}}}

\newtheorem{thm}{Theorem}
\newtheorem{lemma}[thm]{Lemma}

\newtheorem{rem}{Remark}

\newcommand{\indfun}[1]{\mathbbmss{1}\lefto\{#1\right\}}

\safemath{\matA}{\mathsf{A}}
\safemath{\matB}{\mathsf{B}}
\safemath{\matC}{\mathsf{C}}
\safemath{\matD}{\mathsf{D}}
\safemath{\matE}{\mathsf{E}}
\safemath{\matF}{\mathsf{F}}
\safemath{\matG}{\mathsf{G}}
\safemath{\matH}{\mathsf{H}}
\safemath{\matI}{\mathsf{I}}
\safemath{\matJ}{\mathsf{J}}
\safemath{\matK}{\mathsf{K}}
\safemath{\matL}{\mathsf{L}}
\safemath{\matM}{\mathsf{M}}
\safemath{\matN}{\mathsf{N}}
\safemath{\matO}{\mathsf{O}}
\safemath{\matP}{\mathsf{P}}
\safemath{\matQ}{\mathsf{Q}}
\safemath{\matR}{\mathsf{R}}
\safemath{\matS}{\mathsf{S}}
\safemath{\matT}{\mathsf{T}}
\safemath{\matU}{\mathsf{U}}
\safemath{\matV}{\mathsf{V}}
\safemath{\matW}{\mathsf{W}}
\safemath{\matX}{\mathsf{X}}
\safemath{\matY}{\mathsf{Y}}
\safemath{\matZ}{\mathsf{Z}}
\safemath{\matSigma}{\mathsf{\Sigma}}
\safemath{\matPhi}{\mathsf{\Phi}}
\safemath{\matLambda}{\mathsf{\Lambda}}
\safemath{\matDelta}{\mathsf{\Delta}}

\safemath{\randveca}{\bm{A}}
\safemath{\randvecb}{\bm{B}}
\safemath{\randvecc}{\bm{C}}
\safemath{\randvecd}{\bm{D}}
\safemath{\randvece}{\bm{E}}
\safemath{\randvecf}{\bm{F}}
\safemath{\randvecg}{\bm{G}}
\safemath{\randvech}{\bm{H}}
\safemath{\randveci}{\bm{I}}
\safemath{\randvecj}{\bm{J}}
\safemath{\randveck}{\bm{K}}
\safemath{\randvecl}{\bm{L}}
\safemath{\randvecm}{\bm{M}}
\safemath{\randvecn}{\bm{N}}
\safemath{\randveco}{\bm{O}}
\safemath{\randvecp}{\bm{P}}
\safemath{\randvecq}{\bm{Q}}
\safemath{\randvecr}{\bm{R}}
\safemath{\randvecs}{\bm{S}}
\safemath{\randvect}{\bm{T}}
\safemath{\randvecu}{\bm{U}}
\safemath{\randvecv}{\bm{V}}
\safemath{\randvecw}{\bm{W}}
\safemath{\randvecx}{\bm{X}}
\safemath{\randvecy}{\bm{Y}}
\safemath{\randvecz}{\bm{Z}}

\safemath{\randvecLambda}{\bm{\Lambda}}

\safemath{\randmatA}{\amsbb{A}}
\safemath{\randmatB}{\amsbb{B}}
\safemath{\randmatC}{\amsbb{C}}
\safemath{\randmatD}{\amsbb{D}}
\safemath{\randmatE}{\amsbb{E}}
\safemath{\randmatF}{\amsbb{F}}
\safemath{\randmatG}{\amsbb{G}}
\safemath{\randmatH}{\amsbb{H}}
\safemath{\randmatI}{\amsbb{I}}
\safemath{\randmatJ}{\amsbb{J}}
\safemath{\randmatK}{\amsbb{K}}
\safemath{\randmatL}{\amsbb{L}}
\safemath{\randmatM}{\amsbb{M}}
\safemath{\randmatN}{\amsbb{N}}
\safemath{\randmatO}{\amsbb{O}}
\safemath{\randmatP}{\amsbb{P}}
\safemath{\randmatQ}{\amsbb{Q}}
\safemath{\randmatR}{\amsbb{R}}
\safemath{\randmatS}{\amsbb{S}}
\safemath{\randmatT}{\amsbb{T}}
\safemath{\randmatU}{\amsbb{U}}
\safemath{\randmatV}{\amsbb{V}}
\safemath{\randmatW}{\amsbb{W}}
\safemath{\randmatX}{\amsbb{X}}
\safemath{\randmatY}{\amsbb{Y}}
\safemath{\randmatZ}{\amsbb{Z}}
\safemath{\randmatSigma}{\mathbb{\Sigma}}
\safemath{\randmatPhi}{\mathbb{\Phi}}

\safemath{\pdff}{f}
\safemath{\pdfp}{p}
\safemath{\pdfq}{q}

\safemath{\cdfF}{F}
\safemath{\cdfP}{P}
\safemath{\cdfQ}{Q}

\safemath{\veca}{\bm{a}}
\safemath{\vecb}{\bm{b}}
\safemath{\vecc}{\bm{c}}
\safemath{\vecd}{\bm{d}}
\safemath{\vece}{\bm{e}}
\safemath{\vecf}{\bm{f}}
\safemath{\vecg}{\bm{g}}
\safemath{\vech}{\bm{h}}
\safemath{\veci}{\bm{i}}
\safemath{\vecj}{\bm{j}}
\safemath{\veck}{\bm{k}}
\safemath{\vecl}{\bm{l}}
\safemath{\vecm}{\bm{m}}
\safemath{\vecn}{\bm{n}}
\safemath{\veco}{\bm{o}}
\safemath{\vecp}{\bm{p}}
\safemath{\vecq}{\bm{q}}
\safemath{\vecr}{\bm{r}}
\safemath{\vecs}{\bm{s}}
\safemath{\vect}{\bm{t}}
\safemath{\vecu}{\bm{u}}
\safemath{\vecv}{\bm{v}}
\safemath{\vecw}{\bm{w}}
\safemath{\vecx}{\bm{x}}
\safemath{\vecy}{\bm{y}}
\safemath{\vecz}{\bm{z}}

\safemath{\veclambda}{\bm{\lambda}}
\safemath{\vecpi}{\bm{\pi}}
\safemath{\vecsigma}{\bm\sigma}              			

\safemath{\setA}{\mathcal{A}}
\safemath{\setB}{\mathcal{B}}
\safemath{\setC}{\mathcal{C}}
\safemath{\setD}{\mathcal{D}}
\safemath{\setE}{\mathcal{E}}
\safemath{\setF}{\mathcal{F}}
\safemath{\setG}{\mathcal{G}}
\safemath{\setH}{\mathcal{H}}
\safemath{\setI}{\mathcal{I}}
\safemath{\setJ}{\mathcal{J}}
\safemath{\setK}{\mathcal{K}}
\safemath{\setL}{\mathcal{L}}
\safemath{\setM}{\mathcal{M}}
\safemath{\setN}{\mathcal{N}}
\safemath{\setO}{\mathcal{O}}
\safemath{\setP}{\mathcal{P}}
\safemath{\setQ}{\mathcal{Q}}
\safemath{\setR}{\mathcal{R}}
\safemath{\setS}{\mathcal{S}}
\safemath{\setT}{\mathcal{T}}
\safemath{\setU}{\mathcal{U}}
\safemath{\setV}{\mathcal{V}}
\safemath{\setW}{\mathcal{W}}
\safemath{\setX}{\mathcal{X}}
\safemath{\setY}{\mathcal{Y}}
\safemath{\setZ}{\mathcal{Z}}
\safemath{\emptySet}{\varnothing}

\safemath{\veczero}{\mathbf{0}} 

\safemath{\diag}{\mathrm{diag}}

\safemath{\jpg}{\mathcal{CN}}			
\safemath{\complexset}{\amsbb{C}}
\safemath{\realset}{\amsbb{R}}
\safemath{\natunum}{\mathbb{N}}
\safemath{\posrealset}{\realset_{+}}
\safemath{\integerset}{\amsbb{N}}

\newcommand{\given}{\,\vert\,}				
\safemath{\define}{\triangleq}			

\usepackage{bm}
\usepackage{upgreek}
\usepackage{amssymb}
\usepackage{amsfonts}
\usepackage{mathrsfs}
\usepackage{xspace}

\safemath{\mi}{I}
\safemath{\difent}{h}

\safemath{\constrm}{\mathrm{const}}

\safemath{\NonnegReal}{\mathbb{R}^{+}}
\safemath{\re}{\mathrm{re}}
\safemath{\Real}{\mathrm{Re}} 
\safemath{\gradient}{\nabla}
\safemath{\genericpdf}{f}

\safemath{\bl}{n} 
\safemath{\error}{\epsilon} 

\safemath{\cohtime}{n_\mathrm{c}}
\safemath{\rxant}{m_\mathrm{r}}
\safemath{\txant}{m_\mathrm{t}}

\safemath{\snr}{P}
\safemath{\const}{k}

\safemath{\spanm}{\mathrm{span}}

\safemath{\altg}{\tilde{g}}

\safemath{\altk}{\tilde{k}}
\safemath{\altconst}{\altk}
\safemath{\altvecLambda}{\widetilde{\randvecLambda}}

\safemath{\altveclambda}{\tilde{\veclambda}}
\safemath{\altvecsigma}{\tilde{\vecsigma}}              

\safemath{\altLambda}{\widetilde{\Lambda}}

\safemath{\altgamma}{\tilde{\gamma}}
\safemath{\altsigma}{\tilde{\sigma}}      				
\safemath{\altdelta}{\tilde{\delta}}
\safemath{\altrho}{\tilde{\rho}}
\safemath{\altlambda}{\tilde{\lambda}}

\safemath{\altsnr}{\altrho}						

\safemath{\altmatSigma}{\matDelta}
\safemath{\altmatPhi}{\widetilde{\matPhi}} 		
\safemath{\inpdist}{Q_{\randmatX}}

\def\eb{E_{\mathrm{b}}}

\def\cohtime{n_{\mathrm{c}}}
\def\varr{\mathbb{V}\mathrm{ar}}

\def\der{\mathrm{d}}

\begin{document}
\IEEEoverridecommandlockouts

\title{Beta-Beta Bounds: Finite-Blocklength Analog of the Golden Formula}

\author{Wei Yang,~\IEEEmembership{Member,~IEEE}, Austin Collins, Giuseppe Durisi~\IEEEmembership{Senior Member,~IEEE}, \\
Yury Polyanskiy~\IEEEmembership{Senior Member,~IEEE}, and H. Vincent Poor~\IEEEmembership{Fellow,~IEEE}
\thanks{This work was supported in part by the US  National Science Foundation (NSF) under Grants  CCF-1420575, CCF-1717842, and ECCS-1647198, by the Swedish Research Council under grant  2012-4571, by the Center for Science of Information (CSoI), an NSF Science and Technology Center, under grant agreement CCF-09-39370, and by the NSF CAREER award CCF-1253205. The material of this paper was presented in part at the International Zurich Seminar, Zurich, Switzerland, March 2016, and at the IEEE International Symposium on Information Theory (ISIT), Barcelona, Spain, July 2016. }
\thanks{W. Yang is with Qualcomm Technologies, Inc., San Diego, CA 92121 USA (email:weiyang@qti.qualcomm.com).}
\thanks{A. Collins and Y. Polyanskiy are with    the Department of Electrical Engineering and Computer Science, Massachusetts Institute of Technology, Cambridge, MA 02139
USA (e-mail: austinc@mit.edu; yp@mit.edu).}
\thanks{G. Durisi is with the Department of Electrical Engineering, Chalmers University of Technology, Gothenburg 41296, Sweden (e-mail: durisi@chalmers.se).}
\thanks{H. V. Poor is with the Department of Electrical Engineering, Princeton University, Princeton, NJ 08544 USA (e-mail: poor@princeton.edu).}
}

\maketitle
\begin{abstract}
It is well known that the mutual information between two random variables can be expressed as the difference of two relative entropies that depend on an auxiliary distribution, a relation sometimes referred to as the golden formula.
This paper is concerned with a finite-blocklength extension of this relation.
This extension consists of two elements: 1) a finite-blocklength channel-coding converse bound by Polyanskiy and Verd\'{u} (2014), which involves the ratio of two Neyman-Pearson $\beta$ functions (beta-beta converse bound); and 2) a novel beta-beta channel-coding achievability bound, expressed again as the   ratio of two Neyman-Pearson $\beta$ functions.

To demonstrate the usefulness of this finite-blocklength extension of  the golden formula,  the beta-beta achievability and converse bounds are used to obtain a finite-blocklength extension of Verd\'{u}'s (2002) wideband-slope approximation.
The proof parallels the derivation of the latter, with the beta-beta bounds used in place of  the golden formula.

The beta-beta (achievability) bound is also shown to be useful in cases where the capacity-achieving output distribution is not a product distribution due to, e.g., a cost constraint  or structural constraints on the codebook, such as orthogonality or constant composition.
As an example, the bound is used to characterize the channel dispersion of the additive exponential-noise channel and to obtain a finite-blocklength achievability bound (the tightest to date) for multiple-input multiple-output Rayleigh-fading channels with perfect channel state information at the receiver.
\end{abstract}

\begin{IEEEkeywords}
Channel coding, achievability bound, hypothesis testing, golden formula,  finite-blocklength regime.
\end{IEEEkeywords}

\section{Introduction}

\subsection{Background}

In his landmark 1948  paper~\cite{shannon48}, Shannon established the noisy channel coding theorem, which expresses the fundamental limit of reliable data transmission over a noisy channel  in terms of the \emph{mutual information} $I(X;Y)$ between the input~$X$ and the output~$Y$ of the channel.
More specifically, for stationary memoryless channels, the maximum rate at which data can be transmitted reliably   is the channel capacity
 \begin{IEEEeqnarray}{rCl}
 C = \sup_{P_X} I(X;Y)  .
 \label{eq:channel-capacity}
 \end{IEEEeqnarray}
 Here, reliable transmission means that the probability of error can be made arbitrarily small by mapping the information bits into sufficiently long codewords.
In the nonasymptotic regime of fixed blocklength (fixed codeword size), there exists a tension between rate and error probability, which is partly captured by the many nonasymptotic  (finite-blocklength) bounds available in the literature~\cite{feinstein54a,shannon59,verdu94-07a,polyanskiy10-05}.
In many of these bounds, the role of the mutual information is taken by the so-called \emph{(mutual) information density}\footnote{In this paper, $\log$ and $\exp$ functions are taken with respect to the same arbitrary  basis.}
\begin{IEEEeqnarray}{rCl}
i(X;Y)\define \log\frac{\der P_{XY}}{\der (P_{X} P_Y)} (X,Y)
\end{IEEEeqnarray}
or information spectrum~\cite{pinsker1964,han93-03}.
While the various properties enjoyed by the mutual information make the evaluation of capacity relatively simple, computing the finite-blocklength bounds  that involve the information density (e.g., the information-spectrum bounds~\cite{feinstein54a,shannon59,verdu94-07a}) can be very challenging.

One well-known property of the mutual information is that it can be expressed as a difference of relative entropies involving an arbitrary output distribution $Q_{Y}$~\cite[Eq.~(8.7)]{csiszar11}:
\begin{equation}
I(X;Y) =  D(P_{XY} \| P_XQ_{Y}) - D(P_{Y}\| Q_{Y}).
\label{eq:golden-formula-intro}
\end{equation}
Here, $D(\cdot\|\cdot)$ stands for the relative entropy.
This identity---also known as the \emph{golden formula}~\cite[Th. 3.3]{polyanskiy12} or Tops\o e's identity~\cite{Topsoe1967}---has found many applications in  information theory.
One canonical application  is to establish upper bounds on channel capacity~\cite{lapidoth03-10a}.
Indeed, substituting~\eqref{eq:golden-formula-intro} into~\eqref{eq:channel-capacity}, we get  an alternative expression for channel capacity
\begin{IEEEeqnarray}{rCl}
 C &=& \max_{P_X}  \big\{D(P_{XY} \| P_XQ_{Y}) - D(P_{Y}\| Q_{Y}) \big\}  \label{eq:channel-capacity-alt}
\end{IEEEeqnarray}
from which an upper bound on $C$ can be obtained by dropping the term $-D(P_Y\|Q_Y)$.
The golden formula is also used in the derivation of the capacity per unit cost~\cite{verdu90-09} and the wideband slope~\cite{verdu02-06}, in the Blahut-Arimoto algorithm~\cite{arimoto1972-01a,blahut1972-07a},  and in Gallager's formula for the minimax redundancy in universal source coding~\cite{gallager79-u}.
Furthermore, it is useful for characterizing the properties of good channel codes~\cite{shamai97-03,polyanskiy14-01}, and it is often used  in  statistics  to prove lower bounds on the minimax risk via Fano's inequality (see~\cite{yang1999-05az} and~\cite{haussler1997-06a}).

The main purpose of this paper is to provide a finite-blocklength analog of~\eqref{eq:channel-capacity-alt} that is helpful in deriving nonasymptotic results in the same way in which~\eqref{eq:channel-capacity-alt} is helpful in the asymptotic case.\footnote{With a slight abuse in terminology, we shall refer to both~\eqref{eq:golden-formula-intro} and~\eqref{eq:channel-capacity-alt}  as the golden formula.}
Note that a na\"ive way to derive such a finite-blocklength analog would be to rewrite the information density in the information-spectrum   bounds as follows:
\begin{IEEEeqnarray}{rCl}
i(X;Y) = \log \frac{\der P_{X Y } }{\der (P_{X }Q_{Y })} (X ;Y ) - \log \frac{\der P_{Y }}{\der Q_{Y }} (X ;Y ). \IEEEeqnarraynumspace
\label{eq:two-random-varialbe-info-den}
\end{IEEEeqnarray}
However, the resulting bounds are not very useful, because it is difficult to decouple the two random variables on the right-hand side (RHS) of~\eqref{eq:two-random-varialbe-info-den}.
Instead of tweaking the information-spectrum bounds via~\eqref{eq:two-random-varialbe-info-den}, we   derive a finite-blocklength analog of~\eqref{eq:channel-capacity-alt} from first principles.

To summarize our contribution, we need to first introduce some notation.
We consider an abstract channel that consists of an input set~$\setA$, an output set~$\setB$, and a random transformation $P_{Y|X}: \setA \to \setB$.
 An $(M,\error)$ code for the channel $(\setA, P_{Y|X},\setB)$ comprises a message set $\setM \define \{1,\ldots, M\}$, an encoder  $f: \setM \to \setA$, and a decoder $g: \setB \to \setM \cup \{e\}$ ($e$ denotes the error event) that satisfies the \emph{average}  error probability constraint
\begin{IEEEeqnarray}{rCl}
 \frac{1}{M} \sum\limits_{j=1}^{M}  \Big(1- P_{Y\given X} \mathopen{}\big( g^{-1}(j) \given f(j) \big)\Big) \leq \error.
\label{eq:avg-prob-error-def}
\end{IEEEeqnarray}
Here, $g^{-1}(j) \define \{y\in\setY: g(y) =j\}$.
In practical applications, we often take $\setA$ and $\setB$ to be $n$-fold Cartesian products of two alphabets $\setX$ and $\setY$, and the channel to be a sequence of conditional probabilities $P_{Y^n\given X^n}: \setX^n \to \setY^n$.
We shall refer to an $(M,\error)$ code for the channel $\{\setX^n, P_{Y^n\given X^n}, \setY^n\}$ as an $(n,M,\error)$ code.

Binary hypothesis testing, which we introduce next, will play an important role.
Given two probability measures $P$ and $Q$ on a common measurable space $\setX$, we define a randomized test between $P$ and $Q$ as a random transformation $P_{Z\given X} :\setX \to\{0,1\}$, where $0$ indicates that the test chooses $Q$.
The optimal performance achievable among all such randomized tests is given by the Neyman-Pearson function $\beta_\alpha(P,Q)$, which is defined as
\begin{equation}
\label{eq:def-beta-intro}
\beta_\alpha(P,Q) \define  \min \int P_{Z\given X}(1\given x)  Q(\mathrm{d} x)
\end{equation}
where the minimum is over all  tests satisfying
\begin{equation}
 \int P_{Z\given X} (1\given x) P(\mathrm{d} x) \geq \alpha.
\end{equation}
The Neyman-Pearson lemma~\cite{neyman33a} provides the optimal test $P_{Z|X}$ that attains the minimum in~\eqref{eq:def-beta-intro}. This test, which we shall refer to as  the Neyman-Pearson test, involves thresholding the  Radon-Nikodym derivative of $P$ with respect to $Q$.

\subsection{Finite-Blocklength Analog of the Golden Formula}
\label{sec:intro-finite-analog}
A first step towards establishing a finite-blocklength analog of the golden formula was recently provided by Polyanskiy and Verd\'{u}, who proved that  every $(n,M,\error)$ code satisfies the following converse bound~\cite[Th.~15]{polyanskiy14-01}:
\begin{IEEEeqnarray}{rCl}
M \leq \inf\limits_{0\leq \delta < 1-\error}  \frac{\beta_{1-\delta }(P_{Y^n},Q_{Y^n})}{  \beta_{1-\error -\delta} (P_{X^nY^n} , P_{X^n}Q_{Y^n})},\quad \forall \, Q_{Y^n}. \IEEEeqnarraynumspace
\label{eq:prop-lb-beta-p-q-intro}
\end{IEEEeqnarray}
Here, $P_{X^n}$ and  $P_{Y^n}$  denote the empirical input and output distributions induced by the code for the case of uniformly distributed messages.
The proof of~\eqref{eq:prop-lb-beta-p-q-intro} is a refinement of the meta-converse theorem~\cite[Th.~26]{polyanskiy10-05}, in which the decoder is used as a suboptimal   test for discriminating $P_{X^nY^n}$ against $P_{X^n}Q_{Y^n}$.
%
%
%
We shall refer to the converse bound~\eqref{eq:prop-lb-beta-p-q-intro}  as the $\beta\beta$ converse bound.
Note that the special case of $\delta = 0$, which is known as the minimax meta-converse bound, has a long history in information theory as surveyed in~\cite[Sec. II.D]{polyanskiy10-05} for the classical case
and~\cite{nagaoka2001} for quantum channels.

\begin{figure*}[t]
 \centering
\begin{tikzpicture}
\draw (2.8,2.9) node{\small Asymptotic};
\draw (9.8,2.9) node{\small Nonasymptotic};

\draw[rounded corners=3mm] (0, -0.2) rectangle (5.6,1.9);
\draw(2.8, 1.5) node{\small Golden formula:};
\draw (2.8,0.8) node{\small $I(X;Y)=D(P_{XY}||P_XQ_Y)$};
\draw (3.5,0.2) node{\small $-D(P_Y||Q_Y)$};

\draw[rounded corners=3mm] (7, -0.9) rectangle (5.6+7.5,2.6);
\draw(3+7, -1.6+3.8) node{\small $\beta\beta$ bounds:};
\draw (3+7,-2.6+3.8) node{\small $\dfrac{M}{2} \geq \dfrac{\beta_{\tau}(P_{Y^n} , Q_{Y^n})}{\beta_{1-\epsilon+\tau}(P_{X^nY^n} ,P_{X^n}Q_{Y^n}) }$};
\draw (3+7,-3.85+3.8) node{\small $M\leq \dfrac{\beta_{1-\delta}(P_{Y^n} , Q_{Y^n})}{\beta_{1-\epsilon-\delta}(P_{X^nY^n} ,P_{X^n}Q_{Y^n}) } $};

\draw (6.3,0.8) node{{\Large$\Leftarrow$}};

\draw[rounded corners=3mm] (0, -4.2) rectangle (5.6, -2);
\draw(2.8, -2.4) node{\small Duality bound~\cite{lapidoth03-10a}:};
\draw (2.8,-3.4) node{\small $C \le \max\limits_{P_X} D(P_{XY}||P_XQ_Y)$};

\draw (6.3,-3.1) node{{\Large$\Leftarrow$}};

\draw[rounded corners=3mm] (6.9, -4.2) rectangle (13.1,-2);
\draw(10, -2.4) node{\small  Minimax converse bound:};
\draw (10,-3.4) node{\small $M \le \sup\limits_{P_{X^n}} \dfrac{1}{\beta_{1-\epsilon}(P_{X^nY^n} ,P_{X^n}Q_{Y^n})} $};

\draw (9.8, -1.5) node{ {\Large $\Downarrow$}};

\draw (2.8, -1) node{ {\Large $\Downarrow$}};

	\end{tikzpicture}
\caption{Connections between the golden formula and nonasymptotic $\beta\beta$ bounds\label{fig:connections}.}
\end{figure*}
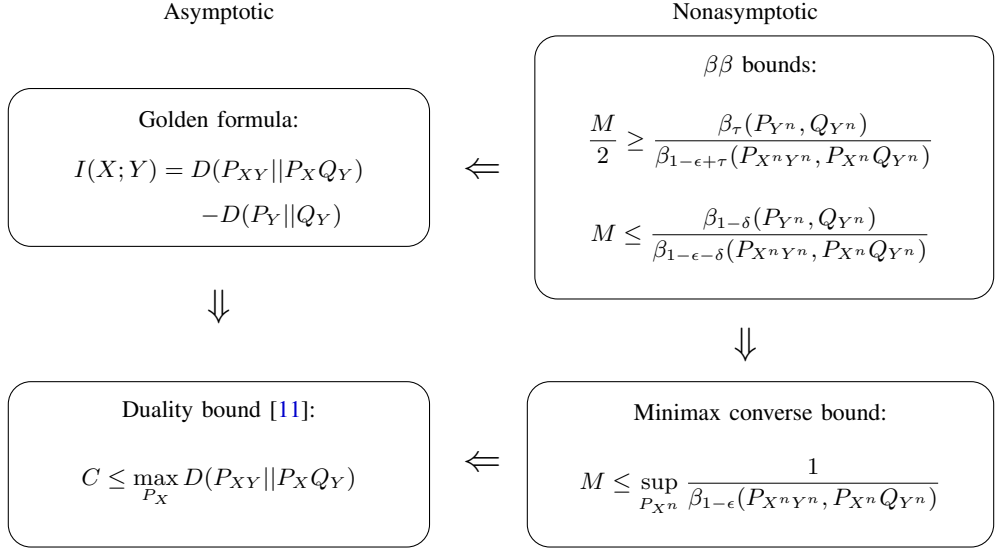

In this paper, we provide the following achievability counterpart of~\eqref{eq:prop-lb-beta-p-q-intro}: for every $0<\error<1$ and every input distribution $P_{X^n}$, there exists an $(n, M, \error)$ code  such that
\begin{equation}
\frac{M}{2} \geq \sup\limits_{0<\tau<\error} \frac{\beta_{\tau}(P_{Y^n} , Q_{Y^n} )}{\beta_{1-\error+\tau}(P_{X^nY^n}, P_{X^n} Q_{Y^n})},\quad \forall \, Q_{Y^n}
\label{eq:kappa-beta-intro-avg-intro}
\end{equation}
where  $P_{Y^n}$ denotes the distribution of $Y^n$ induced by $P_{X^n}$ through $P_{Y^n|X^n}$.
The proof of the $\beta\beta$ achievability bound above relies on Shannon's random coding technique and on a suboptimal decoder that is based on the Neyman-Pearson test between $P_{X^nY^n}$ and $P_{X^n}Q_{Y^n}$.
Hypothesis testing is used twice in the proof: to relate the decoding error probability to $\beta_{1-\error+\tau}(P_{X^nY^n}, P_{X^n} Q_{Y^n})$, and to perform a change of measure from $P_{Y^n}$ to $Q_{Y^n}$.
Fig.~\ref{fig:connections} gives a pictorial summary of the connections between the $\beta\beta$ bounds and various capacity and nonasymptotic bounds.
The analogy between the $\beta\beta$ bounds~\eqref{eq:prop-lb-beta-p-q-intro}--\eqref{eq:kappa-beta-intro-avg-intro} and the golden formula follows because, for product distributions $P^n$ and $Q^n$,
\begin{IEEEeqnarray}{c}
-\frac{1}{n} \log \beta_{\alpha}(P^n,Q^n) = D(P\|Q) + o(1), \,\, \forall \alpha\in(0,1) \IEEEeqnarraynumspace
\label{eq:stein-lemma}
\end{IEEEeqnarray}
as $n\to\infty$ by Stein's lemma~\cite[Th.~11.8.3]{cover06-a}.
For example, one can prove that~\eqref{eq:channel-capacity-alt} is achievable using~\eqref{eq:kappa-beta-intro-avg-intro} as follows: set  $P_{X^n}= (P_X)^n$ and $Q_{Y^n}=(Q_Y)^n$, take the log of both sides of~\eqref{eq:kappa-beta-intro-avg-intro}, use Stein's lemma and optimize  over $P_X$.

 \subsection{Applications}

To demonstrate that the $\beta\beta$ bounds~\eqref{eq:prop-lb-beta-p-q-intro} and~\eqref{eq:kappa-beta-intro-avg-intro} are the natural nonasymptotic equivalent of the golden formula,
we use them to provide a finite-blocklength extension of the \emph{wideband slope} approximation of Verd\'{u}~\cite{verdu02-06}.
More specifically, we derive a second-order characterization of the minimum energy per bit $\eb^*(k, \error, R)$ required to transmit $k$ bits with error probability $\error$ and rate $R$ on  additive white Gaussian noise (AWGN) channels and also on Rayleigh-fading channels with perfect channel state information available at the receiver (CSIR).
Our result implies that $\eb^*(k, \error, R)$ (expressed in $\mathrm{dB}$) can be approximated by an affine function of the rate~$R$.
Furthermore, the slope of this function coincides with the wideband slope by Verd\'{u}.
Our proof   parallels the derivation of the latter in~\cite{verdu02-06}, except that the role of the golden formula is taken by the $\beta\beta$ bounds~\eqref{eq:prop-lb-beta-p-q-intro} and~\eqref{eq:kappa-beta-intro-avg-intro}. Numerical evaluations demonstrate the accuracy of the resulting approximation.

The $\beta\beta$ achievability bound~\eqref{eq:kappa-beta-intro-avg-intro} is also useful in situations where~$P_{Y^n}$ is not a product distribution (although the underlying channel law $P_{Y^n|X^n}$ is stationary and memoryless), for example  due to a cost constraint, or a structural constraint  on the channel inputs, such as orthogonality or constant composition.
In such cases, traditional achievability bounds such as Feinstein's bound~\cite{feinstein54a} or the dependence-testing (DT) bound~\cite[Th.~18]{polyanskiy10-05}, which are expressed in terms of the information density, become difficult to evaluate.
In contrast, the~$\beta\beta$ bound~\eqref{eq:kappa-beta-intro-avg-intro}  requires the evaluation of  $\mathrm{d}P_{Y^n|X^n}/\mathrm{d}Q_{Y^n}$, which factorizes  when~$Q_{Y^n}$ is chosen to be a product distribution.
This allows for an analytical computation of~\eqref{eq:kappa-beta-intro-avg-intro}.
Furthermore, the term $\beta_{\tau}(P_{Y^n}, Q_{Y^n})$, which captures the cost of the change of measure from $P_{Y^n}$ to $Q_{Y^n}$, can be evaluated or bounded even when a closed-form expression for $P_{Y^n}$ is not available.
To illustrate these points:
\begin{itemize}

\item We use the $\beta\beta$ achievability bound to characterize the channel dispersion~\cite[Def. 1]{polyanskiy10-05} of the  additive exponential noise channel introduced in~\cite{verdu96-01a}.
\item We evaluate~\eqref{eq:kappa-beta-intro-avg-intro}  for  the case of multiple-input multiple-output (MIMO) Rayleigh-fading channels with  perfect CSIR.  In this case, \eqref{eq:kappa-beta-intro-avg-intro} yields the tightest known achievability bound.
\end{itemize}


\subsection{Notation}

For an input distribution $P_{X}$ and a channel $P_{Y|X}$, we let $P_X \circ P_{Y|X}$ denote the distribution of~$Y$ induced by $P_X$ through $P_{Y|X}$.
The distribution of a circularly symmetric complex Gaussian random vector with covariance matrix $\mathsf{A}$ is denoted by $\jpg(0,\mathsf{A})$.
We denote by $\chi_{k}^2(\lambda)$ the noncentral chi-squared distribution with $k$ degrees of freedom and noncentrality parameter $\lambda$; furthermore,  $\mathrm{Exp}(\mu)$ stands for the exponential distribution with mean $\mu$.
The Frobenius norm of a matrix $\matA$ is denoted by $\|\matA \|_{\mathsf{F}}$.
For a vector $\vecx=(x_1,\ldots,x_d)$, we let $\|\vecx\|$, $\|\vecx\|_4$, and $\|\vecx\|_{\infty}$ denote the $\ell_2$, $\ell_4$, and $\ell_{\infty}$ norms of $\vecx$, respectively.
 The real part and the complex conjugate of a complex number $x$ are denoted by $\mathrm{Re}(x)$ and $x^*$, respectively.
 For a set $\setA$, we use $|\setA|$ and~$\setA^c$ to denote the set cardinality and the set complement, respectively.
Finally, $\lceil \cdot \rceil$ denotes the ceiling function, and the superscript~$\herm{}$ stands for Hermitian transposition.

\section{The $\beta\beta$ Achievability Bound}
\label{sec:proof-thm-betabeta}
In this section, we formally state our $\beta\beta$ achievability bound.
 \begin{thm}[$\beta\beta$ achievability bound]
\label{thm:betabeta-bound}
For every $0<\error<1$ and every input distribution $P_{X}$, there exists an $(M, \error)$ code for the channel $(\setA, P_{Y|X}, \setB)$ satisfying
\begin{equation}
\frac{M}{2} \geq \sup\limits_{0<\tau<\error} \sup\limits_{Q_{Y}} \frac{\beta_{\tau}(P_{Y} , Q_Y )}{\beta_{1-\error+\tau}(P_{XY}, P_X Q_{Y})}
\label{eq:kappa-beta-intro-avg}
\end{equation}
where  $P_Y = P_X \circ P_{Y|X} $.
\end{thm}

\begin{IEEEproof}
Fix $\error \in(0,1)$, $\tau\in(0,\error)$, and let $P_X$ and $Q_{Y}$ be two arbitrary probability measures on $\setA$ and~$\setB$, respectively. Furthermore, let
\begin{IEEEeqnarray}{rCl}
M= \left\lceil \frac{2\beta_{\tau}(P_{Y} , Q_Y )}{\beta_{1-\error+\tau}(P_{XY}, P_X Q_{Y})} \right\rceil.
\label{eq:def-M-ceil-ratio}
\end{IEEEeqnarray}
Finally, let $P_{Z\given XY}:\setA\times\setB \to\{0,1\}$ be the Neyman-Pearson test that satisfies
\begin{IEEEeqnarray}{rCl}
P_{XY}[Z(X,Y) =1] &\geq& 1-\error +\tau \label{eq:def-z-under-p-intro}\\
P_XQ_{Y}[Z(X,Y)=1] &=& \beta_{1-\error+\tau}(P_{XY}, P_X Q_{Y}).  \label{eq:def-z-under-q-intro} \IEEEeqnarraynumspace
\end{IEEEeqnarray}
In words, $(X,Y)$ may be distributed either according to $P_{XY}$ or according to $P_XQ_Y$, and $Z(X,Y)=1$ means that the test chooses $P_{XY}$.
For a given codebook $\{c_1,\ldots, c_{M}\}$ and a received signal~$y$, the decoder computes the values of $Z(c_j, y)$ and returns the smallest index $j$ for which $Z(c_j, y)=1$. If no such index is found, the decoder declares an error.
The average probability of error of the given codebook $\{c_1,\ldots, c_{M}\}$, under the assumption of uniformly distributed messages, is given by
\begin{IEEEeqnarray}{rCl}
\IEEEeqnarraymulticol{3}{l}{
P_{\mathrm{e}}(c_1,\ldots,c_M)
}\notag\\
&&\quad  = \prob\mathopen{}\bigg[\big\{Z(c_{W}, Y) =0 \big\} \cup \Big( \bigcup\limits_{m < W}  \big\{  Z(c_m,Y)=1 \big\}\Big) \bigg] \IEEEeqnarraynumspace
\label{eq:rand-coding-error-ana}
\end{IEEEeqnarray}
where $W$ is equiprobable on $\{1, \ldots, M\}$ and $Y\sim P_{Y|W}$.

Following Shannon's random coding idea, we next average~\eqref{eq:rand-coding-error-ana} over  all  codebooks $\{C_1,\ldots, C_M\}$ whose codewords are generated as pairwise independent random variables with distribution $P_X$. This yields
\begin{IEEEeqnarray}{rCl}
\IEEEeqnarraymulticol{3}{l}{\Ex{}{P_{\mathrm{e}}(C_1,\ldots,C_M) }}\notag\\
\quad &\leq& \prob\mathopen{}\big[Z(X, Y) =0 \big] +
\prob\mathopen{}\bigg[ \! \max_{m <  W} Z(C_m,Y)=1   \bigg] \label{eq:double-beta-union-bound} \IEEEeqnarraynumspace\\
&\leq & \error -\tau + \prob\mathopen{}\bigg[  \max_{m < W} Z(C_m,Y)=1  \bigg] \label{eq:double-beta-ht-bound}.
\end{IEEEeqnarray}
Here,~\eqref{eq:double-beta-union-bound} follows from the union bound and~\eqref{eq:double-beta-ht-bound} follows from~\eqref{eq:def-z-under-p-intro}.

To conclude the proof of~\eqref{eq:kappa-beta-intro-avg}, it suffices to show that
\begin{IEEEeqnarray}{rCl}
\prob\mathopen{}\bigg[ \max_{m < W} Z(C_m,Y)=1  \bigg]&\leq& \tau
\label{eq:exists-m-z-1-py}
\end{IEEEeqnarray}
for the $M$ defined in~\eqref{eq:def-M-ceil-ratio}, where the probability is computed with respect to $Y\sim P_Y$.
The idea of the proof is to relate this probability to a probability computed with respect to $Y\sim Q_{Y}$ via binary hypothesis testing.
Consider the randomized test $P_{\widetilde{Z}\given Y}: \setY \to \{0,1\}$
\begin{IEEEeqnarray}{rCl}
\widetilde{Z}(y) \define \max_{m < W} Z(C_m, y)
\label{eq:def-z-widetilde-intro}
\end{IEEEeqnarray}
where, as before, $W$ is uniformly distributed on  $\{1,\ldots, M\}$, and the $\{C_m\}$ are pairwise independent random variables with distribution $P_X$.
Here, the random variable $Y$ may be distributed either according to $P_Y$ or according to $Q_Y$, and $\widetilde{Z}(Y)=1$ indicates that the test chooses $P_Y$.
It follows that
\begin{IEEEeqnarray}{rCl}
\IEEEeqnarraymulticol{3}{l}{
\beta_{P_{Y}[\widetilde{Z} =1] } (P_{Y}, Q_{Y})}\notag\\
\qquad &\leq & Q_{Y}[\widetilde{Z} (Y)=1]  \label{eq:def-beta-implies-betap}\\
&\leq&\frac{1}{M} \sum\limits_{j=1}^{M} (j-1) P_XQ_{Y}[Z(X,Y) =1 ]
 \label{eq:split-qy-tildez-1} \\
&=& \frac{M-1}{2} P_X Q_{Y}[Z( X, Y)=1]\label{eq:split-qy-tildez-300}  \\
&=& \frac{M-1}{2} \beta_{1-\error+\tau} (P_{XY}, P_X Q_{Y}) \label{eq:split-qy-tildez-3} \IEEEeqnarraynumspace \\
&\leq & \beta_{\tau} (P_{Y}, Q_{Y}).\label{eq:split-qy-tildez-4}
\end{IEEEeqnarray}
Here,~\eqref{eq:def-beta-implies-betap} follows from~\eqref{eq:def-beta-intro};~\eqref{eq:split-qy-tildez-1} follows from~\eqref{eq:def-z-widetilde-intro} and  from the union bound;~\eqref{eq:split-qy-tildez-3} follows from~\eqref{eq:def-z-under-q-intro}; and~\eqref{eq:split-qy-tildez-4} follows from~\eqref{eq:def-M-ceil-ratio}.
Since the function $\alpha \mapsto \beta_{\alpha}(P_{Y},Q_{Y})$ is  nondecreasing, we conclude that
\begin{IEEEeqnarray}{rCl}
P_{Y}[\widetilde{Z} =1] \leq \tau
\label{eq:exists-m-z-1-py-alt}
\end{IEEEeqnarray}
which is equivalent to~\eqref{eq:exists-m-z-1-py}.
\end{IEEEproof}

\section{Relation to Existing Achievability Bounds}

\label{sec:relation-to-existing-bounds}
We next discuss the relation between Theorem~\ref{thm:betabeta-bound} and the achievability bounds available in the literature.
\subsubsection{ The $\kappa\beta$ bound~\cite[Th.~25]{polyanskiy10-05}} The $\kappa\beta$ bound is based on Feinstein's maximal coding approach and on a suboptimal threshold decoder.
By further lower-bounding the $\kappa$ term in the $\kappa\beta$ bound using~\cite[Lemma~4]{polyanskiy11-08a}, we can relax it to the following bound:
\begin{IEEEeqnarray}{rCl}
M \geq \sup\limits_{ 0<\tau < \error }\sup\limits_{Q_Y} \frac{\beta_{\tau}(P_X \circ P_{Y|X}, Q_Y)}{\sup\nolimits_{x\in \setF} \beta_{1-\error + \tau}(P_{Y|X=x} , Q_{Y})} \IEEEeqnarraynumspace
\label{eq:kappa-beta-relax}
\end{IEEEeqnarray}
which holds under the \emph{maximum} error probability constraint (cf.~\eqref{eq:avg-prob-error-def})
\begin{equation}
 \max\limits_{j\in\{1,\ldots, M\}}  \Big\{1- P_{Y\given X} \mathopen{}\big( g^{-1}(j) \given f(j) \big)\Big\} \leq \error.
\end{equation}
Here, $\setF\subset \setA$ denotes the permissible set of  codewords, and $P_X$ is an arbitrary distribution on $\setF$.
Because of the similarity between~\eqref{eq:kappa-beta-relax} and~\eqref{eq:kappa-beta-intro-avg}, one can interpret the $\beta\beta$ bound as the average-error-probability counterpart of the $\kappa\beta$ bound.\footnote{In fact, the analogy between the $\kappa\beta$ bound and the $\beta\beta$ bound is also clear without applying the relaxation~\eqref{eq:kappa-beta-relax}.
Indeed, the $\kappa$ term in the $\kappa\beta$ bound~\cite[Th.~25]{polyanskiy10-05} defines a relative measure of performance for composite hypothesis testing between the collection $\{P_{Y|X=x}\}_{x\in\setF}$  and $Q_{Y}$. The $\beta_{\tau}(P_{Y},Q_Y)$ term in the $\beta\beta$ bound measures the performance of a binary hypothesis test between $P_Y$ and $Q_Y$, where the distribution $P_Y$ is an average of the collection of distributions $\{P_{Y|X=x}\}_{x\in\setF}$.}
For the case in which $\beta_{\alpha}(P_{Y|X=x},Q_Y)$ does not depend on $x\in \setF$,
by relaxing $M/2$ to $M$ in~\eqref{eq:kappa-beta-intro-avg} and by using~\cite[Lemma~29]{polyanskiy10-05} we recover~\eqref{eq:kappa-beta-relax}
under the weaker average-error-probability formalism.
However, for the case in which $\beta_{\alpha}(P_{Y|X=x},Q_Y)$  does depend on $x\in \setF$, the $\beta\beta$ achievability bound~\eqref{eq:kappa-beta-intro-avg}  can be both easier to analyze and numerically tighter than the $\kappa\beta$ bound~\eqref{eq:kappa-beta-relax} (see Section~\ref{sec:mimo-bf} for an example).

\subsubsection{The dependence-testing (DT) bound~\cite[Th.~18]{polyanskiy10-05}} Setting $Q_Y =P_Y$ in~\eqref{eq:kappa-beta-intro-avg}, using that $\beta_{\tau}(P_Y,P_Y) = \tau$, and rearranging terms, we obtain
\begin{equation}
\error \leq   \inf_{\alpha \in(0,1)} \Big\{ 1-\alpha + \frac{M}{2} \beta_{\alpha}(P_{XY},P_XP_Y) \Big\}.
\label{eq:dt-example}
\end{equation}
Further setting~$\alpha=P_{XY}[\log \big( \mathrm{d}P_{XY}/\mathrm{d}(P_XP_Y) \big)\geq \log (M/2)]$ and using the Neyman-Pearson lemma, we conclude that~\eqref{eq:dt-example} is equivalent to a   weakened version of the DT bound where $(M-1)/2$ is  replaced by $M/2$.
Since this weakened version of the DT bound implies both Shannon's bound~\cite{shannon57} and the bound in~\cite[Th.~2]{wang2009-07a}, the $\beta\beta$ achievability bound~\eqref{eq:kappa-beta-intro-avg}  implies these two bounds as well.

A cost-constrained version of the DT  bound in which all codewords  belong to a given set~$\setF$ can be found in~\cite[Th.~20]{polyanskiy10-05}.
A slightly weakened version of~\cite[Th.~20]{polyanskiy10-05} (with $(M-1)/2$ replaced by $M/2$) is
\begin{IEEEeqnarray}{rCl}
\error &\leq& Q_{XY}\lefto[\log\frac{\der Q_{XY}}{\der (Q_{X}Q_{Y})}(X,Y) \leq \log \frac{M}{2}\right] +  Q_X[\setF^c] \notag \\
\quad&& +\, \frac{M}{2}Q_{X}Q_{Y}\lefto[\log\frac{\der Q_{XY}}{\der(Q_{X}Q_{Y})}(X,Y) \geq \log \frac{M}{2}\right]   . \IEEEeqnarraynumspace
\label{eq:cost-constrained-DT}
\end{IEEEeqnarray}
Here, $Q_{XY} = Q_X P_{Y|X}$, and $Q_Y =  Q_X \circ P_{Y|X}$.
For $0<\error<1/2$, this bound can be derived from~\eqref{eq:kappa-beta-intro-avg} by choosing $P_X[\,\cdot\,] = Q_X[\, \cdot \, \cap \setF]/Q_X[\setF]$ and by using the following bounds:
\begin{IEEEeqnarray}{rCl}
&&\quad \beta_{\tau}(P_Y,Q_Y) \geq \beta_{\tau}(P_X,Q_X)  = \tau Q_X[\setF] \label{eq:beta-num-app1} \\
&&\beta_{1-\error +\tau}(P_{XY} , P_XQ_Y) \notag\\
  && \qquad\qquad\leq {1\over Q_X[\setF] } \beta_{1-(\error -\tau)Q_X[\setF]}(Q_{XY} , Q_XQ_Y) . \IEEEeqnarraynumspace
\label{eq:beta-denom-app1}
\end{IEEEeqnarray}
Here,~\eqref{eq:beta-num-app1} follows from the data-processing inequality for $\beta_{\tau}(\cdot, \cdot)$ (see, e.g.,~\cite[Sec. V]{polyanskiy10-09a}) and straightforward computations, and~\eqref{eq:beta-denom-app1} follows from~\cite[Lemma~25]{polyanskiy13}.

 \subsubsection{Refinements of the DT/Feinstein bound through change of measure}
The idea of bounding the probability $P_{Y}[\cdot]$ via a simpler-to-analyze $Q_Y[\cdot]$ has been applied previously in the literature to evaluate the DT bound and Feinstein's bound.
  %
For example, the following achievability bound is suggested in~\cite[Sec.~II]{molavianJazi15-12a}:
\begin{IEEEeqnarray}{rCl}
\error &\leq& \inf_{\gamma>0}   \bigg\{P_{XY}\lefto[  \frac{\der P_{XY}}{ \der (P_XQ_Y)} (X,Y) \leq  \gamma    \right] \notag\\
&& +\, M \left(\sup_{y} \frac{\mathrm{d}P_Y}{\mathrm{d}Q_Y}(y) \right)P_XQ_Y\lefto[ \frac{\der P_{XY}}{ \der (P_XQ_Y)} (X,Y) \geq  \gamma   \right]\bigg\}  \notag\\
\end{IEEEeqnarray}
which is equivalent to
 \begin{IEEEeqnarray}{rCl}
M \geq \sup\limits_{0<\tau<\error} \frac{\tau}{ \beta_{1-\error +\tau}(P_{XY}, P_XQ_Y)}  \left(\sup_{y} \frac{\mathrm{d}P_Y}{\mathrm{d}Q_Y}(y) \right)^{-1}. \IEEEeqnarraynumspace
\label{eq:bound-molavianjazi}
\end{IEEEeqnarray}
This bound can be obtained from the $\beta\beta$ achievability bound~\eqref{eq:kappa-beta-intro-avg} by using that
\begin{IEEEeqnarray}{rCl}
\beta_{\tau} (P_Y,Q_Y)  \geq \tau \left(\sup_y \frac{\mathrm{d}P_Y}{\mathrm{d}Q_Y}(y) \right)^{-1} .
\label{eq:lower-bound-PY-QY}
\end{IEEEeqnarray}
The following variation of the  Feinstein bound is used in~\cite[Lemma~2]{hoydis15-12} to deal with inputs subject to a cost constraint:
\begin{IEEEeqnarray}{rCl}
\error \leq \inf_{\gamma>0 ,\eta>0 } \bigg\{ P_{XY}\lefto[  \frac{\der P_{XY}}{ \der (P_XQ_Y)} (X,Y) \leq  \gamma \eta  \right]  + \frac{M}{\gamma}\notag\\
 + \,P_Y\lefto[\frac{\der P_Y }{ \der Q_Y}(Y) \geq \eta \right]\bigg\}.\,\,\qquad\quad  \IEEEeqnarraynumspace
\label{eq:var-feinstein}
\end{IEEEeqnarray}
 This bound can be obtained from~\eqref{eq:kappa-beta-intro-avg} by using~\cite[Eq.~(103)]{polyanskiy10-05}  to lower-bound $\beta_\tau(P_Y, Q_Y)$ and by using~\cite[Eq.~(102)]{polyanskiy10-05} to upper-bound~$\beta_\tau(P_{XY}, P_XQ_Y)$.

\section{Wideband Slope at Finite Blocklength}

\subsection{Background}
 Many communication systems (such as deep-space communication and sensor networks)  operate in the low-power regime, where both the spectral efficiency and the energy per bit are low.
As shown in~\cite{verdu02-06},  the key asymptotic performance metrics for additive noise channels  in the low-power regime are the normalized minimum energy per bit $\frac{\eb}{N_0}_{\min}$ (where~$N_0$ denotes the noise power per channel use) and the slope\footnote{The unit of $S_0$ is  $\mathrm{bits}/\mathrm{ch.}\,\mathrm{use}/(3\, \mathrm{dB})$.} $S_0$ of the function spectral efficiency versus energy per bit (in $\mathrm{dB}$)   at $\frac{\eb}{N_0}_{\min}$ (known as the \emph{wideband slope}).
These two quantities determine the first-order behavior of the minimum energy per bit $\eb^*(R)$ as a function of the spectral efficiency\footnote{We shall use the terms spectral efficiency and rate interchangeably.}~$R$ (bits/ch. use) in the limit $R\to 0$. Specifically,~\cite[Eq.~(28)]{verdu02-06}
\begin{IEEEeqnarray}{rCl}
10\log_{10} \frac{\eb^*(R) }{N_0}= 10\log_{10} \frac{\eb}{N_0}_{\min} + \frac{R}{S_0} 10\log_{10} 2 + \littleo(R),\notag\\
 \quad R\to 0 . \IEEEeqnarraynumspace
\label{eq:wideband-slope-capacity}
\end{IEEEeqnarray}
For a wide range of power-constrained channels including the AWGN channel and the fading channel with/without CSIR, it is well known that the minimum energy per bit satisfies~\cite{verdu02-06,shannon49,kennedy69}
\begin{equation}
 \frac{\eb}{N_0}_{\min} = \log_e 2 =-1.59 \, \mathrm{dB}.
 \label{eq:min-energy-per-bit-min}
\end{equation}
Furthermore, it is shown in~\cite{verdu02-06} that $S_0 = 2$ for AWGN channels, and $S_0 = 2\Ex{}{|H|^2}^2/\Ex{}{|H|^4}$ for stationary ergodic fading channels   with perfect CSIR, where $H$ is distributed as one of the fading coefficients.

The asymptotic expansion~\eqref{eq:wideband-slope-capacity} is derived in~\cite{verdu02-06} under the assumption that $n, k\to \infty$ with $k/n \to R$ and $\error \to 0$, where~$k$ denotes the number of information bits $k$ (i.e., $k = \log_2 M$).
Recently, it was shown in~\cite{polyanskiy11-08b} that the minimum energy per bit $\eb^*(k,\error)$ necessary to transmit a finite number $k$ of information bits over an AWGN channel  with error probability~$\error$ and with no constraint on the blocklength satisfies
\begin{IEEEeqnarray}{rCl}
\frac{\eb^*(k,\epsilon)}{N_0} =  \frac{\eb}{N_0}_{\min} + \sqrt{\frac{2\log_e 2}{k} }  Q^{-1}(\error) + \littleo\lefto(\frac{1}{\sqrt{k}}\right). \IEEEeqnarraynumspace
\label{eq:minimum-epb-awgn-yury}
\end{IEEEeqnarray}
%
%
Furthermore, it was shown in \cite{yang2016-to-appear}   that the expansion~\eqref{eq:minimum-epb-awgn-yury} is valid also for  block-memoryless Rayleigh-fading channels (perfect CSIR), provided that  $0<\error<1/2$.

In this section, we study  the tradeoff between energy per bit and spectral efficiency in the regime where not only  $k$ and~$\error$, but also the blocklength $n$ is finite.
In particular, we are interested in the minimum energy per bit $\eb^*(k,\error,R)$ necessary to transmit $k$ information bits with rate~$R$~bits/ch. use and with error probability not exceeding~$\error$.
The quantity  $\eb^*(R)$ in~\eqref{eq:wideband-slope-capacity} and  $\eb^*(k,\epsilon)$ in~\eqref{eq:minimum-epb-awgn-yury} can be obtained from $\eb^*(k,\error,R)$ as follows:
\begin{IEEEeqnarray}{rCl}
\eb^*(R) &=&  \lim\limits_{\error \to 0} \lim\limits_{k\to\infty} \eb^*(k,\error,R)\\
\eb^*(k,\epsilon) &=& \lim\limits_{R \to 0} \eb^*(k,\error,R).
\end{IEEEeqnarray}

\subsection{AWGN Channel}
We consider the complex-valued AWGN channel
\begin{equation}
Y_i = X_i + Z_i,\quad i=1,\ldots, n
\label{eq:channel-io-awgn}
\end{equation}
where $Z^n \sim  \jpg(0, N_0 \matI_n)$.
%
We assume that every codeword~$x^n$ satisfies the power constraint
\begin{equation}
\|x^n\|^2 =  \sum\limits_{i=1}^{n} x_i^2 \leq n \snr .
\label{eq:power-constraint-nongaussian}
\end{equation}
For notational convenience, we shall set $N_0 =1$.
Hence, $P$ in~\eqref{eq:power-constraint-nongaussian} can be interpreted as  the SNR at the receiver.

We next evaluate $\eb^*(k,\epsilon, R)$ in the asymptotic regime $k\to\infty$ and $R\to 0$.
Motivated by~\eqref{eq:wideband-slope-capacity}, we shall approximate $\eb^*(k,\epsilon, R)$ (expressed in $\mathrm{dB}$) by an affine function of~$R$.
The~$\beta\beta$ bounds  turn out to be key tools to derive the asymptotic approximation given in Theorem~\ref{thm:energy-per-bit} below.

\begin{thm}
\label{thm:energy-per-bit}
Consider the AWGN channel~\eqref{eq:channel-io-awgn}.
The following expansion for $\eb^*(k,\epsilon, R)$ holds in the asymptotic limit~$k\to\infty$, $R\to 0$, and $k R \to \infty$:
\begin{IEEEeqnarray}{rCl}
\frac{\eb^*(k,\epsilon, R)}{N_0} =  \log_e 2 + \sqrt{\frac{2\log_e 2}{k} }  Q^{-1}(\error) + \frac{\log_e^2 2}{2} R\IEEEeqnarraynumspace \notag\\
 +\, \littleo\lefto(\!\frac{1}{\sqrt{k}}\!\right)+  o(R). \IEEEeqnarraynumspace
\label{eq:wideband-awgn-thm-linear}
\end{IEEEeqnarray}
Equivalently,
\begin{IEEEeqnarray}{rCl}
\IEEEeqnarraymulticol{3}{l}{
10\log_{10} \frac{\eb^*(k,\epsilon, R)}{N_0} }\notag\\
\qquad &=&  \underbrace{ 10\log_{10} {\frac{\eb}{N_0}}_{\min}+ \frac{R}{S_0} 10\log_{10} 2 }_{\text{wideband slope approximation}}{}  \notag\\
&&+\,    \frac{10\log_{10} e}{\sqrt{\log_e 2}}  \sqrt{\frac{2}{k  } }  Q^{-1}(\error) + \littleo\lefto(\!\frac{1}{\sqrt{k}}\!\right)+  o(R)   \IEEEeqnarraynumspace \label{eq:wideband-awgn-thm-db}
\end{IEEEeqnarray}
where  $10\log_{10} {\frac{\eb}{N_0}}_{\min} = -1.59\,\mathrm{dB}$, and $S_0 = 2$.
If $k\to\infty$, $R\to 0$, but  $kR\to c <\infty$, then we have
\begin{IEEEeqnarray}{rCl}
\frac{\eb^*(k,\epsilon, R)}{N_0} =  \log_e 2 + \sqrt{\frac{2\log_e 2}{k} }  Q^{-1}(\error) + \littleo\lefto(\!\frac{1}{\sqrt{k}}\!\right) . \IEEEeqnarraynumspace
\label{eq:wideband-awgn-thm-linear-boundedkr}
\end{IEEEeqnarray}
Equivalently,
\begin{IEEEeqnarray}{rCl}
\IEEEeqnarraymulticol{3}{l}{
10\log_{10} \frac{\eb^*(k,\epsilon, R)}{N_0} }\notag\\
&=&  10\log_{10} {\frac{\eb}{N_0}}_{\min}   +    \frac{10\log_{10} e}{\sqrt{\log_e 2}}  \sqrt{\frac{2}{k  } }  Q^{-1}(\error) + \littleo\lefto(\!\frac{1}{\sqrt{k}}\!\right) . \IEEEeqnarraynumspace
\end{IEEEeqnarray}
\end{thm}
\begin{IEEEproof}
See Appendix~\ref{app:proof-thm-beta-awgn-asy}.
\end{IEEEproof}

 \begin{rem}
The expansion~\eqref{eq:wideband-awgn-thm-db} can be seen as the finite-blocklength counterpart of~\eqref{eq:wideband-slope-capacity}.
Indeed, comparing~\eqref{eq:wideband-awgn-thm-db} with~\eqref{eq:wideband-slope-capacity}, we see that
\begin{IEEEeqnarray}{rCl}
  \IEEEeqnarraymulticol{3}{l}{
  10\log_{10} \frac{\eb^*(k, \error, R )}{\eb^*(R)} }\notag\\
     \quad &=&  \frac{10\log_{10} e}{\sqrt{ \log_e 2}}  \sqrt{\frac{2}{k} }  Q^{-1}(\error) + \littleo(R) + \littleo\lefto(\!\frac{1}{\sqrt{k}}\!\right) . \IEEEeqnarraynumspace
\end{IEEEeqnarray}
Thus, in the low spectral efficiency regime,  the gap in $\mathrm{dB}$ between $\eb^*(k, \error, R) $ and the asymptotic limit $\eb^*(R) $ (obtained as $k\to\infty$ and $\error \to 0$) is---up to first order---proportional to $1/\sqrt{k}$ and independent of $R$.
Furthermore,~\eqref{eq:wideband-awgn-thm-linear} and~\eqref{eq:minimum-epb-awgn-yury} imply that
\begin{IEEEeqnarray}{rCl}
10\log_{10} \frac{\eb^*(k, \error, R )}{\eb^*(k, \error )} =  \frac{R}{S_0} 10\log_{10} 2 +  \littleo(R) + \littleo\lefto(\!\frac{1}{\sqrt{k}}\!\right). \IEEEeqnarraynumspace
\end{IEEEeqnarray}
\end{rem}
Thus, in the regime of large $k$, the gap in $\mathrm{dB}$ between ${\eb^*(k, \error, R )} $ and the asymptotic limit ${\eb^*(k, \error)}$ is---up to first order---proportional to $R$ and is independent of $k$.

 \begin{rem}
 The result~\eqref{eq:wideband-awgn-thm-linear-boundedkr} implies that the minimum energy per bit $\eb^*(k, \error) $ in~\eqref{eq:minimum-epb-awgn-yury} with no blocklength constraint can be achieved with codes of rate $1/k$, or equivalently, with codes of blocklength $k^2$.
 For comparison,  the code used in~\cite{polyanskiy11-08b}  to achieve~\eqref{eq:minimum-epb-awgn-yury} has blocklength $2^k$.
\end{rem}

We now provide some intuition for the expansion~\eqref{eq:wideband-awgn-thm-linear} (a rigorous proof is given in Appendix~\ref{app:proof-thm-beta-awgn-asy}).
The  asymptotic expression~\eqref{eq:wideband-slope-capacity} relies on the following Taylor-series expansion  of the channel capacity $C(\snr)$ as a function of the SNR $\snr$ when $\snr \to 0$:
\begin{IEEEeqnarray}{rCl}
C(P) = C'(0) P \log e + \frac{C''(0)}{2} P^2 \log e +o(P^2).\IEEEeqnarraynumspace
\label{eq:capacity-expansion-snr}
\end{IEEEeqnarray}
Here, $C'(0)$ and $C''(0)$ denote the first and second derivative, respectively, of the function $C(P)$ (in nats per channel use).
In particular, the first term in~\eqref{eq:capacity-expansion-snr}  determines  the minimum energy per bit $ \frac{\eb}{N_0}_{\min}$ and the second term in~\eqref{eq:capacity-expansion-snr} yields the wideband slope $S_0$~\cite[Eq.~(35) and Th.~9]{verdu02-06}:
\begin{equation}
 \frac{\eb}{N_0}_{\min} = \frac{\log_e 2}{C'(0)} ,\qquad
 S_0 = \frac{2\big[C'(0)\big]^2}{-C''(0)}. \label{eq:compute-ebno-s0}
\end{equation}
Both $ \frac{\eb}{N_0}_{\min} $ and $S_0$ in~\eqref{eq:compute-ebno-s0}  can be computed directly, without the knowledge of $C(\snr)$.
Indeed, set
$Q_{Y} = P_{Y|X=0}$  in the  golden formula~\eqref{eq:channel-capacity-alt}.
One can show that~\cite[Eq.~(41)]{verdu02-06}
\begin{IEEEeqnarray}{rCl}
C'(0) =  \frac{1}{\log e} \lim\limits_{P\to 0}\max_{P_X: \Ex{}{|X|^2}= P} \frac{D(P_{Y|X} \| Q_Y\| P_X)}{P} \label{eq:asy-expan-dpyqy-00} \IEEEeqnarraynumspace
\end{IEEEeqnarray}
and that for both AWGN channels and for fading channels with perfect CSIR,
\begin{equation}
  C''(0)   =  -\frac{2}{\log e}  \lim\limits_{P \to 0} \min_{P_X} \frac{ D(P_{Y} \| Q_{Y}) }{ P^2 }\label{eq:asy-expan-dpyqy}
\end{equation}
where the minimization in~\eqref{eq:asy-expan-dpyqy} is over all $P_X$ that achieve $C'(0)$ in~\eqref{eq:asy-expan-dpyqy-00} and that satisfy $\Ex{}{|X|^2} = P$.
In other words, $C'(0)$ is determined solely by $D(P_{Y|X} \| Q_{Y} |  P_{X})$, and  $C''(0)$ is determined solely by $D(P_{Y} \| Q_{Y}) $.

Moving to the nonasymptotic case, let $R^*(n,\error,\snr)$ be the maximum coding rate for a given blocklength $n$, error probability $\error$, and SNR $\snr$.
Then,~\eqref{eq:wideband-awgn-thm-linear} turns out to be equivalent to the following asymptotic expression (see Appendix~\ref{app:equivalence-rate-energybit} for the proof of this equivalence):
\begin{IEEEeqnarray}{rCl}
\frac{R^*(n,\epsilon,\snr)}{\log e} &=&  \snr - \sqrt{ \frac{2 \snr}{n}}Q^{-1}(\error) - \frac{1}{2}\snr^2
  \notag  \\
&& +\, \littleo\mathopen{}\bigg(\sqrt{\frac{\snr}{n}}\bigg) + \littleo\lefto(\snr^2\right) \IEEEeqnarraynumspace
\label{eq:asy-expansion-r-awgn-eq}
\end{IEEEeqnarray}
as $n\to\infty$, $\snr \to 0$, and $nP^2 \to \infty$.
 In view of~\eqref{eq:asy-expan-dpyqy-00} and \eqref{eq:asy-expan-dpyqy}, it is natural to use  the $\beta\beta$ bounds~\eqref{eq:prop-lb-beta-p-q-intro} and~\eqref{eq:kappa-beta-intro-avg}, since they are nonasymptotic versions of the golden formula. Indeed, we obtain from~\eqref{eq:prop-lb-beta-p-q-intro} and~\eqref{eq:kappa-beta-intro-avg} that
\begin{IEEEeqnarray}{rCl}
R^*(n,\epsilon,P) \approx \max_{P_{X^n}} \frac{1}{n} \Big( && -\log \beta_{1-\epsilon}(P_{X^nY^n}, P_{X^n}Q_{Y^n} ) \notag\\
&&  \qquad\quad\,\, + \,\log \beta_{\alpha} (P_{Y^n},Q_{Y^n}) \Big). \IEEEeqnarraynumspace
\label{eq:beta-beta-rate-exp-11}
\end{IEEEeqnarray}
Next, we choose $Q_{Y^n} = (P_{Y |X=0})^{n}$.
The analysis in~\cite[pp.~4882--4883]{polyanskiy11-08b} implies that
\begin{IEEEeqnarray}{rCl}
\IEEEeqnarraymulticol{3}{l}{
 \max_{P_{X^n}}  \Big(\!-\frac{1}{n}  \log \beta_{1-\epsilon}(P_{X^nY^n} , P_{X^n} Q_{Y^n} )\Big)}\notag\\
 \quad & \approx& \snr \log e - \sqrt{ \frac{2 \snr}{n}}Q^{-1}(\error)  \log e \IEEEeqnarraynumspace
\label{eq:beta-tau-heur-2}
\end{IEEEeqnarray}
which yields the first two terms in~\eqref{eq:asy-expansion-r-awgn-eq}.

Furthermore, one can show through a large-deviation analysis that,
\begin{equation}
\max_{P_{X^n}} \frac{1}{n} \log \beta_{\alpha} (P_{Y^n},Q_{Y^n}) \approx  - D(P_{Y}^*\| Q_Y) \approx  -\frac{\log e}{2} P^2  .
\label{eq:beta-tau-heur}
\end{equation}
Here, the maximization in~\eqref{eq:beta-tau-heur} is taken with respect to all input distributions $P_{X^n}$ for which   $-\frac{1}{n}  \log \beta_{1-\epsilon}(P_{X^nY^n} , P_{X^n} Q_{Y^n} ) $ is close to the RHS of~\eqref{eq:beta-tau-heur-2}.
Substituting~\eqref{eq:beta-tau-heur-2} and~\eqref{eq:beta-tau-heur} into~\eqref{eq:beta-beta-rate-exp-11}, we  recover the dominant terms in~\eqref{eq:asy-expansion-r-awgn-eq}.

One may attempt to establish~\eqref{eq:asy-expansion-r-awgn-eq} by using the normal approximation~\cite{polyanskiy10-05}
\begin{IEEEeqnarray}{rCl}
R^*(n,\error,P) = C(P) - \sqrt{\frac{V(P)}{n}}Q^{-1}(\error) + \littleo\lefto(\frac{1}{\sqrt{n}}\right) \IEEEeqnarraynumspace
\label{eq:na-awgn}
\end{IEEEeqnarray}
and  then by Taylor-expanding $C(P)$ and $V(P)$ for $P\to 0$.
However, there are two major drawbacks to this approach.
First, establishing the normal approximation~\eqref{eq:na-awgn}  is challenging for fading channels (see~\cite{collins2016-07a}  and the remarks after Theorem~\ref{thm:wideband-slope-fading}). So this approach would work only in the AWGN case.
Second,  one   needs to verify that the $ \littleo(1/\sqrt{n})$ term in~\eqref{eq:na-awgn} is uniform in $P$, which is nontrivial.

\subsection{Rayleigh-Fading Channels With Perfect CSIR}
\label{sec:Rayleigh-ebno}
We next consider the Rayleigh-fading channel
\begin{equation}
Y_i= H_i X_i + Z_i,\quad i=1,\ldots,n
\label{eq:rayleigh-block-fading-io-siso}
\end{equation}
where both $\{H_i\}$ and $\{Z_i\}$ are independent and identically distributed (i.i.d.)  $\jpg(0,1)$ random variables. We assume that the channel coefficients  $\{H_i\}$ are  known to the receiver but not to the transmitter.
Furthermore, we assume that  every codeword  $x^n$ satisfies the power constraint~\eqref{eq:power-constraint-nongaussian}.
The wideband slope of this channel is~\cite[Eq.~(208)]{verdu02-06}
\begin{IEEEeqnarray}{rCl}
S_0 = \frac{2 \Ex{}{|H|^2}^2}{\Ex{}{|H|^4} }= 1
 \end{IEEEeqnarray}
where $H\sim \jpg(0,1)$.

Theorem~\ref{thm:wideband-slope-fading} below characterizes the minimum energy per bit $\eb^*(k,\epsilon, R)$ for the Rayleigh-fading channel in the asymptotic limit $k\to\infty$ and $R\to 0$.


\begin{thm}
\label{thm:wideband-slope-fading}
Consider the Rayleigh block-fading channel~\eqref{eq:rayleigh-block-fading-io-siso}. The following expansion for $\eb^*(k,\epsilon, R)$ holds in the asymptotic limit~$k\to\infty$, $R\to 0$, and $k R \to \infty$:
\begin{IEEEeqnarray}{rCl}
\frac{\eb^*(k,\epsilon, R)}{N_0} &=&  \log_e 2 + \sqrt{\frac{2\log_e 2}{k} }  Q^{-1}(\error) + (\log_e^2 2)  R \IEEEeqnarraynumspace \notag\\
&&+ \, \littleo\lefto(\!\frac{1}{\sqrt{k}}\!\right)+  o(R)
\label{eq:wideband-fading-thm-linear}
\end{IEEEeqnarray}
or, equivalently,
\begin{IEEEeqnarray}{rCl}
\IEEEeqnarraymulticol{3}{l}{
10\log_{10} \frac{\eb^*(k,\epsilon, R)}{N_0}}\notag\\
 &=&  \underbrace{  10 \log_{10} \frac{\eb}{N_0}_{\min} + \frac{R}{S_0} 10\log_{10} 2 }_{\text{wideband slope approximation}}{} \notag\\
&&+\,    \frac{10\log_{10} e}{\sqrt{\log_e 2}}  \sqrt{\frac{2}{k  } }  Q^{-1}(\error) + \littleo\lefto(\!\frac{1}{\sqrt{k}}\!\right)+  o(R)   \IEEEeqnarraynumspace
\label{eq:wideband-fading-thm-db}
\end{IEEEeqnarray}
where  $10\log_{10} {\frac{\eb}{N_0}}_{\min} = -1.59\,\mathrm{dB}$, and $S_0 = 1$.
If $k\to\infty$, $R\to 0$, but $kR\to c <\infty$, then we have
\begin{IEEEeqnarray}{rCl}
\frac{\eb^*(k,\epsilon, R) }{N_0}=  \log_e 2 + \sqrt{\frac{2\log_e 2}{k} }  Q^{-1}(\error) + \littleo\lefto(\!\frac{1}{\sqrt{k}}\!\right) \IEEEeqnarraynumspace
\label{eq:wideband-fading-thm-linear-boundedkr}
\end{IEEEeqnarray}
or, equivalently,
\begin{IEEEeqnarray}{rCl}
\IEEEeqnarraymulticol{3}{l}{
10\log_{10} \frac{\eb^*(k,\epsilon, R)}{N_0} }\notag\\
&=&  10 \log_{10} \frac{\eb}{N_0}_{\min}   \! +    \frac{10\log_{10} e}{\sqrt{\log_e 2}}  \sqrt{\frac{2}{k  } }  Q^{-1}(\error) + \littleo\lefto(\!\frac{1}{\sqrt{k}}\!\right) . \IEEEeqnarraynumspace
\end{IEEEeqnarray}
\end{thm}
\begin{IEEEproof}
See Appendix~\ref{app:wideband-slope-rayleigh}.
\end{IEEEproof}

A few remarks are in order:
\begin{itemize}
\item As in the AWGN case,  the minimum energy per bit $\eb^*(k,\error, R)$ over the Rayleigh-fading channel~\eqref{eq:rayleigh-block-fading-io-siso} with perfect CSIR satisfies
\begin{IEEEeqnarray}{rCl}
 10\log_{10} \frac{\eb^*(k, \error, R )}{\eb^*(R)}  &=&    \frac{10\log_{10} e}{\sqrt{ \log_e 2}}  \sqrt{\frac{2}{k} }  Q^{-1}(\error) \notag\\
  && +\, \littleo(R) + \littleo\lefto(\!\frac{1}{\sqrt{k}}\!\right) . \IEEEeqnarraynumspace
\label{eq:wide-band-slope-sergio-fading-diff}
\end{IEEEeqnarray}
%
Again we observe that, in the low spectral efficiency regime, the gap in $\mathrm{dB}$ between $\eb^*(k, \error, R)$ and the asymptotic limit $\eb^*(R) $ is---up to first order---proportional to $1/\sqrt{k}$ and is independent of~$R$.
Furthermore, the gap in the fading case coincides with that  in the AWGN case  up to $\littleo(R) + \littleo(1/{\sqrt{k}})$ terms.

\item Unlike the asymptotic wideband approximation~\eqref{eq:wideband-slope-capacity}, which holds for all fading distributions (see~\cite{verdu02-06}), our result in Theorem~\ref{thm:wideband-slope-fading} relies on the Gaussianity of the fading coefficients, and does not necessarily hold for other fading distributions. In fact,  as observed in~\cite[Sec.~III.D]{yang2016-to-appear}, there are fading distributions for which the minimum energy per bit $\eb^*(k,\epsilon,R)$ does not converge to $-1.59\, \mathrm{dB}$ when $k\to\infty$, $R\to 0$, and $\error$ is fixed.

\item For the case of nonvanishing rate (or, equivalently, nonvanishing SNR $P$), a normal approximation for the maximum rate $R^*(n,P,\error)$ achievable over the channel~\eqref{eq:rayleigh-block-fading-io-siso} when CSIR is available is reported in~\cite{collins2016-07a}. This approximation relies on  the additional constraint that every codeword $x^n$ satisfies $\|x^n\|_\infty = \littleo(n^{1/4})$.
In contrast, Theorem~\ref{thm:wideband-slope-fading} does not require this additional constraint.

\item  One of the challenges that one has to address when establishing a nonasymptotic converse bound on $\eb^*(k,\epsilon,R)$ is that the variance of the information density $i(x^n;Y^nH^n)$ depends on $\|x^n\|_4$ (see~\cite[Eqs.~(47)--(52)]{polyanskiy11-08a}).
In order to obtain a tight converse bound  on $\eb^*(k,\epsilon,R)$ for   fixed   $R$, one needs to expurgate the codewords  whose $\ell_4$ norm $\|x^n\|_4$  is far from that of the codewords of a  Gaussian code (see~\cite{polyanskiy11-08a} and~\cite{collins2016-07a}).
However, in the limit $R\to 0$ of interest in this paper, this expurgation procedure is not needed since the dominant term in the asymptotic expansion of the variance of  $i(x^n;Y^nH^n)$ does not depend on $\|x^n\|_4$. Furthermore, the wideband slope is also insensitive to $\|x^n\|_4$.
Indeed, to achieve the wideband slope of a fading channel with perfect CSIR, QPSK inputs are as good as Gaussian inputs~\cite{verdu02-06}.
\end{itemize}

\begin{figure*}
\centering
	\begin{tikzpicture}
		\begin{axis}[
			xlabel={\footnotesize $E_{\mathrm{b}}/N_0 $, $\mathrm{dB}$},
			ylabel={\footnotesize Spectral efficiency, bits/ch. use},
xmin=-1.7, xmax=0,
ymin=0, ymax=0.4,
grid=major
            ]
            \addplot[blue, thick] table {rate_ach.dat};
		\addplot[red,thick] table {rate_conv.dat};
\addplot[black,thick] table {wsfbl.dat};

\addplot[black,thick] table {wbs.dat};
\addplot[black,thick] table {capa.dat};
\draw (-1.04, 0.35) node[right]{\footnotesize Capacity};
\draw (-1.2, 0.37) node[left]{\footnotesize  Wideband};
\draw (-1.08, 0.35) node[left]{\footnotesize approximation~\eqref{eq:wideband-slope-capacity}};
%
\draw (-1.08,0.07) node[left]{\footnotesize Converse};
\draw[line width = 0.2mm, <-] (-0.98, 0.09) -- (-1.09,0.07);
\draw (-0.83, 0.05) node[right]{\footnotesize Achievability};
\draw[line width = 0.2mm, <-] (-0.96, 0.07) -- (-0.82, 0.05);
\draw (-0.78, 0.1) node[right]{\footnotesize Approximation~\eqref{eq:wideband-awgn-thm-db}};
\draw[line width = 0.2mm, <-] (-0.91, 0.121) -- (-0.76, 0.1);

		\end{axis}
	\end{tikzpicture}
\caption{Minimum energy per bit versus spectral efficiency of the AWGN channel; here, $k=2000$ bits, and $\error =10^{-3}$.\label{fig:minimum-energy-per-bit}}
\end{figure*}
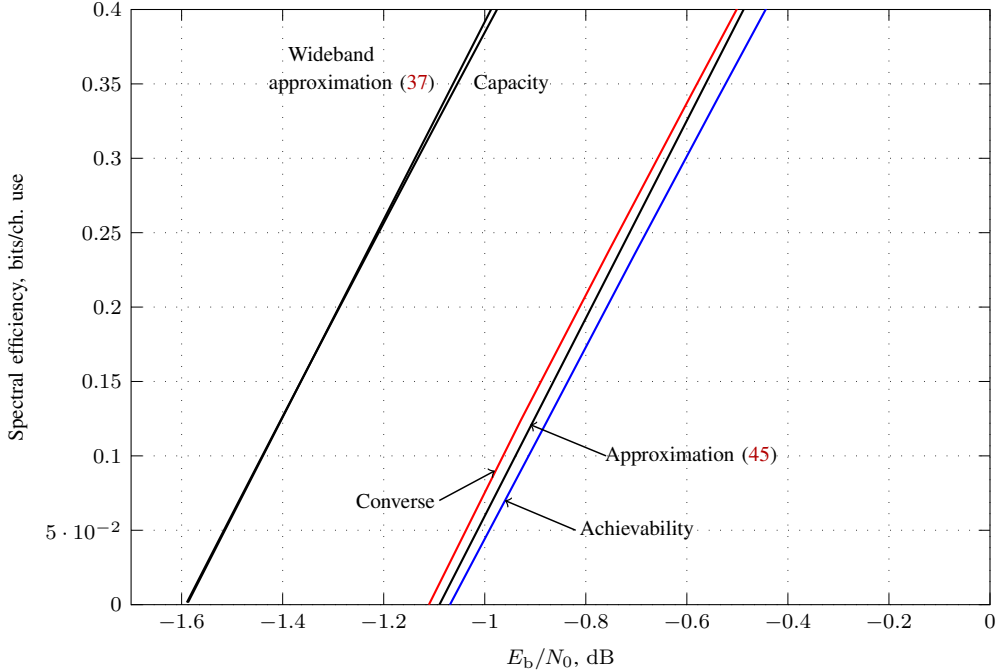

\subsection{Numerical results}
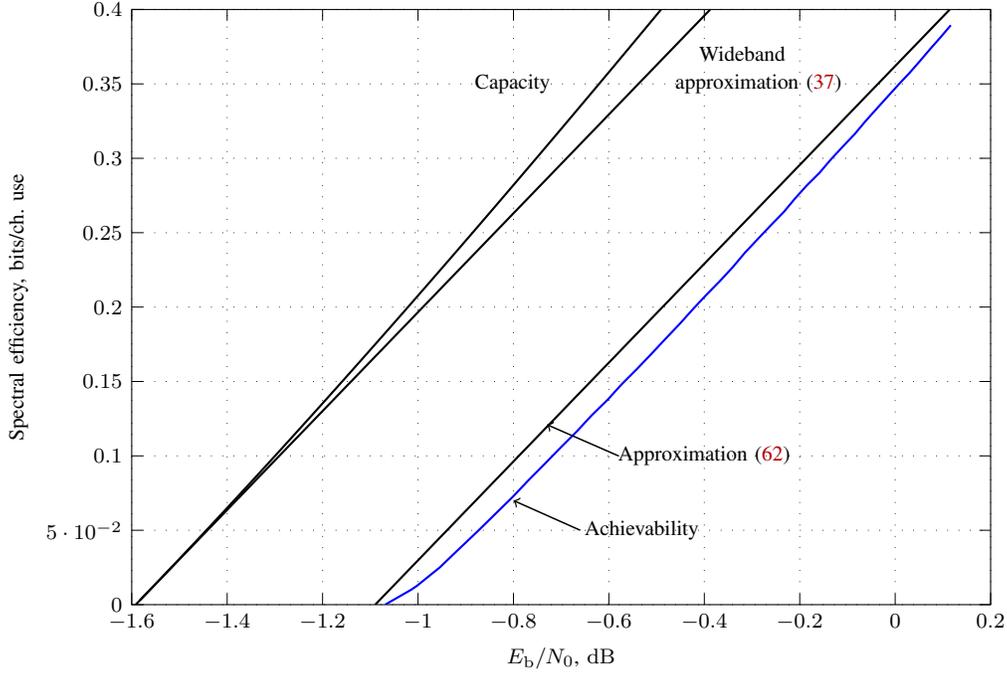
\begin{figure*}[t]
\centering

	\begin{tikzpicture}
		\begin{axis}[
			xlabel={\footnotesize $E_{\mathrm{b}}/N_0 $, $\mathrm{dB}$},
			ylabel={\footnotesize Spectral efficiency, bits/ch. use},
xmin=-1.6, xmax=0.2,
ymin=0, ymax=0.4,
grid=major,
            ]

\addplot[black,thick] table {capa_fading.dat};

\addplot[blue, thick] table {ach_fad.dat};
\addplot[black, thick] coordinates {
(-1.5917, 0)
( -0.9897, 0.2)
(-0.3876, 0.4)
};

\addplot[black, thick] coordinates {
(-1.0897, 0)
( -0.4876, 0.2)
(0.1144, 0.4)
};

\draw (-0.9, 0.35) node[right]{\footnotesize Capacity};
\draw (-0.43, 0.37) node[right]{\footnotesize Wideband};
\draw (-0.48, 0.35) node[right]{\footnotesize approximation~\eqref{eq:wideband-slope-capacity}};
\draw (-0.67, 0.05) node[right]{\footnotesize Achievability};
\draw[line width = 0.2mm, <-] (-0.8, 0.07) -- (-0.66, 0.05);
\draw (-0.6, 0.1) node[right]{\footnotesize Approximation~\eqref{eq:wideband-fading-thm-db}};
\draw[line width = 0.2mm, <-] (-0.73, 0.121) -- (-0.58, 0.1);

		\end{axis}
	\end{tikzpicture}

\caption{Minimum energy per bit versus spectral efficiency of the Rayleigh-fading channel~\eqref{eq:rayleigh-block-fading-io-siso}  with perfect CSIR; here, $k=2000$ bits, $\error =10^{-3}$. \label{fig:minimum-energy-per-bit-Rayleigh}}
\end{figure*}

In Fig.~\ref{fig:minimum-energy-per-bit}, we present a comparison between the approximation~\eqref{eq:wideband-awgn-thm-db} (with the two $\littleo(\cdot)$ terms omitted),  the $\beta\beta$ achievability bound~\eqref{eq:kappa-beta-intro-avg}, and the $\beta\beta$ converse bound~\eqref{eq:prop-lb-beta-p-q-intro}.
 In both the achievability and the converse bound, $Q_{Y^n}$ is chosen to be the product distribution obtained from the capacity-achieving output distribution of the channel~\eqref{eq:channel-io-awgn} (i.e., $Q_{Y^n} = \jpg(0, (1+P)\matI_n)$).
For the parameters considered in Fig.~\ref{fig:minimum-energy-per-bit},  i.e., $k=2000$ bits and $\error=10^{-3}$, the approximation~\eqref{eq:wideband-awgn-thm-db} is accurate.
Fig.~\ref{fig:minimum-energy-per-bit-Rayleigh} provides a similar comparison for the Rayleigh fading channel~\eqref{eq:rayleigh-block-fading-io-siso}.
 In this case, the $\beta\beta$ converse bound is difficult to compute (due to the need to perform an optimization over all  input distributions), and is not plotted.

\section{Other Applications of Theorem~\ref{thm:betabeta-bound}}

 \subsection{The Additive Exponential-Noise Channel}
Consider the additive exponential-noise channel
\begin{equation}
Y_i = X_i + Z_i, \quad i=1, \ldots, n
\label{eq:channel-io-nong}
\end{equation}
where $\{Z_i\}$ are i.i.d. $\mathrm{Exp}(1)$-distributed.
As in~\cite{verdu96-01a}, we require each codeword $x^n \in \realset^n$ to satisfy
\begin{equation}
 x_i \geq 0, \,\, i=1,\ldots,n,   \,\,\, \text{ and } \sum\limits_{i=1}^{n} x_i \leq n\sigma.
\label{eq:exp-cost-constraint}
\end{equation}
The additive exponential-noise channel~\eqref{eq:channel-io-nong} can be used to model a communication system where information is conveyed through the arrival times of packets, and where each packet goes through a  single-server queue with exponential service time~\cite{anantharam1996-01a}.
  It also models a rapidly-varying phase-noise channel combined with an energy detector at the receiver~\cite{martinez2011-06a}.
The capacity of the channel~\eqref{eq:channel-io-nong} under the input constraints specified in~\eqref{eq:exp-cost-constraint} is given by~\cite[Th.~3]{verdu96-01a}
\begin{equation}
C(\sigma) = \log(1+\sigma)
\end{equation}
and is achieved by the input distribution
$P^*_X$ according to which $X$ takes the value zero with probability $1/(1+\sigma)$ and follows an $\mathrm{Exp}(1+\sigma)$ distribution conditioned on it being positive.
Furthermore, the capacity-achieving output distribution is $\mathrm{Exp}(1+\sigma)$.
A discrete counterpart of the exponential-noise channel is studied in~\cite{riedl2011-10a}, where a lower bound on the maximum  coding rate is derived.

Theorem~\ref{thm:exp-noise} below characterizes the   dispersion of the channel~\eqref{eq:channel-io-nong}.

\begin{thm}
\label{thm:exp-noise}
Consider the additive exponential-noise channel~\eqref{eq:channel-io-nong}  subject to the  constraint~\eqref{eq:exp-cost-constraint}. For every $0<\error<1$, the maximum coding rate $R^*(n,\error)$ admits the following expansion:
\begin{equation}
R^*(n,\error) = \log(1+\sigma) -\sqrt{\frac{V(\sigma)}{n}}Q^{-1}(\error) + \bigO\lefto(\frac{\log n}{n}\right)
\label{eq:rate-thm-exp}
\end{equation}
where
\begin{equation}
V(\sigma) = \frac{\sigma^2}{(1+\sigma)^2} \log^2 e.
\end{equation}
\end{thm}

\begin{IEEEproof}
We first prove that~\eqref{eq:rate-thm-exp} is achievable using the $\beta\beta$ achievability bound in Theorem~\ref{thm:betabeta-bound}.
Let $Q_{X^n}= (P_{X}^*)^n$, where $P_X^*$ is the capacity-achieving input distribution.
Let  $P_{X^n}$ be the conditional distribution of $X^n\sim Q_{X^n}$ given that $X^n$ belongs to the set $\setF$ specified below:
\begin{equation}
\setF = \big\{ x^n\in \realset^n: x_i \geq 0, \, n\sigma - \log n \leq  \sum_{i=1}^{n} x_i \leq n\sigma \Big\}.
\label{eq:cost-constraint22}
\end{equation}
By construction, $X^n\sim P_{X^n}$ satisfies the constraint~\eqref{eq:exp-cost-constraint}.
Finally, let $P_{Y^n } \define  P_{X^n} \circ P_{Y^n\given X^n} $ and let $Q_{Y^n} \define Q_{X^n} \circ P_{Y^n\given X^n}$. %
We apply the $\beta\beta$ bound in Theorem~\ref{thm:betabeta-bound}  with $\tau = 1/\sqrt{n}$ and with $P_{X^n}$ and $Q_{Y^n}$ chosen as above.
The term $\beta_{\tau}(P_{Y^n},Q_{Y^n})$ can be evaluated as follows:
\begin{IEEEeqnarray}{rCl}
\IEEEeqnarraymulticol{3}{l}{
 \log \beta_{\tau}(P_{Y^n},Q_{Y^n}) } \notag\\
&\geq &  \log Q_{X^n}\mathopen{}\left[ \setF \right] + \log \tau  \label{eq:lower-bound-beta-2} \IEEEeqnarraynumspace\\
&=& \log \lefto(Q\mathopen{}\left(\frac{ - \log n}{\sqrt{n \mathbb{V}\mathrm{ar}_{P_X^*}[X]}}\right)  - Q(0) - \bigO\lefto(\frac{1}{\sqrt{n}} \right)\!\right) +\log \tau \notag\\
&&\label{eq:lower-bound-beta-3} \IEEEeqnarraynumspace\\
&=&  \bigO(\log n ) \label{eq:lower-bound-beta-4}.
\end{IEEEeqnarray}
Here,~\eqref{eq:lower-bound-beta-2}  follows from~\eqref{eq:beta-num-app1},
and~\eqref{eq:lower-bound-beta-3} follows from the Berry-Ess\'{e}en theorem (see, e.g.,~\cite[Sec. XVI.5]{feller70a}).

We next evaluate $\beta_{1-\error+\tau} (P_{X^n Y^n},P_{X^n} Q_{Y^n})$. It follows from~\cite[Eq.~(103)]{polyanskiy10-05} that
\begin{IEEEeqnarray}{rCl}
\beta_{1-\error+\tau} (P_{X^n Y^n},P_{X^n} Q_{Y^n}) \leq \exp(-\gamma_n)
\label{eq:ub-beta-exp-noise}
\end{IEEEeqnarray}
where $\gamma_n$ satisfies
\begin{equation}
P_{X^nY^n} \lefto[ \log \frac{\der P_{X^n Y^n} }{\der (P_{X^n}Q_{Y^n})} (X^n, Y^n) \geq \gamma_n\right] \geq 1-\error +\tau.
\label{eq:exp-noise-beta-ach}
\end{equation}
Note that, under $P_{X^nY^n}$, the random variable $\log\frac{\der P_{X^n Y^n} }{\der (P_{X^n}Q_{Y^n})} (X^n, Y^n)$ has the same distribution as
\begin{IEEEeqnarray}{rCl}
n\log (1+ \sigma)   + \frac{\log e}{1+\sigma}\sum\limits_{i=1}^{n} X_i - \frac{\sigma\log e}{1+\sigma}\sum\limits_{i=1}^{n} Z_i  \IEEEeqnarraynumspace
\label{eq:info-density-dist-exp}
\end{IEEEeqnarray}
where $Z_i$ are i.i.d. $\mathrm{Exp}(1)$-distributed.
This random variable depends on the codeword $X^n$ only through $\sum\nolimits_{i=1}^{n} X_i$.
Furthermore, given $\sum\nolimits_{i=1}^{n} X_i$, this random variable is the sum of $n$ i.i.d. random variables.
Using  the Berry-Ess\'{e}en theorem and~\eqref{eq:cost-constraint22} to evaluate the left-hand side (LHS) of~\eqref{eq:exp-noise-beta-ach}, we conclude that
\begin{IEEEeqnarray}{rCl}
\IEEEeqnarraymulticol{3}{l}{
P_{X^nY^n} \lefto[\log\frac{\der P_{X^n Y^n} }{\der P_{X^n}Q_{Y^n}} (X^n, Y^n)    \geq  \gamma_n\right] }\notag\\
&\geq & \prob\lefto[\frac{\sigma \log e }{1+ \sigma } \!\sum\limits_{i=1}^{n} (1-Z_i) \geq \gamma_n - n \log(1+\sigma) -\frac{\log e }{1\!+\!\sigma} \log n \right] \notag\\
\\
&\geq & Q\lefto( \frac{\gamma_n + (\log e)(\log n)/(1+\sigma)  -n\log(1+\sigma) }{\sqrt{nV(\sigma)}}  \right) \notag \\
&& -\, \bigO\lefto(\frac{1}{\sqrt{n}}\right) .
\label{eq:exp-noise-beta-ach-bound1}
\end{IEEEeqnarray}
Equating the  RHS of~\eqref{eq:exp-noise-beta-ach-bound1} to $1-\error +\tau$, and solving it for $\gamma_n$, we conclude that
\begin{IEEEeqnarray}{rCl}
\gamma_n &=& n\log(1+\sigma) -\sqrt{nV(\sigma)}Q^{-1}(\error)  + \bigO(\log n). \IEEEeqnarraynumspace
\label{eq:CLT-beta-deno-exp-n}
\end{IEEEeqnarray}
Substituting~\eqref{eq:CLT-beta-deno-exp-n} into~\eqref{eq:ub-beta-exp-noise}, and then~\eqref{eq:ub-beta-exp-noise} and~\eqref{eq:lower-bound-beta-4} into~\eqref{eq:kappa-beta-intro-avg}, we establish that~\eqref{eq:rate-thm-exp} is achievable.

To prove the converse part of Theorem~\ref{thm:exp-noise}, we first notice that by~\cite[Lemma~39]{polyanskiy10-05}, we can assume without loss of generality that all codewords $x^n$ belong to  the simplex
\begin{IEEEeqnarray}{rCl}
\setF_{n} \define \Big\{x^n \in \realset^n: \sum\limits_{i=1}^{n} x_i = n\sigma,\,\, x_i \geq 0\Big\}.
\label{eq:power-constraint-exp-equal}
\end{IEEEeqnarray}
 Let $Q_{Y^n} =  Q_{X^n} \circ P_{Y^n|X^n}$, where, as before, $Q_{X^n} = (P_{X}^*)^n$.
By the meta-converse theorem~\cite[Th.~27]{polyanskiy10-05},  every $(n,M,\error)$ code for the channel~\eqref{eq:channel-io-nong} satisfies
\begin{IEEEeqnarray}{rCl}
M \leq \sup\limits_{P_{X^n}}\frac{1}{\beta_{1-\error}(P_{X^nY^n}, P_{X^n}Q_{Y^n})}
\label{eq:meta-converse-for-exp-noise}
\end{IEEEeqnarray}
where the supremum is   over all probability distributions supported on $\setF_n$.
We next note that the function $\beta_{1-\error}(P_{Y^n|X^n =x^n} ,Q_{Y^n})$ takes the same value for all $x^n \in \setF_n$.
Indeed, by using the Neyman-Pearson lemma we observe that, for every $x^n \in \setF_n$,
\begin{equation}
\beta_{1-\error}(P_{Y^n|X^n =x^n} ,Q_{Y^n}) = e^{- n\sigma/(1+\sigma)} \prob\lefto[ \sum_{i=1}^{n} Z_i \leq  \frac{n - \xi_n }{1+\sigma} \right]
\end{equation}
where $\xi_n$ satisfies
\begin{IEEEeqnarray}{rCl}
\prob\lefto[  \sum_{i=1}^{n} Z_i \leq n    - \xi_n \right] = 1-\error.
\end{IEEEeqnarray}
Using~\cite[Lemma~29]{polyanskiy10-05},  we conclude that for every $x^n \in \setF_n$,
\begin{equation}
\sup_{P_{X^n}}\beta_{1-\error}(P_{X^nY^n}, P_{X^n}Q_{Y^n}) = \beta_{1-\error}( P_{Y^n|X^n=x^n} , Q_{Y^n} ).
\label{eq:symmetry-exp-beta}
\end{equation}
For convenience, we shall set $x^n = \bar{x}^n \define [\sigma, \ldots, \sigma]$.
This choice makes $P_{Y^n|X^n=x^n}$ a distribution of i.i.d. random variables. Using~\cite[Lemma~58]{polyanskiy10-05} and performing straightforward algebraic manipulations, we obtain
\begin{IEEEeqnarray}{rCl}
\IEEEeqnarraymulticol{3}{l}{
- \log \beta_{1-\error}( P_{Y^n|X^n=\bar{x}^n} , Q_{Y^n})}\notag\\
\quad & =&  n\log(1+\sigma) -\sqrt{nV(\sigma)}Q^{-1}(\error)  + \bigO(\log n).
\label{eq:beta-x-exp-conv}
\end{IEEEeqnarray}
We conclude the proof by substituting~\eqref{eq:symmetry-exp-beta} and~\eqref{eq:beta-x-exp-conv} into~\eqref{eq:meta-converse-for-exp-noise}.
\end{IEEEproof}

\subsection{MIMO Block-Fading Channel With Perfect CSIR}
\label{sec:mimo-bf}
In this section, we  use the $\beta\beta$ achievability bound~\eqref{eq:kappa-beta-intro-avg} to characterize the maximum coding rate achievable over an $\txant\times \rxant$ Rayleigh MIMO block-fading channel.
The channel is assumed to  stay constant over~$\cohtime$ channel uses (a coherence interval) and to change independently across coherence intervals.
The input-output relation within the $k$th coherence interval is given by
\begin{equation}
\label{eq:channel-io-mimo}
\randmatY_k =\randmatX_k \randmatH_k + \randmatZ_k.
\end{equation}
Here, $\randmatX_k\in \complexset^{\cohtime \times\txant}$ and $\randmatY_k\in\complexset^{\cohtime \times \rxant}$ are the transmitted and received matrices, respectively;
the entries of the fading matrix $\randmatH_k\in\complexset^{\txant\times\rxant}$ and of the noise matrix $\randmatZ_k \in\complexset^{\cohtime\times\rxant}$ are i.i.d. $\jpg(0,1)$.
We assume that $\{\randmatH_k\}$ and $\{\randmatZ_k\}$ are independent, that they take on independent realizations over successive coherence intervals, and that they do not depend on the matrices $\{\randmatX_k\}$.
The channel matrices  $\{\randmatH_k\}$ are assumed to be known to the receiver but not to the transmitter.
We shall also assume that each codeword spans $\ell \in\amsbb{N}$ coherence intervals, i.e., that the blocklength of the code is $n=\ell  \cohtime$.
Finally, each codeword~$\matX^\ell $ is constrained to satisfy
\begin{equation}
\label{eq:power-mimo-constraint}
\fnorm{\matX^\ell }^2 \leq nP.
\end{equation}

\subsubsection{Capacity and dispersion}
In the asymptotic limit $\ell \to \infty$ for fixed $\cohtime$,
the capacity of~\eqref{eq:channel-io-mimo} is given by~\cite{telatar99-11a}
\begin{IEEEeqnarray}{rCl}
C(P) = \Ex{\randmatH}{\log\mathrm{det} \lefto(\matI_{\txant} + \sqrt{P/\txant} \randmatH \herm{\randmatH} \right)}.
\label{eq:capacity-csir-mimo}
\end{IEEEeqnarray}
If either $\txant=\rxant =1$ or $\rxant \geq 2$, the capacity is achieved by a unique input distribution, under which the matrix $\randmatX$ has i.i.d. $\jpg(0, P/\txant)$ entries~\cite{telatar99-11a}.
In this case, we denote the capacity-achieving input distribution by $P_{\randmatX}^*$.
If $\txant>1$ and $\rxant =1$ (i.e., a multiple-input single-output channel), the capacity-achieving input distribution is not unique~\cite{collins14-07}.
The capacity-achieving output distribution is always unique and is denoted by $P_{\randmatY\randmatH}^*$.\footnote{Since the channel matrix $\randmatH$ is known at the receiver, the channel output consists of the pair $(\randmatY , \randmatH )$.}
More specifically, $ P_{\randmatY\randmatH}^* = P_{\randmatH}  P_{\randmatY|\randmatH}^* $, where  under $P_{\randmatY|\randmatH =\matH}^* $, the column vectors of $\randmatY$ are i.i.d. $\jpg(\mathbf{0}, \herm{\matH_i}\matH_i + P/\sqrt{\txant}\matI_{\rxant})$ distributed.
The channel dispersion for the  single-antenna case with perfect CSIR was derived  in~\cite{polyanskiy11-08a}.
This result was extended to multiple-antenna block-fading channels  in~\cite{collins2016-07a} and~\cite{collins14-07}.
In particular, it was shown in~\cite{collins2016-07a} that\footnote{It is also shown in~\cite{collins2016-07a} that the bound~\eqref{eq:na-csir-mimo-austin} is tight under the additional constraint $\|\matX^\ell\|_\infty = \littleo(n^{1/4})$ on each codeword $\matX^\ell$.}  for every $0< \error < 1/2$
\begin{IEEEeqnarray}{rCl}
R^*(n,\epsilon) \geq  C(P) -\sqrt{\frac{V(P)}{n}} Q^{-1}(\error) + \littleo(1/\sqrt{n}).
\label{eq:na-csir-mimo-austin}
\end{IEEEeqnarray}
Here,
\begin{IEEEeqnarray}{rCl}
V(P) = \inf \mathbb{E} \mathopen{}\bigg[ \mathbb{V}\mathrm{ar}\lefto[\log \frac{\der P_{\randmatY\randmatH\given \randmatX}}{ \der P^*_{\randmatY\randmatH}} (\randmatX,\randmatY,\randmatH) \, \Big| \,  \randmatX\right]\bigg]
\end{IEEEeqnarray}
where the infimum is over the set of capacity-achieving input distributions.
For the case $\txant=\rxant =1$ or $\rxant \geq 2$, the infimum is over the singleton $\{P_{\randmatX}^*\}$.

\subsubsection{Evaluation of the $\beta\beta$ achievability bound~\eqref{eq:kappa-beta-intro-avg}}\label{sec:mimo-compute-beta}
We now turn our attention to the computation of the $\beta\beta$ achievability bound~\eqref{eq:kappa-beta-intro-avg} for the channel~\eqref{eq:channel-io-mimo}.
For simplicity, we shall focus on the case in which the capacity-achieving input distribution is unique.
To compute~\eqref{eq:kappa-beta-intro-avg}, we choose $P_{\randmatX^\ell}$ as the uniform distribution on $\setS_n \define \{\matX^\ell: \fnorm{\matX^\ell}^2 = n \snr \}$, and set  $Q_{\randmatY^\ell \randmatH^\ell} = (P_{\randmatY\randmatH}^*)^\ell$.
With these choices, we have
\begin{equation}
R^*(n,\error) \geq \frac{1}{n} \sup_{0<\tau<\error} \log \frac{\beta_{\tau}(P_{\randmatY^\ell\randmatH^\ell}, Q_{\randmatY^\ell\randmatH^\ell})}{\beta_{1-\error+\tau}(P_{\randmatX^\ell\randmatY^\ell\randmatH^\ell},P_{\randmatX^\ell}Q_{\randmatY^\ell\randmatH^\ell})}.
\label{eq:mimo-betabeta-app}
\end{equation}
The denominator  $\beta_{1-\error+\tau}(P_{\randmatX^\ell\randmatY^\ell\randmatH^\ell},P_{\randmatX^\ell}Q_{\randmatY^\ell\randmatH^\ell})$ in~\eqref{eq:mimo-betabeta-app} can be computed using the Neyman-Pearson lemma and standard Monte Carlo techniques.
However, computing~$\beta_{\tau}(P_{\randmatY^\ell\randmatH^\ell}, Q_{\randmatY^\ell\randmatH^\ell})$ in the numerator is more involved, since there is no simple expression for $P_{\randmatY^\ell\randmatH^\ell }$.
 To circumvent this, we further lower-bound $\beta_{\tau}(P_{\randmatY^\ell\randmatH^\ell}, Q_{\randmatY^\ell\randmatH^\ell})$ using the data-processing inequality for $\beta_{\alpha}(\cdot,\cdot)$ as follows.
Let~$\widetilde{\randmatX}^\ell$ be a sequence of i.i.d. random matrices with $\widetilde{\randmatX}_k \sim P_{\randmatX}^*$, $ k=1,\ldots, \ell$. Then, $P_{\randmatX^\ell}$ can be obtained via
$\widetilde{\randmatX}^\ell$ through $\randmatX^\ell = \sqrt{nP}  \widetilde{\randmatX}^\ell/ \big\|\widetilde{\randmatX}^\ell\big\|_{\mathsf{F}}$.
Let $P^{(\mathrm{s})}_{\randmatY^\ell \randmatH^\ell\given \randmatX^\ell } \define P_{\randmatH^\ell} P^{(\mathrm{s})}_{\randmatY^\ell \given \randmatX^\ell \randmatH^\ell}$, where $P^{(\mathrm{s})}_{\randmatY^\ell \given \randmatX^\ell \randmatH^\ell }$ denotes  the channel law defined by
\begin{equation}
\label{eq:equivalent-channel}
\randmatY_k =  \randmatX_k\randmatH_k  \frac{\sqrt{nP}}{\fnorm{\randmatX^\ell}} +\randmatZ_k,\qquad k=1,\ldots, \ell.
 \end{equation}
We have that $P_{\randmatY^\ell\randmatH^\ell} =  P_{\randmatX^\ell} \circ P_{\randmatY^\ell \randmatH^\ell \given \randmatX^\ell} =  (P_{\randmatX}^*)^{\ell} \circ P^{(\mathrm{s})}_{\randmatY^\ell \randmatH^\ell\given \randmatX^\ell}$.
Furthermore, $Q_{\randmatY^\ell \randmatH^\ell } =  (P_{\randmatX}^*)^{\ell}  \circ P_{\randmatY^\ell \randmatH^\ell \given \randmatX^\ell}$.
Now, by the data-processing inequality,
\begin{equation}
\label{eq:data-processing-beta-app}
\beta_{\tau}(P_{\randmatY^\ell\randmatH^\ell}, Q_{\randmatY^\ell \randmatH^\ell}) \geq \beta_{\tau}\mathopen{}\big( (P_{\randmatX}^*)^{\ell} P^{(\mathrm{s})}_{\randmatY^\ell \randmatH^\ell\given \randmatX^\ell} ,  (P_{\randmatX}^*)^{\ell}  P_{\randmatY^\ell \randmatH^\ell\given \randmatX^\ell} \big).
\end{equation}
Since the Radon-Nikodym derivative $\frac{\der  P^{(\mathrm{s})}_{\randmatY^\ell \randmatH^\ell\given \randmatX^\ell}  }{\der    P_{\randmatY^\ell \randmatH^\ell\given \randmatX^\ell} } $ can be computed in closed form, the RHS of~\eqref{eq:data-processing-beta-app} can be computed using the Neyman-Pearson lemma and Monte Carlo techniques.

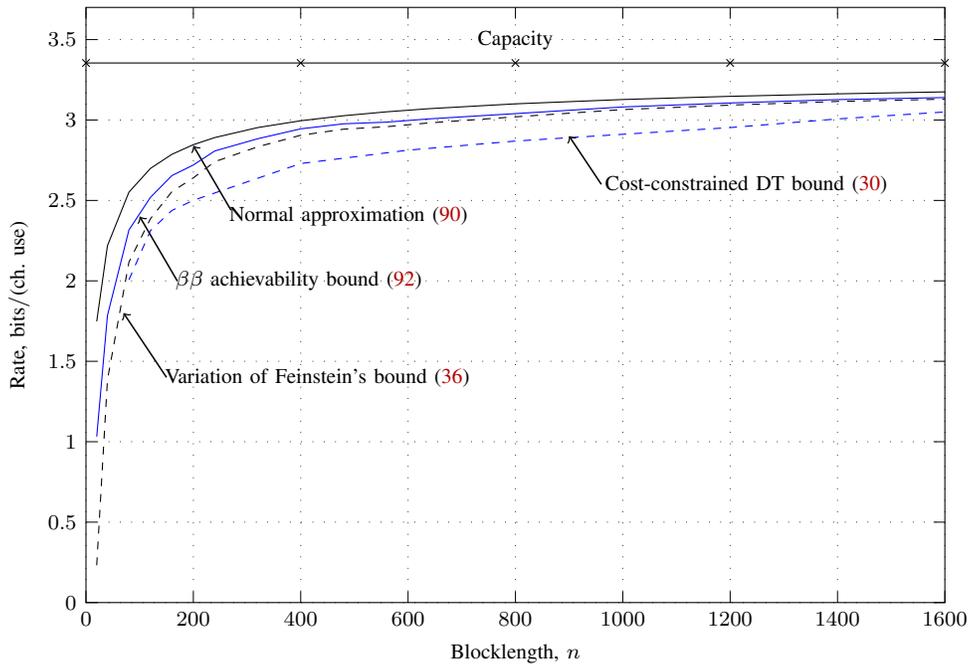
\begin{figure*}
\centering

	\begin{tikzpicture}
		\begin{axis}[
			xlabel={\footnotesize  Blocklength, $n$},
			ylabel={\footnotesize  Rate, bits$/$(ch. use)},
xmin=0, xmax=1600,
ymin=0, ymax=3.7,
grid=major,
xtick={0,200,...,1600},
xticklabels={%
$0$,
$200$,
$400$,
$600$,
$800$,
$1000$,
$1200$,
$1400$,
$1600$,}
            ]
		\addplot[black] table {na.dat};
\addplot[blue] table {ach.dat};
\addplot[black,dashed] table {rate_fein.dat};
\addplot[blue,dashed] table {rate_ddt.dat};

\addplot[black,mark=x] coordinates {
(0, 3.3546)
(400, 3.3546)
(800, 3.3546)
(1200, 3.3546)
(1600, 3.3546)
};

\draw (800, 3.5) node[]{\footnotesize Capacity};
\draw[line width = 0.2mm, <-] (200,2.84) -- (270,2.44);
\draw (250, 2.4) node[right]{\footnotesize  Normal approximation~\eqref{eq:na-csir-mimo-austin}};
\draw[line width = 0.2mm, <-] (100,2.4) -- (170,2);
\draw (150, 2) node[right]{\footnotesize  $\beta\beta$ achievability bound~\eqref{eq:mimo-betabeta-app}};
\draw[line width = 0.2mm, <-] (70,1.8) -- (150,1.4);
\draw (130, 1.4) node[right]{\footnotesize  Variation of Feinstein's bound~\eqref{eq:var-feinstein}};

\draw[line width = 0.2mm, <-] (900,2.9) -- (960,2.6);
\draw (950, 2.6) node[right]{\footnotesize  Cost-constrained DT bound~\eqref{eq:cost-constrained-DT}};

		\end{axis}
	\end{tikzpicture}
 \caption{Bounds on the maximum rate for a $4\times 4$ MIMO Rayleigh block-fading channel; here SNR=$0$ dB, $\error=0.001$, and $\cohtime =4$.\label{fig:mimo}}
 \end{figure*}

The resulting bound\footnote{The numerical routines used to obtain these results are available at https://github.com/yp-mit/spectre}
 is shown in Fig.~\ref{fig:mimo}  for a $4\times 4$ Rayleigh block-fading channel with $\cohtime =4$, $\error=0.001$, and SNR = $0$~dB.
For comparison, we have also plotted the normal approximation~\eqref{eq:na-csir-mimo-austin} (with the $\littleo(1/\sqrt{n})$ term omitted), the cost-constrained DT bound~\eqref{eq:cost-constrained-DT}, and the variation  of Feinstein's bound provided in~\eqref{eq:var-feinstein}.
 More specifically,~\eqref{eq:cost-constrained-DT} is computed with
 \begin{IEEEeqnarray}{rCl}
 \setF =\{\matX^\ell: \|\matX^\ell\|^2_{\mathsf{F}} \leq n P \}
 \end{IEEEeqnarray}
 and  with $\randmatX^\ell \sim Q_{\randmatX^\ell} $ having i.i.d. $\jpg(0, P'/\txant)$ entries. Here, $P'$ is chosen such that $Q_{\randmatX^\ell}[\setF^c] = \error/2$.
To compute~\eqref{eq:var-feinstein}, we chose the same $P_{\randmatX^\ell}$ and $Q_{ \randmatY^\ell \randmatH^\ell}$ that we used to compute~\eqref{eq:mimo-betabeta-app}.
Not surprisingly, the $\beta\beta$ achievability bound~\eqref{eq:kappa-beta-intro-avg} is uniformly tighter than both~\eqref{eq:cost-constrained-DT} and~\eqref{eq:var-feinstein}.
Note also that~\eqref{eq:var-feinstein} is better than the cost-constrained DT bound~\eqref{eq:cost-constrained-DT} mainly because it uses a better input distribution.

We also note that the $\kappa\beta$ bound~\cite[Th.~25]{polyanskiy10-05} (with $\setF = \setS_n$) is much more difficult to compute because one needs to maximize over all codewords $\matX^\ell\in \setS_n$. Furthermore, for  blocklength values of practical interest,  we expect that
\begin{IEEEeqnarray}{rCl}
\IEEEeqnarraymulticol{3}{l}{
\max\limits_{\matX^\ell \in \setS_n}  \beta_{1-\error+\tau}( P_{ \randmatY^\ell  \randmatH^\ell\given \randmatX^\ell =\matX^\ell}, Q_{ \randmatY^\ell \randmatH^\ell})} \notag\\
\qquad &\gg& \beta_{1-\error+\tau}(P_{\randmatX^\ell \randmatY^\ell \randmatH^\ell},P_{\randmatX^\ell}Q_{ \randmatY^\ell \randmatH^\ell}) \IEEEeqnarraynumspace
\label{eq:kappa-beta-versus-beta-beta}
\end{IEEEeqnarray}
which means that the resulting $\kappa\beta$ bound may be much looser than~\eqref{eq:mimo-betabeta-app}.
To validate this claim, we evaluate $\beta_{1-\error+\tau}( P_{ \randmatY^\ell \randmatH^\ell\given \randmatX^\ell =\matX^\ell}, Q_{ \randmatY^\ell \randmatH^\ell})$ for a specific codeword $\hat{\matX}^\ell \in \setS_n$ constructed as follows: the entries $\hat{x}_{j,k}^{(i)}$ of the matrix $\hat{\matX}_i$ are $\hat{x}_{1,1}^{(1)} = \sqrt{ n P}$, and $\hat{x}_{j,k}^{(i)} = 0$,  $\forall (i,j,k) \neq (1,1,1)$.
By construction, $\beta_{1-\error+\tau}( P_{ \randmatY^\ell  \randmatH^\ell\given \randmatX^\ell =\hat{\matX}^\ell})$ is  a lower bound on the LHS of~\eqref{eq:kappa-beta-versus-beta-beta}.
It can be shown that the ratio between $\beta_{1-\error+\tau}( P_{\randmatY^\ell \randmatH^\ell\given \randmatX^\ell = \hat{\matX}^\ell}, Q_{ \randmatY^\ell \randmatH^\ell})$ and $\beta_{1-\error+\tau}(P_{\randmatX^\ell \randmatY^\ell \randmatH^\ell},P_{\randmatX^\ell}Q_{ \randmatY^\ell \randmatH^\ell})$ grows exponentially in~$\ell$.
Numerically,  we observe that for the   parameters   in Fig.~\ref{fig:mimo} and the choices $\tau = \error/2 = 5\times 10^{-4}$ and $n = \cohtime l = 400$,
\begin{IEEEeqnarray}{rCl}
\IEEEeqnarraymulticol{3}{l}{
- \log_2    \beta_{1-\error+\tau}( P_{ \randmatY^\ell \randmatH^\ell \given \randmatX^\ell = \hat{\matX}^\ell}, Q_{ \randmatY^\ell \randmatH^\ell})}\notag\\
\quad  & \approx &  - \log_2 \beta_{1-\error+\tau}(P_{\randmatX^\ell \randmatY^\ell \randmatH^\ell},P_{\randmatX^\ell}Q_{ \randmatY^\ell \randmatH^\ell} )  + 735 \text{ bits}. \IEEEeqnarraynumspace
\end{IEEEeqnarray}

\section{Concluding Remarks}
We have developed a novel channel coding achievability bound (which we refer to as the $\beta\beta$ achievability bound) that involves two binary hypothesis tests, one for the joint distribution of the input and the output, and the other for the output distribution.
Our bound together with a  $\beta\beta$ converse bound established earlier by Polyanskiy and Verd\'{u}~\cite[Th.~15]{polyanskiy14-01}  yields a  nonasymptotic version of the golden formula~\eqref{eq:golden-formula-intro}.
Connections between the $\beta\beta$ bounds and various other nonasymptotic bounds as well as their asymptotic counterparts are summarized in Fig.~\ref{fig:connections}.
%

 The analogy between the $\beta\beta$ bounds and  the golden formula allows us to extend to the nonasymptotic regime   asymptotic analyses in which the golden formula plays a key role.
To demonstrate this point, we have used the $\beta\beta$ bounds to obtain a finite-blocklength extension of Verd\'{u}'s  wideband-slope approximation~\cite{verdu02-06}.
%
%
Our proof parallels the derivation of the latter, except that the beta-beta bounds are used in place of the golden formula.
We have also used the  $\beta\beta$ achievability bound to characterize  the channel dispersion of the additive exponential-noise channel and to obtain a finite-blocklength  achievability bound for MIMO Rayleigh-fading channels with perfect CSIR. In both cases, the $\beta\beta$ achievability bound proves to be very useful: in the former case, it simplifies the asymptotic analysis; in the latter case, it yields the tightest achievability bound known to date.

A crucial step in evaluating both the $\beta\beta$ achievability and converse  bounds is to choose an appropriate output distribution $Q_Y$.
A good choice of $Q_Y$ yields a bound that is both analytically tractable and numerically tight.
Some general principles for choosing $Q_Y$ are discussed in~\cite[Ch.~I.3.4]{yang15-08} and~\cite{polyanskiy13-07a}.


\appendices

\section{Proof of Theorem~\ref{thm:energy-per-bit}}
\label{app:proof-thm-beta-awgn-asy}

The proof is divided into four parts. (i) We first show in Appendix~\ref{app:equivalence-rate-energybit} that, when $k\to\infty$, $R\to 0$, and $kR \to \infty$, the expansion \eqref{eq:asy-expansion-r-awgn-eq} implies~\eqref{eq:wideband-awgn-thm-linear}.
We then prove~\eqref{eq:asy-expansion-r-awgn-eq} by providing (ii)   achievability and (iii) converse results in Appendices~\ref{app:proof-awgn-ach} and~\ref{app:proof-awgn-conv}, respectively.
(iv) Finally, we prove~\eqref{eq:wideband-awgn-thm-linear-boundedkr} in Appendix~\ref{app:proof-awgn-bdd}.

\subsection{The asymptotic expansion \eqref{eq:asy-expansion-r-awgn-eq}  implies \eqref{eq:wideband-awgn-thm-linear}}
\label{app:equivalence-rate-energybit}

Observe that the blocklength $n$, the transmit power $P$, the data rate $R$, the energy per bit $\eb$, and the number of information bits $k$  are related as follows:
\begin{equation}
k = n R \quad \text{ and } \quad P = \eb R
\label{eq:design-parameter-all}
\end{equation}
where $R$ is measured in bits per channel use.
Hence,  $\eb^*(k,\error,R)$ can be obtained from $R^*(n,\error, P)$  by solving the equation
\begin{IEEEeqnarray}{rCl}
R= R^*(k /R, \epsilon,\eb^*(k,\error,R)R ).
\label{eq:relation-between-r-ebno}
\end{IEEEeqnarray}
Substituting~\eqref{eq:asy-expansion-r-awgn-eq} into~\eqref{eq:relation-between-r-ebno} after setting the base of the $\log$ in~\eqref{eq:asy-expansion-r-awgn-eq} to $2$, we obtain
\begin{IEEEeqnarray}{rCl}
R \log_e 2 &=&   \eb^*\cdot R - \sqrt{ \frac{2 \eb^* \cdot R^2 }{k}}Q^{-1}(\error) - \frac{1}{2}(\eb^* \cdot R )^2
  \notag  \\
  &&+\, \littleo\mathopen{}\bigg(\sqrt{\frac{\eb^* R^2 }{k }}\bigg) + \littleo\lefto( (\eb^* \cdot R )^2\right)
  \label{eq:rate-energy-relation1}
\end{IEEEeqnarray}
in the asymptotic limit  $ n\to\infty$, $P\to 0$, and $nP^2\to \infty$. Here,  we set $\eb^* = \eb^*(k,\error,R)$ for notational convenience.
Dividing both sides of~\eqref{eq:rate-energy-relation1} by~$R$, and rearranging, we obtain (recall that $N_0=1$)
 \begin{IEEEeqnarray}{rCl}
 \eb^* &=&   \log_e 2   - \sqrt{ \frac{2 \eb^* }{k}}Q^{-1}(\error) - \frac{(\eb^*)^2  R}{2} \notag\\
 && +\, \littleo\mathopen{}\bigg(\sqrt{\frac{\eb^*   }{k }}\bigg) + \littleo\lefto( (\eb^*)^2 R\right).
  \label{eq:rate-energy-relation2}
\end{IEEEeqnarray}
This implies that
\begin{equation}
\eb^* =  \log_e 2 + \littleo(1).
\label{eq:asy-eb-star}
\end{equation}
Substituting~\eqref{eq:asy-eb-star} into the RHS of~\eqref{eq:rate-energy-relation2}, we recover~\eqref{eq:wideband-awgn-thm-linear}.

In the next two sections we shall establish that~\eqref{eq:asy-expansion-r-awgn-eq} holds in the asymptotic limit  $n\to \infty$, $P \to 0$, and $nP^2 \to \infty$.
Note that by~\eqref{eq:design-parameter-all} and~\eqref{eq:asy-eb-star}, these limits are equivalent to the limits $k\to \infty$, $R\to 0$, and $kR \to \infty$ stated in Theorem~\ref{thm:energy-per-bit}.
 Unless otherwise specified, all asymptotic expansions in the next two sections hold in the limit $n\to\infty$, $P\to 0$, and $nP^2 \to \infty$.

\subsection{Proof of~\eqref{eq:asy-expansion-r-awgn-eq}: Achievability}
\label{app:proof-awgn-ach}

We shall apply the $\beta\beta$ achievability bound~\eqref{eq:kappa-beta-intro-avg} with  $P_{X^n}$ chosen as the uniform distribution over the power sphere
\begin{IEEEeqnarray}{rCl}
\setF_n \define \{ x^n\in \complexset^n: \|x^n\|^2 =nP \}
\label{eq:define-power-sphere-sn}
\end{IEEEeqnarray}
and with $Q_{Y^n} = \jpg(\mathbf{0}, \matI_n)$.
Furthermore, we shall set
\begin{equation}
\tau = (nP^2)^{-1/2} .
\label{eq:def-taun-appendix-gau-ach}
\end{equation}
Since we are interested in the asymptotic regime $n\to\infty$, we can assume without loss of generality that $\tau <\error$.
By~\eqref{eq:kappa-beta-intro-avg},
\begin{IEEEeqnarray}{rCl}
R^*(n,\error,\snr) &\geq& \frac{1}{n} \log \frac{\beta_{\tau }(P_{Y^n}, Q_{Y^n})}{ \beta_{1-\error + \tau }(P_{X^nY^n}, P_{X^n}Q_{Y^n})  } .\IEEEeqnarraynumspace
\label{eq:beta-beta-awgn-app1}
\end{IEEEeqnarray}

The denominator on the RHS of~\eqref{eq:beta-beta-awgn-app1} can be computed as follows.
Due to the spherical symmetry of $\setF_n$ and $Q_{Y^n}$, we have that $\beta_{\alpha}(P_{Y^n|X^n =x^n}, Q_{Y^n} )$ takes the same value for all $x^n\in\setF_n$.
Let $\bar{x}^n \define [\sqrt{nP},0,\ldots,0]$. We next apply~\cite[Lemma~29]{polyanskiy10-05} to conclude that for every $P_{X^n}$ supported on $\setF_n$,
\begin{IEEEeqnarray}{rCl}
\IEEEeqnarraymulticol{3}{l}{
\beta_{1-\epsilon + \tau}(P_{X^nY^n},P_{X^n} Q_{Y^n})}\notag\\
\quad  &=& \beta_{1-\epsilon + \tau}(P_{Y^n\given X^n =\bar{x}^n }, Q_{Y^n}) \\
&=& \beta_{1-\epsilon + \tau}(\jpg(\sqrt{nP}, 1), \jpg(0,1)). \IEEEeqnarraynumspace
\label{eq:beta-pq-gaussian-01}
\end{IEEEeqnarray}
It follows from  the Neyman-Pearson lemma~\cite{neyman33a} that
\begin{equation}
\beta_{1-\epsilon + \tau}(P_{X^nY^n},P_{X^n} Q_{Y^n})  =   Q\lefto(\sqrt{2n \snr} + Q^{-1}(1-\epsilon + \tau) \right) .
\label{eq:q-func-beta-awgn}
\end{equation}
 To evaluate the RHS of~\eqref{eq:q-func-beta-awgn} we use that~\cite[Eq.~(3.53)]{verdu98}
\begin{equation}
\log Q(x) = -\frac{x^2\log e}{2} -\log x -\frac{1}{2}\log (2\pi) + o(1), \, x\to\infty \label{eq:verdu-expan-qfunc}
\end{equation}
and obtain
\begin{IEEEeqnarray}{rCl}
\IEEEeqnarraymulticol{3}{l}{
 \log \beta_{1-\epsilon + \tau }(P_{X^nY^n},P_{X^n} Q_{Y^n}) }\notag \\
 \quad  &=&  - n \snr\log e +  \sqrt{2n\snr}Q^{-1}(\error )\log e + \littleo\lefto(\sqrt{n \snr}\right). \IEEEeqnarraynumspace
 \label{eq:denominator-beta-beta-awgn}
\end{IEEEeqnarray}
For latter use, we notice that the expansion~\eqref{eq:denominator-beta-beta-awgn}  holds not only for the uniform distribution over the power sphere~$\setF_n$; rather it holds   for every probability distribution $P_{X^n}$ supported on~$\setF_n$.

To evaluate $\beta_{\tau }(P_{Y^n}, Q_{Y^n}) $, we shall make  use of the following lemma, which provides a variational characterization of the $\beta$ function.
\begin{lemma}
\label{lemma:properties-of-beta}
For every pair of probability measures $P$ and $Q$ on $\setX$ and every $\alpha \in (0,1)$, we have that
\begin{equation}
\beta_{\alpha}(P, Q) = \max_{R} \beta_{\beta_\alpha (P, R) } (R, Q)
\label{eq:variational-characterization-beta}
\end{equation}
where the maximum is over all probability measures $R$ on $\setX$.
Furthermore, the maximum is achieved by all probability measures contained  in the one-parameter exponential family $\{R_{\lambda}: \lambda \in (0,1)\}$ connecting $P$ and $Q$. Specifically, $R_\lambda $ is defined as
\begin{IEEEeqnarray}{rCl}
\frac{\mathrm{d}R_\lambda}{\mathrm{d}\mu}(x) \define e^{-\lambda D_{1-\lambda}(P\|Q)} \left(\frac{\mathrm{d}P}{\mathrm{d}\mu} (x) \right)^{1 - \lambda} \left(\frac{\mathrm{d}Q}{\mathrm{d}\mu}(x)\right)^\lambda .  \IEEEeqnarraynumspace
\label{eq:exponential-family-R-lambda}
\end{IEEEeqnarray}
Here, $D_{1-\lambda}(P\|Q)$ denotes the R\'{e}nyi divergence of order $1-\lambda$~\cite{renyi1961}, and $\mu$ is a measure on $\setX$ that satisfies $P\ll \mu$ and $Q\ll \mu$.
\end{lemma}
\begin{IEEEproof}
See Appendix~\ref{app:proof-lemma-variation}.
\end{IEEEproof}

Let $(P_{Y}^*)^n=\jpg(\mathbf{0}, (1+P)\matI_n)$ be the product distribution obtained from the capacity-achieving output distribution of an AWGN channel with SNR $P$.  It follows from Lemma~\ref{lemma:properties-of-beta} that
\begin{equation}
\beta_{\tau }(P_{Y^n}, Q_{Y^n}) \geq \beta_{\beta_{\tau }(P_{Y^n} , (P_{Y}^*)^n)}((P_{Y}^*)^n , Q_{Y^n}).
\label{eq:lb-beta-pq-p-star}
\end{equation}
By~\eqref{eq:lower-bound-PY-QY} and~\cite[Prop.~2]{molavianJazi15-12a},
\begin{IEEEeqnarray}{rCl}
\beta_{\tau }(P_{Y^n} , (P_{Y}^*)^n) &\geq& c_1 \tau  \frac{\sqrt{1+2P} }{1+P}  \define \hat{\tau}
\label{eq:lb-beta-pq-p-star-2}
\end{IEEEeqnarray}
where $c_1$ is a positive constant independent of $P$. Since $\tau \mapsto \beta_{\tau}$ is nondecreasing, it follows from~\eqref{eq:lb-beta-pq-p-star} and~\eqref{eq:lb-beta-pq-p-star-2} that
\begin{IEEEeqnarray}{rCl}
\beta_{\tau }(P_{Y^n}, Q_{Y^n}) \geq \beta_{\hat{\tau} } ((P_{Y}^*)^n , Q_{Y^n}).
\label{eq:lb-beta-pq-n-awgn-2}
\end{IEEEeqnarray}
To evaluate the RHS of~\eqref{eq:lb-beta-pq-n-awgn-2}, we use the Neyman-Pearson lemma and that both $(P_{Y}^*)^n$ and $Q_{Y^n} $  are product distributions.
Specifically, under $ (P_{Y}^*)^n$, the random variable $\log \frac{\der (P_{Y}^*)^n}{\der Q_{Y^n} }(Y^n)$ has the same distribution as
\begin{IEEEeqnarray}{rCl}
\sum\limits_{i=1}^{n} \Big( | Z_i |^2  P\log e  - \log (1 + P) \Big)
\label{eq:info-density-r-q-ach-awgn}
\end{IEEEeqnarray}
where the $\{Z_i\}$ are i.i.d. $\jpg(0,1)$ random variables.
Furthermore,
 \begin{IEEEeqnarray}{rCl}
&&  \Ex{}{  | Z_i |^2  P\log e - \log (1 + P) \vphantom{\Big(}}   =  \frac{1}{2} P^2\log e + \bigO(P^3) \label{eq:info-density-r-q-ach-awgn-mean}\IEEEeqnarraynumspace \\
&&\varr \mathopen{}\Big[| Z_i |^2  P\log e - \log (1 + P) \Big] = P^2  \log^2 e . \label{eq:info-density-r-q-ach-awgn-var}
 \end{IEEEeqnarray}
Using~\cite[Lemma~59]{polyanskiy10-05}, and~\eqref{eq:info-density-r-q-ach-awgn}--\eqref{eq:info-density-r-q-ach-awgn-var},  we conclude that
\begin{IEEEeqnarray}{rCl}
\IEEEeqnarraymulticol{3}{l}{
\log \beta_{\hat{\tau}} ((P_{Y}^*)^n  , Q_{Y^n})}\notag\\
  &\geq& - \frac{1}{2} n P^2 \log e - \sqrt{\frac{2 n P^2 \log^2 e }{ \hat{\tau} }} + \log \frac{\hat{\tau}}{2} + \bigO(nP^3) \label{eq:bound-cdf-chi-squared2-awgn1}\IEEEeqnarraynumspace\\
&= &  - \frac{1}{2} n P^2 \log e + \littleo(nP^2). \IEEEeqnarraynumspace
 \label{eq:bound-cdf-chi-squared2-awgn2}
\end{IEEEeqnarray}
Here, the second and third terms on the RHS of~\eqref{eq:bound-cdf-chi-squared2-awgn1} behave as $\littleo(nP^2)$ due to~\eqref{eq:def-taun-appendix-gau-ach},~\eqref{eq:lb-beta-pq-n-awgn-2}, and the assumption that $nP^2 \to \infty$.
Substituting~\eqref{eq:bound-cdf-chi-squared2-awgn2} into~\eqref{eq:lb-beta-pq-n-awgn-2}, and then~\eqref{eq:denominator-beta-beta-awgn} and~\eqref{eq:lb-beta-pq-n-awgn-2}  into~\eqref{eq:beta-beta-awgn-app1}, we conclude that~\eqref{eq:asy-expansion-r-awgn-eq} is achievable.

\subsection{Proof of~\eqref{eq:asy-expansion-r-awgn-eq}: Converse}
\label{app:proof-awgn-conv}

As in \cite[Section~III.J]{polyanskiy10-05}, we can assume without loss of generality that each codeword $x^n$ satisfies the power constraint~\eqref{eq:power-constraint-nongaussian} with equality, i.e., $x^n \in \setF_n$.
To prove the converse,  we shall use the $\beta\beta$ converse bound~\eqref{eq:prop-lb-beta-p-q-intro} with $Q_{Y^n}= \jpg(\mathbf{0}, \matI_n)$ and with
\begin{equation}
\delta = 24\sqrt{2}n^{-1/2} + e^{- \sqrt{nP^2}}.
\label{eq:def-delta-n-awgn-conv}
\end{equation}
This yields
\begin{equation}
R^*(n,\error,\snr) \leq \sup_{P_{X^n}} \frac{1}{n} \log \frac{ \beta_{1-\delta } (P_{Y^n} , Q_{Y^n})}{\beta_{1-\error -\delta }(P_{X^n Y^n} , P_{X^n}Q_{Y^n})  }
\label{eq:beta-beta-awgn-app1-conv}
\end{equation}
where the supremum is over all probability distributions $P_{X^n}$ supported on $\setF_n$, and $P_{Y^n} =  P_{X^n}\circ  P_{Y^n| X^n} $.

For this choice of parameters, the denominator on the RHS of~\eqref{eq:beta-beta-awgn-app1-conv} admits the same asymptotic expansion as~\eqref{eq:denominator-beta-beta-awgn}.
Next, we evaluate the numerator $ \beta_{1-\delta} (P_{Y^n} , Q_{Y^n})$.
%
%
%
Consider the following test between $P_{Y^n}$ and $Q_{Y^n}$:
\begin{IEEEeqnarray}{rCl}
T(y^n) = \indfun{2\|y^n\|^2 \geq \gamma }
\end{IEEEeqnarray}
 where $\gamma$ is chosen so that
\begin{equation}
P_{Y^n}\mathopen{}\Big[2\|Y^n\|^2 \geq \gamma  \Big] = 1-\delta.
\label{eq:def-gamma-awgn-conv}
\end{equation}
By the Neyman-Pearson lemma,
\begin{equation}
\beta_{1-\delta}(P_{Y^n} , Q_{Y^n}) \leq  Q_{Y^n}\mathopen{}\Big[ 2\|Y^n\|^2 \geq \gamma\Big] .
\label{eq:beta-pyqy-awgn-conv}
\end{equation}
%
 %
Note that under $P_{Y^n}$ the random variable $2\|Y^n\|^2$ has the same distribution as $\sum\nolimits_{i=1}^{2n} (Z_i + \sqrt{P})^2$ with $\{Z_i\} $ i.i.d. and $\mathcal{N}(0,1)$-distributed, and
regardless of the choice of $P_{X^n}$ (provided that $P_{X^n}$ is supported on $\setF_n$).
Next, we estimate $P_{Y^n}[2\|Y^n\|^2 \geq \gamma   ] $ using the Berry-Ess\'{e}en theorem (see, e.g., \cite[Ch.~XVI.5]{feller70a}) as follows:
\begin{IEEEeqnarray}{rCl}
\IEEEeqnarraymulticol{3}{l}{
\left|P_{Y^n}\mathopen{}\Big[2\|Y^n\|^2 \geq \gamma   \Big]  - Q\lefto( \frac{\gamma - 2n(1+ P) }{ \sqrt{4n(1+ 2P)}} \! \right) \right|}\notag\\
\qquad  &\leq& \frac{6 \Ex{}{|(Z_1 +\sqrt{P})^2 - 1-P|^3}}{(2+4P)^{3/2}\sqrt{2n}} \IEEEeqnarraynumspace \\
& \leq& 24\sqrt{2} n^{-1/2} \label{eq:app-berry-esseen-awgn}.
\end{IEEEeqnarray}
Here, the last step follows because
\begin{IEEEeqnarray}{rCl}
\IEEEeqnarraymulticol{3}{l}{
\Ex{}{|(Z_1 +\sqrt{P})^2 - 1-P|^3}}\notag\\
\quad  &\leq& \lefto(\Ex{}{|(Z_1 +\sqrt{P})^2 - 1-P|^4}\right)^{3/4} \\
&=& (12 (1+2P)^2 + 48(1+4P))^{3/4} \\
&\leq& 64^{3/4}(1+2P)^{3/2}.
\end{IEEEeqnarray}
Using~\eqref{eq:def-gamma-awgn-conv} and~\eqref{eq:app-berry-esseen-awgn}, we obtain
%
%
%
\begin{equation}
\gamma \geq 2n(1+P) - \sqrt{4n(1+2P)} Q^{-1}(\delta- 24\sqrt{2}{n}^{-1/2}) \define \widetilde{\gamma}.
\label{eq:gamma-p}
\end{equation}
Using~\eqref{eq:def-delta-n-awgn-conv} and the expansion
\begin{equation}
Q^{-1}(t) = \sqrt{2\log_e (1/ t) } (1+\littleo(1)) ,\quad t \to 0
\label{eq:inverse-q-expansion}
\end{equation}
which follows from~\eqref{eq:verdu-expan-qfunc},  we conclude that the threshold $ \widetilde{\gamma}$ behaves as
\begin{IEEEeqnarray}{rCl}
\widetilde{\gamma} &=& 2n(1+  P) + \sqrt{8n(1+2P)}(nP^2)^{1/4}(1+\littleo(1)) \label{eq:gamma-n-awgn-conv-1} \IEEEeqnarraynumspace\\
&=&2n(1 +  P )+ \littleo(nP) \label{eq:gamma-n-awgn-conv-2}
\end{IEEEeqnarray}
in the limit $n\to\infty$.
Under $Q_{Y^n}$, the random variable $2\|Y^n\|$ has the same distribution as $\sum\nolimits_{i=1}^{2n}Z_i^2$ with $\{Z_i\} $ i.i.d. and $\mathcal{N}(0,1)$-distributed.
Next, we use the moderate-deviation bound~\cite[Th.~3.7.1]{dembo98} to evaluate $Q_{Y^n}[ 2\|Y^n\|^2 \geq \gamma] $ as follows:
\begin{IEEEeqnarray}{rCl}
\IEEEeqnarraymulticol{3}{l}{
\limsup_{n\to\infty} \frac{2n }{(\widetilde{\gamma} - 2n)^2} \log Q_{Y^n}[2 \|Y^n\|^2 \geq \gamma]  }\notag\\
\quad
&\leq  &  \limsup_{n\to\infty} \frac{2n }{(\widetilde{\gamma} - 2n)^2} \log \prob\lefto[ \sum\limits_{i=1}^{2n} Z_i^2 \geq \widetilde{\gamma} \right] \label{eq:bound-cdf-chi-squared-c32} \\
&= & \limsup_{n\to\infty} \frac{2n }{(\widetilde{\gamma} - 2n)^2} \log \prob\lefto[\frac{1 }{ \widetilde{\gamma}  -2n } \bigg( \sum\limits_{i=1}^{2n} Z_i^2 -2n\bigg)\geq 1 \right]\notag\\
&&\\
&\leq & -\frac{1}{4} \log e.
 \label{eq:bound-cdf-chi-squared-c3}
\end{IEEEeqnarray}
Here,~\eqref{eq:bound-cdf-chi-squared-c32} follows from~\eqref{eq:gamma-p}, and~\eqref{eq:bound-cdf-chi-squared-c3} follows by using~\cite[Th.~3.7.1]{dembo98} with $a_n = 2n(\widetilde{\gamma}_n - 2n)^{-2}$ and $\Gamma=[1,\infty]$.
Combining~\eqref{eq:beta-pyqy-awgn-conv},~\eqref{eq:gamma-n-awgn-conv-2}, and~\eqref{eq:bound-cdf-chi-squared-c3}, we obtain that
\begin{IEEEeqnarray}{rCl}
\log \beta_{1-\delta}(P_{Y^n} , Q_{Y^n}) &\leq & -\frac{\log e}{4} \frac{(\widetilde{\gamma} - 2n)^2}{2n} (1+o(1))\IEEEeqnarraynumspace  \\
&=& -\frac{1}{2} nP^2 \log e + \littleo(nP^2).
\end{IEEEeqnarray}
We conclude the converse proof  of~\eqref{eq:asy-expansion-r-awgn-eq} by substituting~\eqref{eq:denominator-beta-beta-awgn} and~\eqref{eq:bound-cdf-chi-squared-c3} into~\eqref{eq:beta-beta-awgn-app1-conv}.

\subsection{Proof of~\eqref{eq:wideband-awgn-thm-linear-boundedkr}}
\label{app:proof-awgn-bdd}
Note that the converse part of~\eqref{eq:wideband-awgn-thm-linear-boundedkr} follows directly from~\eqref{eq:minimum-epb-awgn-yury} since, by definition,
\begin{IEEEeqnarray}{rCl}
\eb^*(k,\error, R) \geq \eb^*(k,\error).
\end{IEEEeqnarray}
Thus, it remains to show that~\eqref{eq:wideband-awgn-thm-linear-boundedkr} is achievable under the conditions $k\to\infty$, $R\to 0$, and $kR\to c<\infty$, which implies that~\eqref{eq:wideband-awgn-thm-linear-boundedkr} is achievable with codes of blocklength $k^2 / c$.
 Without loss of generality, we can assume that $c>0$.  Indeed, we can always transform a code that satisfies $kR\to \tilde{c}>0$   into a code satisfying $k R \to 0 $ by appending $k^3$ zero symbols to each codeword, without changing the total energy of each codeword.

Applying the $\beta\beta$ bound~\eqref{eq:kappa-beta-intro-avg} with the same $P_{X^n}$ and the same $Q_{Y^n}$ as in Appendix~\ref{app:proof-awgn-ach} but with $\tau =  (n P)^{-1/2}$, we conclude that there exists a sequence of $(n,M ,\error)$ codes that satisfy
\begin{IEEEeqnarray}{rCl}
M \geq  \frac{\beta_{\tau}(P_{Y^n}, Q_{Y^n})}{ \beta_{1-\error + \tau}(P_{X^nY^n}, P_{X^n}Q_{Y^n})  } .\IEEEeqnarraynumspace
\label{eq:beta-beta-awgn-app1-fixedkr}
\end{IEEEeqnarray}
This time, we shall study the asymptotic behavior of~\eqref{eq:beta-beta-awgn-app1-fixedkr} in the asymptotic limit $n\to \infty$, $P\to 0$, and $n P^2 \to  c/\log_2^2 e$ (which is equivalent to the limit $k\to\infty$, $R\to 0$, and $kR\to c$).
The denominator satisfies the same expansion as in~\eqref{eq:denominator-beta-beta-awgn}, whereas the numerator satisfies
\begin{IEEEeqnarray}{rCl}
\log \beta_{\tau}(P_{Y^n}, Q_{Y^n}) &\geq&  - \frac{1}{2} n P^2 \log e - \sqrt{\frac{2 n P^2 \log^2 e }{ \hat{\tau} }} \IEEEeqnarraynumspace\notag\\
&&+\, \log \frac{\hat{\tau}}{2} + \bigO(nP^3) \label{eq:bound-cdf-chi-squared2-awgn1-ccc} \\
&=& \bigO((n P)^{1/4}).
\label{eq:beta-tau-n-fixedkr}
\end{IEEEeqnarray}
Here, $\hat{\tau}$ is defined in~\eqref{eq:lb-beta-pq-p-star-2}, and~\eqref{eq:bound-cdf-chi-squared2-awgn1-ccc} follows from~\eqref{eq:bound-cdf-chi-squared2-awgn1}.
Substituting~\eqref{eq:denominator-beta-beta-awgn} and~\eqref{eq:beta-tau-n-fixedkr} into~\eqref{eq:beta-beta-awgn-app1-fixedkr},  taking the logarithm (with base $2$) on both sides, we conclude that
\begin{equation}
k \geq nP \log_2 e -  \sqrt{2n\snr}Q^{-1}(\error )\log_2 e + \littleo\lefto(\sqrt{n \snr}\right).
\label{eq:app-lb-fixedkr-awgn}
\end{equation}
 Using the relations~\eqref{eq:design-parameter-all} and the bound~\eqref{eq:app-lb-fixedkr-awgn} and proceeding as in~\eqref{eq:relation-between-r-ebno}--\eqref{eq:asy-eb-star}, we conclude that~\eqref{eq:wideband-awgn-thm-linear-boundedkr} is achievable.

\subsection{Proof of Lemma~\ref{lemma:properties-of-beta}}
\label{app:proof-lemma-variation}
  Let~$T: \setX \to \{0,1\}$ be the  (possibly randomized) Neyman-Pearson test that achieves $\beta_{\alpha}(P,Q)$. For every probability measure $R$ on $\setX$, it follows that
\begin{IEEEeqnarray}{rCl}
\beta_{\alpha}(P,Q) &=& Q[T=1] \label{eq:bound-Q-T-0} \\
 &\geq&  \beta_{R[T=1]} (R, Q) \label{eq:bound-Q-T-1}\\
 &\geq& \beta_{\beta_\alpha (P,R)} (R,Q ). \label{eq:bound-Q-T-2}
 \end{IEEEeqnarray}
Here,~\eqref{eq:bound-Q-T-1} follows from the definition of $\beta_{\alpha}(R,Q)$, and~\eqref{eq:bound-Q-T-2} follows because $R[T=1] \geq \beta_{P[T=1]} (P,R)$, because $P[T=1] =\alpha$ by definition of $T$, and because $\alpha \mapsto \beta_{\alpha}(R,Q)$ is monotonically nondecreasing.  Maximizing the RHS of~\eqref{eq:bound-Q-T-2} over all probability measures $R$ on~$\setX$, we obtain
\begin{IEEEeqnarray}{rCl}
\beta_{\alpha} (P,Q) \geq \sup_{R} \beta_{\beta_\alpha (P,R)} (R,Q ).
\end{IEEEeqnarray}
It remains to show that for the $\{R_\lambda$\}, $\lambda \in (0,1)$, defined in~\eqref{eq:exponential-family-R-lambda}, we have
\begin{IEEEeqnarray}{rCl}
\beta_{\alpha} (P,Q) =  \beta_{\beta_\alpha (P,R_\lambda)} (R_\lambda,Q ).
\label{eq:beta-p-q--r-lambda-1}
\end{IEEEeqnarray}
Indeed, we observe that\footnote{In the case in which $P$ is not absolutely continuous with respect to $Q$, we set $\der P/ \der Q = +\infty$ on the singular set.}
 \begin{IEEEeqnarray}{rCl}
\log  \frac{\der P}{\der R_{\lambda}}(x)  &=&  \lambda D_{1-\lambda}(P \| Q) +\lambda \log \frac{\der P}{\der Q}(x)
\label{eq:neyman-pearson-equ1}
 \end{IEEEeqnarray}
and
 \begin{equation}
\log  \frac{\der R_{\lambda}}{\der Q}(x)   =  - \lambda D_{1-\lambda}(P \| Q) +  (1- \lambda)  \log \frac{\der P}{\der Q}(x)
\label{eq:neyman-pearson-equ2}
 \end{equation}
 for every $x$ in the support of $Q$.
The identities~\eqref{eq:neyman-pearson-equ1} and~\eqref{eq:neyman-pearson-equ2} imply that the test $T$ in~\eqref{eq:bound-Q-T-0}--\eqref{eq:bound-Q-T-2} coincides with the Neyman-Pearson test for distinguishing between $P$ and $R_{\lambda}$ and between $R_{\lambda}$ and $Q$. This in turn implies, by the Neyman-Pearson lemma, that both~\eqref{eq:bound-Q-T-1} and~\eqref{eq:bound-Q-T-2} hold with equality. This establishes~\eqref{eq:beta-p-q--r-lambda-1}.

\section{Proof of Theorem~\ref{thm:wideband-slope-fading}}
\label{app:wideband-slope-rayleigh}

As in Appendix~\ref{app:equivalence-rate-energybit}, to prove~\eqref{eq:wideband-fading-thm-linear} it is sufficient to show that the maximum coding rate $R^*(n,\epsilon,\snr)$ satisfies
\begin{equation}
\frac{R^*(n,\epsilon,\snr)}{\log e} =  \snr - \sqrt{ \frac{2 \snr}{n}}Q^{-1}(\error) -   \snr^2 + \littleo\mathopen{}\bigg(\sqrt{\frac{\snr}{n}}\bigg) + \littleo\lefto(\snr^2\right)
\label{eq:asy-expansion-r-fading-eq}
\end{equation}
in the asymptotic limit $n\to\infty$, $P\to 0$, and $nP^2 \to \infty$. The achievability and the converse part of~\eqref{eq:asy-expansion-r-fading-eq} are proved in Appendices~\ref{section:proof-fading-ach-asy} and~\ref{sec:fading-conv-asy}, respectively.   The proof of~\eqref{eq:wideband-fading-thm-linear-boundedkr} follows similar steps as the ones reported in Appendix~\ref{app:proof-awgn-bdd}, and is thus omitted.
 Unless otherwise specified, all asymptotic expansions in Appendices~\ref{section:proof-fading-ach-asy} and~\ref{sec:fading-conv-asy} hold in the limit $n\to\infty$, $P\to 0$, and $nP^2 \to \infty$.

\subsection{Proof of~\eqref{eq:asy-expansion-r-fading-eq}: Achievability}
\label{section:proof-fading-ach-asy}
We shall apply the $\beta\beta$ achievability bound~\eqref{eq:kappa-beta-intro-avg} to the channel~\eqref{eq:rayleigh-block-fading-io-siso} with input $X^n$ and output $(Y^n,H^n)$ (recall that we assumed perfect CSIR) with  the same choices of $P_{X^n}$ and $Q_{Y^n}$ as in Section~\ref{app:proof-awgn-ach}. Namely, $P_{X^n}$ is chosen  as the uniform distribution over the power sphere~$\setF_n$ defined in~\eqref{eq:define-power-sphere-sn}, and $Q_{Y^n} = \jpg(\mathbf{0}, \matI_n)$.
Furthermore, we set
\begin{equation}
\tau = P + (nP^2)^{-1/2}.
\label{eq:fading-taun-asy-ach}
\end{equation}
Since $\tau \to 0$ as $P\to 0$ and $nP^2 \to \infty$, and since we are interested in the asymptotic behavior of $R^*(n,\epsilon, P)$ as $n\to \infty$, $P\to 0$, and $nP^2\to\infty$, we can assume without loss of generality that $\tau<\epsilon$.
It follows from~\eqref{eq:kappa-beta-intro-avg} that
\begin{IEEEeqnarray}{rCl}
nR^*(n,\epsilon,P) &\geq& \log \beta_{\tau }(P_{Y^n H^n} , Q_{Y^n} P_{H^n}) \notag\\
&&-\,\log\beta_{1-\error+\tau }(P_{X^nY^nH^n }, P_{X^n}Q_{Y^n}P_{H^n})\IEEEeqnarraynumspace\notag\\
\label{eq:betabeta-csi-energy}
\end{IEEEeqnarray}
where $P_{Y^n H^n} = P_{X^n} \circ  P_{Y^n H^n | X^n}$.

To evaluate the second term on the RHS of~\eqref{eq:betabeta-csi-energy}, we use~\cite[Eq.~(103)]{polyanskiy10-05} and obtain
\begin{IEEEeqnarray}{rCl}
-\log\beta_{1-\error+\tau }(P_{X^nY^nH^n }, P_{X^n}Q_{Y^n}P_{H^n}) \geq \gamma_0
\label{eq:lb-beta-fading-ach-haha}
\end{IEEEeqnarray}
where $\gamma_0$ satisfies
\begin{IEEEeqnarray}{rCl}
P_{X^nY^nH^n }\lefto[ \log \frac{\der P_{X^nY^nH^n }}{ \der (P_{X^n}Q_{Y^n}P_{H^n})} \leq \gamma_0 \right] =\error-\tau. \IEEEeqnarraynumspace
\label{eq:def-gamma-0-ach-fading-proof}
\end{IEEEeqnarray}
Observe now that, under $P_{X^n Y^nH^n }$, the random variable $\log \frac{ \der P_{X^nY^nH^n} }{\der(P_{X^n}Q_{Y^n}P_{H^n})}(X^n,Y^n,H^n)$ has the same distribution as
\begin{IEEEeqnarray}{rCl}
\log e\sum\limits_{i=1}^{n} \big(|H_iX_i|^2 +  2 \mathrm{Re}(H_iX_i Z_i^*) \big).
\label{eq:info-den-csir-ach}
\end{IEEEeqnarray}
Next, we use the central limit theorem for functions~\cite[Prop.~1]{iri2015-06a} (see also~\cite[Prop.~1]{molavianJazi15-12a}) to derive an asymptotic expansion for $\gamma_0$ in~\eqref{eq:def-gamma-0-ach-fading-proof}.
Specifically, let $\widetilde{X}^n \sim \jpg(\mathbf{0}, P \matI_n)$. It follows that $X^n \sim P_{X^n}$ has the same distribution as $ \sqrt{nP}\widetilde{X}^n/\|\widetilde{X}^n\|$.
Let
\begin{IEEEeqnarray}{rCl}
A_{1,i} &\define& |H_i \widetilde{X}_i|^2 - P\\
A_{2,i} &\define& |\widetilde{X}_i|^2 - P\\
A_{3,i} & \define&  2 \mathrm{Re}(H_i \widetilde{X}_i Z_i^*),\quad i=1,\ldots, n.
\end{IEEEeqnarray}
The random vectors $\{[A_{1,i},A_{2,i},A_{3,i}]\}$ are i.i.d. with zero mean and covariance matrix
\begin{IEEEeqnarray}{rCl}
\matV = \begin{bmatrix}
3P^2 & P^2 & 0\\
P^2 & P^2 & 0\\
0& 0& 2P
\end{bmatrix}.
\end{IEEEeqnarray}
Let $g: \realset^3 \to \realset$ be defined as
\begin{IEEEeqnarray}{rCl}
g(a_1,a_2,a_3) \define \frac{(a_1 +P) P}{P+a_2} + \frac{a_3\sqrt{P}}{\sqrt{P+a_2}} \IEEEeqnarraynumspace
\end{IEEEeqnarray}
and observe that
\begin{IEEEeqnarray}{rCl}
( n\log e) \cdot  g\lefto(\frac{1}{n}\sum\limits_{i=1}^{n}A_{1,i} , \frac{1}{n}\sum\limits_{i=1}^{n}A_{2,i} ,\frac{1}{n}\sum\limits_{i=1}^{n}A_{3,i}  \right) \IEEEeqnarraynumspace
\end{IEEEeqnarray}
has the same distribution as~\eqref{eq:info-den-csir-ach}.
Finally, let $\vecj $ denote the gradient of $g$ at $(0,0,0)$.
It follows that
\begin{IEEEeqnarray}{rCl}
\vecj \matV \tp{\vecj} & =& 2P^2 + 2P .
\end{IEEEeqnarray}
We are now ready to invoke~\cite[Prop.~1]{iri2015-06a}  and conclude that for every $\gamma \in \realset $
\begin{IEEEeqnarray}{rCl}
\IEEEeqnarraymulticol{3}{l}{
 \prob\mathopen{}\bigg[\sum\limits_{i=1}^{n} \big(|H_iX_i|^2 +  2 \mathrm{Re}(H_iX_i Z_i^*) \big)   \leq \gamma  \bigg]
 }\notag\\
 \quad &\leq& Q\lefto( \frac{ nP -\gamma }{\sqrt{2n(P+P^2 )}} \right) + \bigO\lefto(n^{-1/2}\right).
\label{eq:berry-esseen-func-1}
\end{IEEEeqnarray}
Setting the RHS of~\eqref{eq:berry-esseen-func-1} equal to $\error -\tau$ and solving for $\gamma$, we obtain an asymptotic lower bound on $\gamma_0$, which we use to further lower-bound~\eqref{eq:lb-beta-fading-ach-haha} as follows:
%
%
%
\begin{IEEEeqnarray}{rCl}
\IEEEeqnarraymulticol{3}{l}{
-\log\beta_{1-\error+\tau}(P_{X^n}P_{Y^nH^n|X^n}, P_{X^n}Q_{Y^n}P_{H^n}) }\notag\\
&\geq&  n \snr\log e - \sqrt{2n\snr (1+ \snr)}Q^{-1}\lefto(\error - \tau - \bigO(n^{-1/2} ) \right)\log e \notag\\
\\
&=&n \snr \log e - \sqrt{2n \snr }Q^{-1}\lefto(\error \right)\log e + \littleo\lefto(\sqrt{nP}\right) .
\label{eq:ach-fading-betad-asy}
\end{IEEEeqnarray}
Here,~\eqref{eq:ach-fading-betad-asy} follows by Taylor-expanding   $\sqrt{1+P }$ for $P\to 0$ and by Taylor-expanding $Q^{-1}(\cdot)$ around~$\error$ for $\tau\to 0$ and $n\to\infty$.

To evaluate $ \beta_{\tau}(P_{Y^n H^n} , Q_{Y^n} P_{H^n}) $ on the RHS of~\eqref{eq:betabeta-csi-energy},  we again use Lemma~\ref{lemma:properties-of-beta} in Appendix~\ref{app:proof-awgn-ach}.
Let   $(P^*_{Y H})^n$ be the product distribution obtained from the capacity-achieving output distribution of the channel~\eqref{eq:rayleigh-block-fading-io-siso}  with SNR $P$.
%
%
Then, by Lemma~\ref{lemma:properties-of-beta},
\begin{IEEEeqnarray}{rCl}
\IEEEeqnarraymulticol{3}{l}{
\beta_{\tau}(P_{Y^n H^n},Q_{Y^n} P_{H^n} ) }\notag\\
\quad &\geq& \beta_{\beta_{\tau}(P_{Y^n H^n}, (P^*_{Y H})^n)} ( (P^*_{Y H})^n, Q_{Y^n} P_{H^n} ). \IEEEeqnarraynumspace
\label{eq:lb-beta-p-q-asy}
\end{IEEEeqnarray}
We lower-bound $\beta_{\tau}(P_{Y^n H^n}, (P^*_{Y H})^n )$ by following steps similar to those reported in Section~\ref{sec:mimo-bf}.
Specifically, let $P_{Y^nH^n|X^n}^{(s)}$ be the transition probability  of the following channel:
\begin{equation}
\label{eq:equivalent-channel-siso}
Y_i =   \frac{\sqrt{nP} H_i X_i }{\|X^n\|} + Z_i,\qquad i=1,\ldots, n.
 \end{equation}
Let  $(P_{X}^*)^n= \jpg(\mathbf{0}, P \matI_n)$ be the product distribution obtained from the capacity-achieving input distribution for the channel~\eqref{eq:rayleigh-block-fading-io-siso}  under perfect CSIR.
Then, we have that $ P_{Y^n H^n} = (P_{X}^*)^n \circ  P_{Y^nH^n|X^n}^{(s)}$ and $ (P^*_{Y H})^n  =    (P_{X}^*)^n \circ P_{Y^nH^n|X^n}$.
By the data-processing inequality,
\begin{IEEEeqnarray}{rCl}
\IEEEeqnarraymulticol{3}{l}{
\beta_{\tau}(P_{Y^n H^n},  (P^*_{Y H})^n ) }\notag\\
\quad & \geq&  \beta_{\tau}(  (P_{X}^*)^n P_{Y^nH^n|X^n}^{(s)}  , (P_{X}^*)^n  P_{Y^nH^n|X^n}). \IEEEeqnarraynumspace
\label{eq:data-processing-beta-app-fading-siso}
\end{IEEEeqnarray}
To lower-bound the RHS of~\eqref{eq:data-processing-beta-app-fading-siso}, we shall use the following bound~\cite{haroutunian68-04} (see also~\cite[Eqs.~(154)--(156)]{polyanskiy10-05}):
\begin{equation}
\log \beta_{\alpha}(P,Q) \geq -\frac{D(P\|Q ) + h_{\mathrm{b}}(\alpha)}{\alpha}
\end{equation}
where $h_{\mathrm{b}}(\alpha) \define -\alpha\log\alpha -(1-\alpha)\log(1-\alpha)$ denotes the binary entropy function.
This yields
\begin{IEEEeqnarray}{rCl}
\IEEEeqnarraymulticol{3}{l}{
 \beta_{\tau}((P_{X}^*)^n P_{Y^nH^n|X^n}^{(s)}  , (P_{X}^*)^n  P_{Y^nH^n|X^n} ) \geq  }  \notag\\
 && \exp\mathopen{}\bigg(\!\!-\frac{D( (P_{X}^*)^n P_{Y^n\! H^n|X^n}^{(s)}  \|  (P_{X}^*)^n  P_{Y^n \! H^n|X^n}  ) + h_{\mathrm{b}}(\tau)}{\tau}\bigg).\notag\\
\label{eq:data-proce-KL-beta}
\end{IEEEeqnarray}
Note now that
\begin{IEEEeqnarray}{rCl}
\IEEEeqnarraymulticol{3}{l}{
D\lefto((P_{X}^*)^n P_{Y^nH^n|X^n}^{(s)} \| (P_{X}^*)^n  P_{Y^nH^n|X^n}\right) } \notag\\
\quad &=& D\lefto(P_{Y^n| H^n X^n}^{(s)} \|  P_{Y^n| H^n X^n} | P_{H^n}(P_{X}^*)^n   \right)  \label{eq:evaluate-KL-two-gaussian-begin} \\
&=& \Ex{(P_{X}^*)^n }{ \sum\limits_{i=1}^{n} \left| \frac{\sqrt{n P} H_i  X_i}{\| X^n\| }  - H_i X_i \right|^2} \log e  \IEEEeqnarraynumspace \label{eq:evaluate-KL-two-gaussian}\\
&=& \Ex{ (P_{X}^*)^n  }{\big(\sqrt{nP} - \|X^n\| \big)^2} \log e \IEEEeqnarraynumspace\\
&= & 2n P\lefto(1 - \frac{\Gamma(n+1/2)}{\sqrt{n}\Gamma(n)}\right) \log e  \label{eq:evaluate-KL-two-gaussian3}  \\
&\leq& 2n P \lefto(1 - \sqrt{1-\frac{1}{2n+1}}\right) \log e \label{eq:evaluate-KL-two-gaussian4}\\
&\leq &P \frac{2n}{ 2n+1}\log e   \label{eq:evaluate-KL-two-gaussian-end-1} \\
 &  \leq& P \log e.    \label{eq:evaluate-KL-two-gaussian-end}
\end{IEEEeqnarray}
Here,~\eqref{eq:evaluate-KL-two-gaussian3} follows because $\Ex{ ( P_{X}^*)^n}{\|{X}^n\|} = \sqrt{P}\Gamma(n+1/2)/\Gamma(n)$; \eqref{eq:evaluate-KL-two-gaussian4} follows from Wendel's inequality~\cite[Eq.~(7)]{wendel1948-11a}; and~\eqref{eq:evaluate-KL-two-gaussian-end-1} follows because $\sqrt{1-x} \geq 1-x$ for every $ x\in[0,1]$.
 Substituting~\eqref{eq:evaluate-KL-two-gaussian-end} in~\eqref{eq:data-proce-KL-beta}, we obtain
\begin{IEEEeqnarray}{rCl}
\beta_{\tau}(P_{Y^n H^n}, (P^*_{Y H})^n  )    &\geq&  \exp\lefto( - \frac{P\log e + h_{\mathrm{b}}(\tau)}{\tau}\right) \IEEEeqnarraynumspace\\
&\geq & e^{-2}\tau \define \hat{\tau}  .\label{eq:data-processing-beta-app-fading-siso-final}
\end{IEEEeqnarray}
In the first step we used that $\tau \geq P$ and that
\begin{IEEEeqnarray}{rCl}
\frac{h_{\mathrm{b}}(\tau)   }{\tau } = - \log \tau + \frac{1-\tau}{\tau}\log\frac{1}{1-\tau} \leq -\log \tau +\log e. \IEEEeqnarraynumspace
\end{IEEEeqnarray}
Since $\alpha \mapsto \beta_\alpha(P,Q)$ is nondecreasing, we conclude from~\eqref{eq:lb-beta-p-q-asy} and~\eqref{eq:data-processing-beta-app-fading-siso-final} that
\begin{IEEEeqnarray}{rCl}
\beta_{\tau}(P_{Y^n H^n},Q_{Y^n} P_{H^n} ) \geq \beta_{\hat{\tau}} ( (P^*_{Y H})^n , Q_{Y^n} P_{H^n} ).\IEEEeqnarraynumspace
\label{eq:lb-beta-p-q-asy-22}
\end{IEEEeqnarray}

We next lower-bound the RHS of~\eqref{eq:lb-beta-p-q-asy-22} by using the Neyman-Pearson lemma and that  both $(P^*_{Y H})^n$ and $Q_{Y^n} P_{H^n} $  are product distributions.
Specifically, under $ (P^*_{Y H})^n$, the random variable $\log \frac{\der  (P^*_{Y H})^n }{\der (Q_{Y^n} P_{H^n})}(Y^n,H^n)$ has the same distribution as
\begin{IEEEeqnarray}{rCl}
\sum\limits_{i=1}^{n} \underbrace{|Z_i|^2 |H_i|^2 P\log e  - \log(1+|H_i|^2 P) }_{\define B_i} \IEEEeqnarraynumspace
\label{eq:info-density-r-q-ach}
\end{IEEEeqnarray}
where the random variables $\{B_i\}$ defined above are i.i.d.. Let
 \begin{IEEEeqnarray}{rCl}
 I_n &\define& \Ex{}{B_i} ,\quad  V_n \define \mathbb{V}\mathrm{ar}[B_i]  . 
 \end{IEEEeqnarray}
A straightforward computation reveals that
\begin{IEEEeqnarray}{rCl}
I_n  &=& \frac{\Ex{}{|H|^4}}{2}P^2\log e + \littleo(P^2) \label{eq:eva-mean-info-den-csir-ach}
\end{IEEEeqnarray}
and that $V_n$ can be bounded as follows:
\begin{IEEEeqnarray}{rCl}
3P^2\log^2 e &\leq& V_n  \leq  11P^2 \log^2 e \label{eq:eva-var-info-den}.
\end{IEEEeqnarray}
By~\cite[Lemma~59]{polyanskiy10-05},
\begin{IEEEeqnarray}{rCl}
\IEEEeqnarraymulticol{3}{l}{
\log \beta_{\hat{\tau}} ((P^*_{Y H})^n, Q_{Y^n} P_{H^n} ) }\notag\\
\quad &\geq& - n I_n - \sqrt{\frac{2nV_n }{\hat{\tau}}}  + \frac{1}{2} \log \frac{\hat{\tau}}{2} \\
&=& -\frac{\Ex{}{|H|^4}}{2} n P^2\log e +\littleo ( nP^2)  .
\label{eq:evaluate-beta-p-q}
\end{IEEEeqnarray}
Here, in~\eqref{eq:evaluate-beta-p-q} we used~\eqref{eq:fading-taun-asy-ach},~\eqref{eq:eva-mean-info-den-csir-ach}, and~\eqref{eq:eva-var-info-den}.
Finally, substituting~\eqref{eq:evaluate-beta-p-q}  into~\eqref{eq:lb-beta-p-q-asy-22},  then~\eqref{eq:ach-fading-betad-asy} and~\eqref{eq:lb-beta-p-q-asy-22} into~\eqref{eq:betabeta-csi-energy}, and using  that $\Ex{}{|H|^4} =2$, we conclude that~\eqref{eq:asy-expansion-r-fading-eq} is achievable.

\subsection{Proof of~\eqref{eq:asy-expansion-r-fading-eq}: Converse}
\label{sec:fading-conv-asy}

It follows from~\cite[Lemma~39]{polyanskiy10-05} that we can assume without loss of generality that each codeword $x^n$ of a given $(n,M,\error)$ code satisfies the power constraint~\eqref{eq:power-constraint-nongaussian} with equality.
Furthermore, by arguing as in~\cite[Eqs. (284)--(286)]{polyanskiy10-05}, we can assume without loss of generality that the~\emph{maximum} probability of error of the code is upper-bounded by $\error$.
This allows us to use the maximum-error-probability version~\cite[Eq.~(222)]{polyanskiy14-01} of the $\beta\beta$ converse bound~\eqref{eq:prop-lb-beta-p-q-intro}.
Particularizing this bound  to the channel~\eqref{eq:rayleigh-block-fading-io-siso}, we conclude that every $(n,M,\error)$ code $\setC$ (maximum probability of error) satisfies
\begin{IEEEeqnarray}{rCl}
\IEEEeqnarraymulticol{3}{l}{
\beta_{\alpha}(P_{Y^nH^n}, Q_{Y^nH^n}) }\notag\\
\quad &\geq& \frac{\delta M }{1-\alpha+\delta} \inf_{x^n\in \setC} \beta_{\alpha-\error-\delta}(P_{Y^nH^n|X^n=x^n} , Q_{Y^nH^n}) \IEEEeqnarraynumspace
\label{eq:beta-beta-max-error}
\end{IEEEeqnarray}
where $\error+\delta \leq \alpha \leq 1$, $\delta>0$, and $P_{Y^nH^n}$ denotes the   output distribution induced by the code.

To evaluate the bound~\eqref{eq:beta-beta-max-error}, we shall choose $Q_{Y^nH^n} = Q_{Y^n} P_{H^n}$ with $Q_{Y^n} = \jpg( \mathbf{0} , \matI_n)$, and set
$\delta =\delta_n$ and $\alpha=1-\delta_n$ with $\delta_n$ chosen  such that $\delta _n\to 0$ as $n\to \infty$.
With these choices, we obtain
\begin{IEEEeqnarray}{rCl}
\log M &\leq &
\log \beta_{1- \delta_n} (P_{Y^n H^n}, Q_{Y^n}P_{ H^n})  \notag\\
&&-\,  \inf_{x^n \in \setC}  \log \beta_{1-\error-2\delta_n}(P_{Y^nH^n|X^n=x^n}, Q_{Y^n}P_{H^n}) \,\, \notag\\
&& +\,  \log 2.
\label{eq:beta-beta-conv-fading-1}
\end{IEEEeqnarray}
The second term on the RHS of~\eqref{eq:beta-beta-conv-fading-1} is upper-bounded by
\begin{IEEEeqnarray}{rCl}
\IEEEeqnarraymulticol{3}{l}{
- \inf_{x^n \in \setC}  \log\beta_{1-\error-2\delta_n}(P_{Y^nH^n|X^n=x^n}, Q_{Y^n}P_{H^n}) }\notag\\
&\leq &- \inf_{x^\infty }  \log\beta_{1-\error-2\delta_n}(P_{Y^\infty H^\infty|X^\infty=x^\infty}, P_{Y^\infty| X^\infty =\mathbf{0} }P_{H^\infty}) \notag\\
  \label{eq:beta-pq-error-case11} \\
 &\leq& n P \log e -\sqrt{2n P}Q^{-1}(\epsilon-2\delta_n)\log e \notag\\
 &&+\, \bigO(\log (nP))  \label{eq:beta-pq-error-case12} \\
 &=& n P \log e -\sqrt{2n P}Q^{-1}(\epsilon)\log e + \littleo(\sqrt{nP}) .
 \label{eq:beta-pq-error-case1}
\end{IEEEeqnarray}
 Here,  $x^\infty$ denotes the  infinite-dimensional sequence $(x_1,x_2,...)$,  $P_{Y^\infty H^\infty|X^\infty} = \prod_{i=1}^{\infty} P_{Y_iH_i |X_i}$,
  $P_{Y^\infty|X^\infty =\mathbf{0}} = \prod_{i=1}^{n}P_{Y_i|X_i=0}$, and the infimum in~\eqref{eq:beta-pq-error-case11}  is taken over all $x^\infty$ that satisfy $\|x^\infty\|^2 = nP$.
  The inequality~\eqref{eq:beta-pq-error-case11} follows because the feasible region of the optimization problem on the LHS of~\eqref{eq:beta-pq-error-case11} is contained in the feasible region of the optimization problem on the RHS of~\eqref{eq:beta-pq-error-case11};~\eqref{eq:beta-pq-error-case12} follows from~\cite[App.~IV]{yang2016-to-appear} (in particular, see~\cite[Eqs. (243) and (267)]{yang2016-to-appear}); and in~\eqref{eq:beta-pq-error-case1}  we have used that $\delta_n \to 0$ as $n\to \infty$.
Note that, the RHS of~\eqref{eq:beta-pq-error-case1} holds for all  $(n,M,\error)$ codes.

To conclude the proof, it is sufficient  to show that there exists a vanishing sequence $\{\delta_n\} $ such that  for every $(n,M,\error)$ code
\begin{IEEEeqnarray}{rCl}
\IEEEeqnarraymulticol{3}{l}{
\log \beta_{1- \delta_n} (P_{Y^n H^n}, Q_{Y^n}P_{ H^n}) }\notag\\
\quad  & \leq & - \frac{\Ex{}{|H|^4}}{2} n P^2 \log e + \littleo(n P^2).
\label{eq:bound-on-beta-deltan-general}
\end{IEEEeqnarray}
%
The   idea  is to construct a suboptimal test to distinguish between   $P_{Y^n H^n}$ and $Q_{Y^n}P_{ H^n}$.
Coarsely speaking, our test is constructed as follows: if $P_{Y^nH^n}$ is induced by a code whose codewords have suitably small $\ell_4$ norm, then we use as the test the optimal
Neyman-Pearson test between the capacity-achieving output distribution $(P_{YH}^*)^n$ and $Q_{Y^n}P_{ H^n}$.
Indeed, we expect the output distribution induced by such a code to resemble the capacity-achieving output distribution $(P_{YH}^*)^n$.
If instead $P_{Y^nH^n}$ is induced by a code whose codewords are peaky in an $\ell_4$ sense, we distinguish between $P_{Y^n H^n}$ and $Q_{Y^n}P_{ H^n}$  simply by testing the peakiness of $Y^n$.

We proceed now with the proof.
We start by  dividing the codebook $\setC$ into two subcodebooks $\setC_1$ and $\setC_2$ according to the peakiness of the codewords.
More specifically, we set
\begin{IEEEeqnarray}{rCl}
\setC_1 &\define& \{ x^n \in \setC:\,\,  \|x^n\|_4^4 \leq  n\eta_n \}
\label{eq:def-setc1-csir}\\
\setC_2 &\define& \setC \setminus \setC_1 = \{ x^n \in \setC:\,\,  \|x^n\|_4^4 >  n\eta_n \}\label{eq:def-setc2-csir}
\end{IEEEeqnarray}
where the threshold $\eta_n$ is defined as\footnote{When the entries of $X^n$ are independent Gaussian random variables, we expect to have $\|X^n\|_4^4 \approx \bigO (nP^2)$. This means that for a random Gaussian code  $|\setC_1| \gg |\setC_2|$.}
\begin{equation}
\eta_n \define \max\mathopen{}\Big\{ nP^{5/2} , (nP^2)^{3/4} \Big\}.
\end{equation}
Note that $\eta_n \to \infty$ as $n \to \infty$.
Let $\lambda \define |\setC_1|/|\setC|$. Furthermore, let $P_{Y^nH^n}^{(1)}$ and $P_{Y^nH^n}^{(2)}$ denote the output probability distributions induced by the subcodes $\setC_1$ and $\setC_2$, respectively. It follows that
\begin{IEEEeqnarray}{rCl}
P_{Y^n H^n} = \lambda P_{Y^nH^n}^{(1)} + (1-\lambda) P_{Y^nH^n}^{(2)}.
\label{eq:combination-two-output-dist}
\end{IEEEeqnarray}
Lemma~\ref{lemma:bounds-of-beta-12} below   allows us to upper-bound  $\beta_{1- \delta_n} (P_{Y^n H^n}, Q_{Y^n}P_{ H^n})$ by  $\beta_{1- \delta_{1,n}} (P_{Y^n H^n}^{(1)}, Q_{Y^n}P_{ H^n}) + \beta_{1- \delta_{2,n}} (P_{Y^n H^n}^{(2)}, Q_{Y^n}P_{ H^n})$ for some suitably  chosen $\delta_{1,n}$ and $\delta_{2,n}$.
\begin{lemma}
\label{lemma:bounds-of-beta-12}
Let $P = \lambda P_1 +(1-\lambda)P_2$ be a convex combination of $P_1$ and $P_2$. Then, for every probability measure $Q$, and every $\delta_1, \delta_2 \in (0,1)$,  we have
\begin{equation}
\beta_{1-\lambda\delta_1 -(1-\lambda)\delta_2}(P,Q) \leq \beta_{1-\delta_1}(P_1, Q) + \beta_{1-\delta_2}(P_2, Q).
\label{eq:lemma-combine-2-dist}
\end{equation}
\end{lemma}
\begin{IEEEproof}
See Appendix~\ref{app:proof-lemma-properties}.
\end{IEEEproof}

Set now
\begin{IEEEeqnarray}{rCl}
\delta_{1,n} &=& \max\mathopen{}\Big\{P^{1/4} , (nP^2)^{-1/8}\Big\}\\
 \delta_{2,n} &=& 1 -  \exp\lefto(- 4 \tilde{\xi}_n^{1/2}   P /\eta_n  \right)  \Big( 1 - \frac{2n}{\tilde{\xi}_n}\Big)
\end{IEEEeqnarray}
with $Z_1 \sim \mathcal{N}(0,1)$, and
\begin{equation}
\tilde{\xi}_n \define \max\mathopen{}\Big\{ n^2P^{15/4} , n^{9/8}P^{1/4} \big\}.
\label{eq:def-tilde-xi-n-app-fading}
\end{equation}
Furthermore, set
 \begin{IEEEeqnarray}{rCl}
\delta_n  = \lambda \delta_{1,n} + (1-\lambda)\delta_{2,n} .
 \end{IEEEeqnarray}
It can be shown that the sequences $\{\delta_{1,n}\} $, $\{\delta_{2,n}\} $, and  $\{\delta_n\} $ all vanish as $n\to\infty$, $P\to0$, and $nP^2\to \infty$.
We shall prove that
\begin{IEEEeqnarray}{rCl}
\IEEEeqnarraymulticol{3}{l}{
\log \beta_{1- \delta_{1,n}} (P_{Y^n H^n}^{(1)}, Q_{Y^n}P_{ H^n})}\notag\\
\quad &\leq &  -\frac{\Ex{}{|H|^4}}{2} nP^2 \log e + \littleo(n P^2)
\label{eq:bound-on-beta-deltan-case1-intro}
\end{IEEEeqnarray}
and that
\begin{IEEEeqnarray}{rCl}
\IEEEeqnarraymulticol{3}{l}{
\log \beta_{1-\delta_{2,n}} (P_{Y^nH^n}^{(2)} , Q_{Y^n} P_{ H^n}) }\notag\\
\quad   &\leq& -\frac{\Ex{}{|H|^4}}{2} nP^2 \log e + \littleo(n P^2).
 \label{eq:bound-beta-beta-case2-qq-intro}
\end{IEEEeqnarray}
The desired bound~\eqref{eq:bound-on-beta-deltan-general} then follows  from~\eqref{eq:combination-two-output-dist},~\eqref{eq:bound-on-beta-deltan-case1-intro},~\eqref{eq:bound-beta-beta-case2-qq-intro} and Lemma~\ref{lemma:bounds-of-beta-12}.
The proofs of~\eqref{eq:bound-on-beta-deltan-case1-intro} and~\eqref{eq:bound-beta-beta-case2-qq-intro} are provided in Appendices~\ref{app:proof-bound-case1} and~\ref{app:proof-bound-case2}, respectively.

\subsubsection{Proof of~\eqref{eq:bound-on-beta-deltan-case1-intro}}
\label{app:proof-bound-case1}
To upper-bound $\beta_{1- \delta_{1,n}} (P_{Y^n H^n}^{(1)}, Q_{Y^n}P_{ H^n})$, we consider the following test between $P_{Y^n H^n}^{(1)}$ and $Q_{Y^n}P_{H^n}$:
\begin{IEEEeqnarray}{rCl}
T(y^n, h^n) \define \indfun{ \sum\limits_{i=1}^{n} \frac{|h_i|^2( |y_i|^2-1)}{1+|h_i|^2 \sqrt{P} } \geq \xi_n }
\label{eq:test-suboptimal-csir-conv2}
\end{IEEEeqnarray}
where the threshold $\xi_n$ satisfies
\begin{IEEEeqnarray}{rCl}
P_{Y^n H^n}^{(1)}[T=1] \geq 1-\delta_{1,n}.
\label{eq:prob-yh-t1}
\end{IEEEeqnarray}
As mentioned earlier, this test is related to the Neyman-Pearson test between the capacity-achieving output distribution $(P_{YH}^*)^{n}$ and $Q_{Y^n}P_{H^n}$.
The term $(1+|h_i|^2 \sqrt{P})$ in the denominator of~\eqref{eq:test-suboptimal-csir-conv2} is included because the moment generating function of the random variable $|H_i|^2 (|Y_i|^2-1)$  (with $Y^n \sim Q_{Y^n}$) does not exist.
It follows from~\eqref{eq:prob-yh-t1} and from the definition of the $\beta$ function that
\begin{IEEEeqnarray}{rCl}
\beta_{1-\delta_n}(P_{Y^n H^n}^{(1)}, Q_{Y^n}P_{ H^n}) \leq Q_{Y^n}P_{H^n}[T=1]. \IEEEeqnarraynumspace
\label{eq:bound-beta-pyqy-def-csir}
\end{IEEEeqnarray}

To evaluate the RHS of~\eqref{eq:bound-beta-pyqy-def-csir}, we first determine $\xi_n$.
Let
\begin{IEEEeqnarray}{rCl}
A_i \define \frac{|H_i (H_iX_i + Z_i)| ^2-|H_i|^2}{1+|H_i|^2\sqrt{P}}.
\end{IEEEeqnarray}
Then
\begin{IEEEeqnarray}{rCl}
P_{Y^n H^n}^{(1)}[T=1] = \Ex{X^n}{ \prob\lefto[\sum\limits_{i=1}^{n} A_i  \geq \xi_n  \Big| X^n \right]}.
\label{eq:write-prob-as-expectation}
\end{IEEEeqnarray}
Set now
\begin{IEEEeqnarray}{rCl}
\xi_n &=&   \min\limits_{x^n\in \setC_1} \bigg\{ \sum\limits_{i=1}^{n} \Ex{}{A_i|X^n=x^n} \notag \\
&&\qquad \qquad  -\, \sqrt{ \delta_{1,n}^{-1} \sum\nolimits_{i=1}^{n} \varr[A_i|X^n=x^n]} \bigg\}. \IEEEeqnarraynumspace
\label{eq:def-xi-n-csir}
\end{IEEEeqnarray}
We have that for every $x^n\in\setC_1$
\begin{IEEEeqnarray}{rCl}
\IEEEeqnarraymulticol{3}{l}{
 \prob\lefto[\sum\limits_{i=1}^{n} A_i \leq  \xi_n \Big| X^n =x^n \right] }\notag\\
 \quad  &\leq&  \prob\mathopen{}\bigg[\sum\limits_{i=1}^{n} A_i \leq   \sum\limits_{i=1}^{n} \Ex{}{A_i|X^n=x^n} \notag\\
  && \quad -\, \sqrt{\delta_{1,n}^{-1} \sum\nolimits_{i=1}^{n} \varr[A_i|X^n=x^n]}   \bigg| X^n =x^n \bigg] \IEEEeqnarraynumspace  \label{eq:chebyshev-inequality1} \\
 & \leq &  \delta_{1,n}  .
 \label{eq:chebyshev-inequality}
\end{IEEEeqnarray}
Here,~\eqref{eq:chebyshev-inequality1} follows from~\eqref{eq:def-xi-n-csir}, and~\eqref{eq:chebyshev-inequality} follows from Chebyshev's inequality.
Note that~\eqref{eq:write-prob-as-expectation} and~\eqref{eq:chebyshev-inequality} imply  that $\xi_n$ defined in~\eqref{eq:def-xi-n-csir} indeed satisfies~\eqref{eq:prob-yh-t1}.

To characterize the asymptotic behavior of  $\xi_n$, we make use of the following estimates of the conditional mean and the conditional variance of $A_i$ given $X^n=x^n$:
\begin{IEEEeqnarray}{rCl}
\sum\limits_{i=1}^{n} \Ex{}{A_i| X^n=x^n} &=& nP \Ex{}{\frac{|H|^4}{1+|H|^2\sqrt{P}}}\IEEEeqnarraynumspace \\
&=&    nP \Ex{}{|H|^4}  +\littleo(nP)
\label{eq:estimate-mean-ai-given-x}
\end{IEEEeqnarray}
and
\begin{IEEEeqnarray}{rCl}
\IEEEeqnarraymulticol{3}{l}{
\sum\limits_{i=1}^{n} \varr[A_i |X^n=x^n]}\notag\\
  &\leq & 3 \sum\limits_{i=1}^{n}\bigg(
|x_i|^4  \varr \lefto[ \frac{|H_i|^4}{1+|H_i|^2\sqrt{P}} \right] \notag\\
&& + \,\varr \lefto[\frac{2|H_i|^2 \mathrm{Re}(H_ix_i Z_i^* )}{1+|H_i|^2\sqrt{P}}\right] + \varr \lefto[\frac{|H_i|^2(Z_i^2 -1) }{1+|H_i|^2\sqrt{P}}\right]
\bigg)  \label{eq:estimate-variance-ai-given-x-step1}\notag\\
 \\
&\leq & 3\Ex{}{|H|^8 } \|x^n\|_4^4  + \bigO( n) \\
&\leq&   3\Ex{}{|H|^8 } n \eta_n + \bigO( n) \label{eq:estimate-variance-ai-given-x}
\IEEEeqnarraynumspace
\end{IEEEeqnarray}
for every $x^n \in \setC_1$. Here,~\eqref{eq:estimate-variance-ai-given-x-step1}  follows because
\begin{equation}
\varr \lefto[\sum\limits_{i=1}^{K} B_i\right] \leq K \sum\limits_{i=1}^{K}\varr[B_i ]
\end{equation}
for all random variables $B_1,\ldots, B_K$, and in~\eqref{eq:estimate-variance-ai-given-x} we have used~\eqref{eq:def-setc1-csir}.
Substituting~\eqref{eq:estimate-mean-ai-given-x}  and~\eqref{eq:estimate-variance-ai-given-x} into~\eqref{eq:def-xi-n-csir}, we obtain
\begin{equation}
\xi_n \geq nP \Ex{}{|H|^4} - \sqrt{3 \Ex{}{|H|^8} ( \delta_{1,n}^{-1}  n\eta_n )} + \littleo\lefto( \sqrt{ \delta_{1,n}^{-1}  n\eta_n }\right).
\label{eq:asy-exp-xinn-inti}
\end{equation}
To conclude this part of the proof, we next characterize the asymptotic behavior of $  \delta_{1,n}^{-1}  n\eta_n $ as follows:
\begin{IEEEeqnarray}{rCl}
  \delta_{1,n}^{-1}  n\eta_n  &=& n   \frac{\max\{nP^{5/2} , (nP)^{3/4}\} }{\max\{ P^{1/4} , (nP)^{-1/8}\}} \\
  &\leq & n  \max\lefto\{ \frac{nP^{5/2}  }{P^{1/4} } , \frac{ (nP^2)^{3/4}}{ (nP^2)^{-1/8}} \right\}
  \label{eq:estimate-delta-n-inv1} \\
  &=& \max\{ n^2P^{9/4}, n(nP^2)^{7/8}\}  =  \littleo(n^2P^2) . \IEEEeqnarraynumspace \label{eq:estimate-delta-n-inv}
\end{IEEEeqnarray}
Here, \eqref{eq:estimate-delta-n-inv1} follows from  the inequality
\begin{IEEEeqnarray}{rCl}
\frac{\max\{a,b\}}{\max\{c,d\}} \leq \max\lefto\{\frac{a}{c}, \frac{b}{d}\right\}, \,\,\forall a,b,c,d>0. \IEEEeqnarraynumspace
\end{IEEEeqnarray}
Substituting~\eqref{eq:estimate-delta-n-inv} into~\eqref{eq:asy-exp-xinn-inti}, we obtain
\begin{IEEEeqnarray}{rCl}
\xi_n &\geq &  nP \Ex{}{|H|^4} + \littleo(nP).
\end{IEEEeqnarray}
Since $\xi_n \leq  nP \Ex{}{|H|^4} + \littleo(nP)$ as can be inferred from~\eqref{eq:def-xi-n-csir} and~\eqref{eq:estimate-mean-ai-given-x}, we  conclude that
\begin{IEEEeqnarray}{rCl}
\xi_n &= &  nP \Ex{}{|H|^4} + \littleo(nP).
\label{eq:asy-exp-xinn}
\end{IEEEeqnarray}

Next, we evaluate $Q_{Y^n}P_{H^n}[T=1]$. The idea is to use the G\"{a}rtner-Ellis theorem~\cite[Th.~2.3.6]{dembo98}, which characterizes the probability of large deviations of a random variable from its mean.
Let
%
\begin{IEEEeqnarray}{rCl}
D_i \define   \frac{|H_i|^2 (|Z_i|^2-1)}{1+|H_i|^2 \sqrt{P}}
\end{IEEEeqnarray}
where $Z_i \sim \jpg(0,1)$, $i=1,\ldots,n$,  are independent of $\{H_i\}$.
Note that
\begin{IEEEeqnarray}{rCl}
Q_{Y^n}P_{H^n}[T=1] = \prob\lefto[ \sum\limits_{i=1}^{n}D_i \geq \xi_n \right].
\end{IEEEeqnarray}
Let $\Lambda_n(\cdot)$ denote the logarithmic moment-generating function~\cite[Eq.~(2.2.1)]{dembo98} of the random variable $\xi_n^{-1} \sum\nolimits_{i=1}^{n}D_i$.
We shall prove the following result: for every $c\in \realset$,
\begin{IEEEeqnarray}{rCl}
\lim\limits_{n\to\infty} \frac{n}{\xi_n^2} \Lambda_n\lefto( \frac{c \xi_n^2}{n} \right) = \frac{c^2 \Ex{}{|H|^4}}{2} \define \Lambda(c). \IEEEeqnarraynumspace
\label{eq:lmgf-sum-bi}
 \end{IEEEeqnarray}
 Let us assume that~\eqref{eq:lmgf-sum-bi} holds;  then we have
\begin{IEEEeqnarray}{rCl}
\IEEEeqnarraymulticol{3}{l}{
\limsup\limits_{n\to\infty} \frac{n}{ \xi_n^2} \log Q_{Y^n}P_{H^n} [T=1] } \\
&=& \limsup\limits_{n\to\infty} \frac{n}{ \xi_n^2} \log \prob\lefto[\xi^{-1}_n \sum\limits_{i=1}^{n} D_i \geq 1  \right] \IEEEeqnarraynumspace \\
 &\leq& -  \inf_{t\geq 1}  \frac{t^2 \log e}{2 \Ex{}{|H|^4}} \label{eq:app-gartner-ellis} \\
 &=&  - \frac{ \log e}{2 \Ex{}{|H|^4}}. \label{eq:bound-q-betabetanu-conv1}
\end{IEEEeqnarray}
Here,~\eqref{eq:app-gartner-ellis} follows from the G\"{a}rtner-Ellis theorem,\footnote{Note that we have used the G\"{a}rtner-Ellis theorem with $1/n$ replaced by the vanishing sequence $n/\xi_n^2$ (see~\cite[Remark (a), p.~44]{dembo98}).} and because the Fenchel-Legendre transformation~\cite[Def. 2.2.2]{dembo98} of $\Lambda(c)$ defined in~\eqref{eq:lmgf-sum-bi} is
\begin{IEEEeqnarray}{rCl}
\Lambda^*(t)  = \sup_{\alpha \in \realset} \Big\{  t\alpha  - \Lambda(\alpha )\Big\}  = \frac{t^2}{2\Ex{}{|H|^4}}.
\end{IEEEeqnarray}
Using~\eqref{eq:bound-q-betabetanu-conv1} in~\eqref{eq:bound-beta-pyqy-def-csir}, we obtain
\begin{equation}
\log \beta_{1- \delta_{1,n}} (P_{Y^n H^n}^{(1)}, Q_{Y^n}P_{ H^n})
\leq  - \frac{ \log e}{2 \Ex{}{|H|^4}} \frac{\xi_n^2}{n} (1+\littleo(1)) \label{eq:bound-on-beta-deltan-case1}.
\end{equation}
The desired bound~\eqref{eq:bound-on-beta-deltan-case1-intro} follows by substituting~\eqref{eq:asy-exp-xinn} into~\eqref{eq:bound-on-beta-deltan-case1}.

It remains to prove~\eqref{eq:lmgf-sum-bi}.
We have
\begin{IEEEeqnarray}{rCl}
\IEEEeqnarraymulticol{3}{l}{
\Lambda_n\lefto(\frac{ c \xi_n^2}{n}\right)}\notag\\
 &=& \log_e \Ex{}{ \exp\lefto(  \log e \frac{c \xi_n^2}{n \xi_n }   \sum\limits_{i=1}^{n} D_i \right)} \\
&=& n\log_e  \Ex{H}{ \Ex{}{ \exp\lefto(\frac{c\xi_n }{n}\frac{(|H|^2\log e )( |Z|^2-1)}{1+|H|^2 \sqrt{P} }  \right) \Big| H} } . \notag\\
\label{eq:compute-lmgf-sum-bi}
\end{IEEEeqnarray}
Observe now that the following bound holds for every $c>0$,
 every $h\in\complexset$,  and for all sufficiently large $n$:
\begin{IEEEeqnarray}{rCl}
\frac{c \xi_n }{n}  \frac{ |h|^2\log e }{1+|h|^2 \sqrt{P}} &\leq &
\frac{c\xi_n \log e}{n \sqrt{P} }  <1.
\end{IEEEeqnarray}
Here, the second inequality follows from~\eqref{eq:asy-exp-xinn}.
Since $|Z|^2\sim \mathrm{Exp}(1)$, we have that for all sufficiently large $n$,
\begin{IEEEeqnarray}{rCl}
 \IEEEeqnarraymulticol{3}{l}{
   \Ex{H}{ \Ex{Z}{ \exp\lefto(  \frac{c \xi_n }{n} \frac{  |H|^2 \log e }{1+|H|^2 \sqrt{P}}  (|Z|^2-1 )   \right) \Big| H} } }\notag\\
   &=&  \mathbb{E}\mathopen{} \bigg[ \! \left(  1-    \frac{c\xi_n }{n} \frac{  |H|^2 \log e }{1+|H|^2 \sqrt{P}} \right)^{-1} \!\!
   \underbrace{\exp\lefto( \!-\frac{ c \xi_n }{n} \frac{   |H|^2\log e  }{1+|H|^2 \sqrt{P}} \right)}_{\leq 1}\! \bigg] \notag\\
   \\
   &\leq & \left(1- \frac{c\xi_n \log e}{n \sqrt{P} }  \right)^{-1} <\infty.
\label{eq:compute-lmgf-sum-bi-22}
\end{IEEEeqnarray}
Using~\eqref{eq:compute-lmgf-sum-bi-22}, the dominated convergence theorem (see, e.g.,~\cite[Th.~1.34]{rudin87a}), and the Taylor series expansion
\begin{IEEEeqnarray}{rCl}
  e^{x} = 1+ x+ \frac{x^2}{2} + \bigO(x^3), \quad x\to 0
\end{IEEEeqnarray}
we conclude that
\begin{IEEEeqnarray}{rCl}
\IEEEeqnarraymulticol{3}{l}{
\Ex{}{ \exp\lefto(\log e\frac{c \xi_n}{n}\frac{ |H|^2 ( |Z|^2-1)}{1+|H|^2 \sqrt{P}} \right) }}\notag\\
&=&  1+  \frac{c \xi_n}{n} \underbrace{\Ex{}{\frac{|H|^2 ( |Z|^2-1)}{1+|H|^2 \sqrt{P}} } }_{=0} \notag\\
&&+\, \frac{c^2}{2} \frac{\xi_n^2}{n^2} \underbrace{\Ex{}{\left(\frac{|H|^2 ( |Z|^2-1)}{1+|H|^2 \sqrt{P}} \right)^2}}_{= \Ex{}{|H|^4}  + \bigO(\sqrt{P})} + \bigO\lefto((\xi_n/n)^3\right)  \IEEEeqnarraynumspace\\
&=& 1 + \frac{c^2}{2}\frac{\xi_n^2}{n^2} \Big(\Ex{}{|H|^4} +\bigO(\sqrt{P}) \Big) + \bigO\lefto((\xi_n/n)^3\right) . \IEEEeqnarraynumspace \label{eq:evaluating-lmgf-sumbi2}
\end{IEEEeqnarray}
Substituting~\eqref{eq:evaluating-lmgf-sumbi2} into~\eqref{eq:compute-lmgf-sum-bi}, and then using that $\log_e(1+x) = x  +\littleo(x), \,x\to0$, we conclude  that
\begin{IEEEeqnarray}{rCl}
  \frac{n}{\xi_n^2} \Lambda_n\big( c \xi_n^2/n\big) &=& \frac{n^2}{\xi_n^2} \left(\frac{c^2\Ex{}{|H|^4}}{2} \frac{\xi^2_n}{n^2} + \littleo\lefto(\frac{\xi^2_n}{n^2}\right) \right) \notag\\
  &=& \frac{c^2\Ex{}{|H|^4}}{2} + \littleo(1)
\label{eq:lmgf-sum-bi-proof}
 \end{IEEEeqnarray}
which implies~\eqref{eq:lmgf-sum-bi}.

\subsubsection{Proof of~\eqref{eq:bound-beta-beta-case2-qq-intro}}
 \label{app:proof-bound-case2}

Consider the test
\begin{IEEEeqnarray}{rCl}
T(y^n,h^n) \define \indfun{ \|y^n\|_4^4 \geq \tilde{\xi}_n }
\end{IEEEeqnarray}
where $\tilde{\xi}_n$ is defined in~\eqref{eq:def-tilde-xi-n-app-fading}.
%
%
Note that
\begin{equation}
\frac{\tilde{\xi}_n}{n} \geq (nP^2)^{1/8} \to \infty ,\quad n\to\infty.
\label{eq:evaluate-xin-conv}
\end{equation}
We evaluate the probability that $T=1$ under  $P_{Y^nH^n}^{(2)}$ and under $Q_{Y^n}P_{H^n}$ by following closely the proof of~\cite[Th. 9]{polyanskiy2012-10a}.
We start by noting that
\begin{IEEEeqnarray}{rCl}
\IEEEeqnarraymulticol{3}{l}{
P_{Y^nH^n}^{(2)}[T=1] }\notag\\
\quad &=& \prob[ \|H^nX^n + Z^n \|_4 \geq \tilde{\xi}_n^{1/4}] \\
&\geq& \prob[ \|H^nX^n\|_4 -\|Z^n \|_4 \geq \tilde{\xi}_n^{1/4}] \label{eq:bound-p-t-case2-2} \\
&\geq & \prob[\|H^nX^n\|_4 \geq 2\tilde{\xi}_n^{1/4}  ] \cdot \prob[  \|Z^n \|_4\leq  \tilde{\xi}_n^{1/4}] \label{eq:bound-p-t-case2-3}\\
&\geq& \prob[ |H| \|X^n\|_{\infty} \geq 2\tilde{\xi}_n^{1/4}   ]\cdot \prob[ \|Z^n \|_4\leq  \tilde{\xi}_n^{1/4}] \label{eq:bound-p-t-case2}. \IEEEeqnarraynumspace
\end{IEEEeqnarray}
Here,~\eqref{eq:bound-p-t-case2-2} follows by the triangle inequality, and~\eqref{eq:bound-p-t-case2} follows because $\|\cdot\|_4 \geq \|\cdot\|_\infty$.
The first term in the product on the RHS of~\eqref{eq:bound-p-t-case2} can be bounded as
\begin{IEEEeqnarray}{rCl}
\prob[ |H| \|X^n\|_{\infty} \geq   2\tilde{\xi}_n^{1/4}    ] &\geq & \prob\mathopen{}\Big[ |H|   \sqrt{\eta_n/P} \geq 2\tilde{\xi}_n^{1/4}  \Big] \IEEEeqnarraynumspace \label{eq:term1-prod-pt-0}\\
&=& \exp\lefto( -  4 \tilde{\xi}_n^{1/2}    P/\eta_n \right) \label{eq:term1-prod-pt-1}\\
&\geq & 1 - \littleo(1).
\label{eq:term1-prod-pt-2}
\end{IEEEeqnarray}
Here,~\eqref{eq:term1-prod-pt-0} follows because, by H\"{o}lder's inequality and by~\eqref{eq:def-setc2-csir}, $\|x^n\|_{\infty} \geq  \|x^n\|^2_4 / \|x^n\|_2 \geq \sqrt{\eta_n/P}$ for every $x^n \in \setC_2$;~\eqref{eq:term1-prod-pt-2} follows because
\begin{IEEEeqnarray}{rCl}
 \frac{\tilde{\xi}_n^{1/2} P }{ \eta_n} &=&  \frac{\max\{ n P^{23/8} , (nP^2)^{5/8} \}}{ \max\{ nP^{5/2} , (nP^2)^{3/4}\}} \\
&\leq& \max\lefto\{  \frac{n P^{23/8}}{nP^{5/2}} ,  \frac{(nP^2)^{5/8}}{ (nP^2)^{3/4}} \right\}\\
&=& \max\{P^{3/8} , (nP^2)^{-1/8}\} = o(1) . \IEEEeqnarraynumspace
\end{IEEEeqnarray}
The second term  in the product on the RHS of~\eqref{eq:bound-p-t-case2} can be bounded  using Markov's inequality as follows:
\begin{IEEEeqnarray}{rCl}
\prob[ \|Z^n \|_4\leq \tilde{\xi}_n^{1/4}  ]  &=&  1-  \prob[ \|Z^n \|_4^4\geq  \tilde{\xi}_n ] \\
 &\geq& 1 - \frac{n \Ex{}{|Z_1|^4}}{\tilde{\xi}_n} \label{eq:term1-prod-pt-2221}\\
 & = &   1 -\littleo (1).
\label{eq:term1-prod-pt-22}
\end{IEEEeqnarray}
Here, the last step follows from~\eqref{eq:evaluate-xin-conv}.
Substituting~\eqref{eq:term1-prod-pt-1} and~\eqref{eq:term1-prod-pt-2221} into~\eqref{eq:bound-p-t-case2}, we obtain
\begin{IEEEeqnarray}{rCl}
P_{Y^nH^n}^{(2)}[T=1]   &\geq&  \exp\lefto(- 4 \tilde{\xi}_n^{1/2}   P /\eta_n  \right)  \Big( 1 - \frac{n \Ex{}{|Z_1|^4}}{\tilde{\xi}_n}\Big) \IEEEeqnarraynumspace \\
 &=& 1-\delta_{2,n}.
\label{eq:bound-p-t-case-2-p}
\end{IEEEeqnarray}

Next, we evaluate $Q_{Y^n}P_{H^n}[T=1]$.  Since $Y^n \sim \jpg(\mathbf{0}, \matI_n)$ under $Q_{Y^n}$,   by  following the same steps as the ones reported in~\cite[Eqs.~(117)--(122)]{polyanskiy2012-10a}
and by using the Gaussian isoperimetric inequality~\cite[Eq.~(3.4.8)]{raginsky13}, we obtain
\begin{IEEEeqnarray}{rCl}
\IEEEeqnarraymulticol{3}{l}{
 \log Q_{Y^n}[ \|Y^n\|^{4}_4 \geq  \tilde{\xi}_n] }\notag\\
 \quad &\leq&   - \left( \tilde{\xi}_n^{1/4} - 3^{1/4} n^{1/4}\right)^2 \log e - \log 2  \\
&=& -\tilde{\xi}_n^{1/2} \log e + \littleo(\tilde{\xi}_n^{1/2} )  \label{eq:bound-beta-beta-case2-qq-bound-01}\\
&\leq & -\frac{\Ex{}{|H|^4}}{2} nP^2 \log e + \littleo(n P^2).
 \label{eq:bound-beta-beta-case2-qq-bound}
 \end{IEEEeqnarray}
Here, in~\eqref{eq:bound-beta-beta-case2-qq-bound-01} we used~\eqref{eq:evaluate-xin-conv}, and~\eqref{eq:bound-beta-beta-case2-qq-bound}  holds for all sufficiently large $n$, since
\begin{equation}
\tilde{\xi}_n^{1/2}/(nP^2)    \geq P^{-1/8} \to \infty,\quad  n\to\infty.
\end{equation}
Combining~\eqref{eq:bound-p-t-case-2-p} and~\eqref{eq:bound-beta-beta-case2-qq-bound}, we conclude~\eqref{eq:bound-beta-beta-case2-qq-intro}. %

\subsection{Proof of Lemma~\ref{lemma:bounds-of-beta-12}}
\label{app:proof-lemma-properties}

Let $T_1$ and $T_2$ be the Neyman-Pearson tests that achieve $\beta_{1-\delta_1}(P_1,Q)$ and $\beta_{1-\delta_2}(P_2,Q)$, respectively. Consider the following test for distinguishing  between $P$ and $Q$:
\begin{IEEEeqnarray}{rCl}
T \define \indfun{ T_1 =1 \cup T_2 =1 } .
\end{IEEEeqnarray}
Under $P$, we have
\begin{IEEEeqnarray}{rCl}
P[T=1] &=& \lambda P_1[T=1] + (1-\lambda) P_2[T=1]\\
&\geq& \lambda P_1[T_1 =1] + (1-\lambda) P_2[T_2 =1]\\
&=& 1 -\lambda \delta_1 -(1-\lambda)\delta_2. \label{eq:TTT-underP}
\end{IEEEeqnarray}
Under $Q$, we have
\begin{IEEEeqnarray}{rCl}
Q[T=1]
&\leq& Q[T_1 =1 ] + Q[T_2 =1 ]\\
&=& \beta_{1-\delta_1}(P_1,Q) + \beta_{1-\delta_2}(P_2,Q)
\label{eq:TTT-underQ}
\end{IEEEeqnarray}
where in the first step we used the union bound.  Note that~\eqref{eq:TTT-underP} and~\eqref{eq:TTT-underQ} imply~\eqref{eq:lemma-combine-2-dist}.


\begin{thebibliography}{10}
\providecommand{\url}[1]{#1}
\csname url@samestyle\endcsname
\providecommand{\newblock}{\relax}
\providecommand{\bibinfo}[2]{#2}
\providecommand{\BIBentrySTDinterwordspacing}{\spaceskip=0pt\relax}
\providecommand{\BIBentryALTinterwordstretchfactor}{4}
\providecommand{\BIBentryALTinterwordspacing}{\spaceskip=\fontdimen2\font plus
\BIBentryALTinterwordstretchfactor\fontdimen3\font minus
  \fontdimen4\font\relax}
\providecommand{\BIBforeignlanguage}[2]{{%
\expandafter\ifx\csname l@#1\endcsname\relax
\typeout{** WARNING: IEEEtran.bst: No hyphenation pattern has been}%
\typeout{** loaded for the language `#1'. Using the pattern for}%
\typeout{** the default language instead.}%
\else
\language=\csname l@#1\endcsname
\fi
#2}}
\providecommand{\BIBdecl}{\relax}
\BIBdecl

\bibitem{shannon48}
C.~E. Shannon, ``A mathematical theory of communication,'' \emph{Bell Syst.
  Tech. J.}, vol.~27, pp. 379--423 and 623--656, Jul./Oct. 1948.

\bibitem{feinstein54a}
A.~Feinstein, ``A new basic theorem of information theory,'' \emph{IRE Trans.
  Inform. Theory}, vol.~4, no.~4, pp. 2--22, 1954.

\bibitem{shannon59}
C.~E. Shannon, ``Probability of error for optimal codes in a {G}aussian
  channel,'' \emph{Bell Syst. Tech.~J.}, vol.~38, no.~3, pp. 611--656, May
  1959.

\bibitem{verdu94-07a}
S.~Verd\'{u} and T.~S. Han, ``A general formula for channel capacity,''
  \emph{{IEEE} Trans. Inf. Theory}, vol.~40, no.~4, pp. 1147--1157, Jul. 1994.

\bibitem{polyanskiy10-05}
Y.~Polyanskiy, H.~V. Poor, and S.~Verd{\'u}, ``Channel coding rate in the
  finite blocklength regime,'' \emph{{IEEE} Trans. Inf. Theory}, vol.~56,
  no.~5, pp. 2307--2359, May 2010.

\bibitem{pinsker1964}
M.~Pinsker, \emph{Information and information stability of random variables and
  processes}.\hskip 1em plus 0.5em minus 0.4em\relax San Francisco: Holden-Day,
  1964.

\bibitem{han93-03}
T.~S. Han and S.~Verd{\'u}, ``Approximation theory of output statistics,''
  \emph{{IEEE} Trans. Inf. Theory}, vol.~39, no.~3, pp. 752--772, May 1993.

\bibitem{csiszar11}
I.~Csisz\'{a}r and J.~K\"{o}rner, \emph{Information Theory: Coding Theorems for
  Discrete Memoryless Systems}, 2nd~ed.\hskip 1em plus 0.5em minus 0.4em\relax
  Cambridge, U.K.: Cambridge Univ. Press, 2011.

\bibitem{polyanskiy12}
Y. Polyanskiy and Y. Wu, \emph{Lecture Notes on Information Theory}. MIT (6.441), UIUC (ECE 563), Yale (STAT 664), 2012--2017.

\bibitem{Topsoe1967}
F. Tops\o e, ``An information theoretical identity and a problem involving capacity,''
\emph{Studia Scientiarum Math. Hung.}, vol.~2, pp. 291-–292, 1967.

\bibitem{lapidoth03-10a}
A.~Lapidoth and S.~M. Moser, ``Capacity bounds via duality with applications to
  multiple-antenna systems on flat-fading channels,'' \emph{{IEEE} Trans. Inf.
  Theory}, vol.~49, no.~10, pp. 2426--2467, Oct. 2003.

\bibitem{verdu90-09}
S.~Verd\'{u}, ``On channel capacity per unit cost,'' \emph{{IEEE} Trans. Inf.
  Theory}, vol.~36, no.~5, pp. 1019--1030, Sep. 1990.

\bibitem{verdu02-06}
------, ``Spectral efficiency in the wideband regime,'' \emph{{IEEE} Trans.
  Inf. Theory}, vol.~48, no.~6, pp. 1319--1343, Jun. 2002.

\bibitem{arimoto1972-01a}
S.~Arimoto, ``An algorithm for computing the capacity of arbitrary discrete
  memoryless channels,'' \emph{{IEEE} Trans. Inf. Theory}, vol.~IT-18, no.~1, pp.
  14--20, Jan. 1972.

\bibitem{blahut1972-07a}
R.~E. Blahut, ``Computation of channel capacity and rate-distortion function,''
  \emph{{IEEE} Trans. Inf. Theory}, vol.~IT-18, no.~4, pp. 460--473, Jul. 1972.

\bibitem{gallager79-u}
\BIBentryALTinterwordspacing
R.~G. Gallager, ``Source coding with side information and universal coding,''
  1979, unpublished manuscript. [Online]. Available:
  \url{http://web.mit.edu/gallager/www/papers/paper5.pdf}
\BIBentrySTDinterwordspacing

\bibitem{shamai97-03}
S.~{Shamai (Shitz)} and S.~Verd\'{u}, ``The empirical distribution of good
  codes,'' \emph{{IEEE} Trans. Inf. Theory}, vol.~43, no.~3, pp. 836--846, May
  1997.

\bibitem{polyanskiy14-01}
Y.~Polyanskiy and S.~Verd\'{u}, ``Empirical distribution of good channel codes
  with non-vanishing error probability,'' \emph{{IEEE} Trans. Inf. Theory},
  vol.~60, no.~1, pp. 5--21, Jan. 2014.

\bibitem{yang1999-05az}
Y.~Yang and A.~Barron, ``Information-theoretic determination of minimax rates
  of convergence,'' \emph{Ann. Stat.}, vol.~27, no.~5, pp. 1564--1599, 1999.

\bibitem{haussler1997-06a}
D.~Haussler and M.~Opper, ``Mutual information, metric entropy and cumulative
  relative entropy risk,'' \emph{Ann. Stat.}, vol.~25, no.~6, pp. 2451--2492,
  1997.

\bibitem{neyman33a}
J.~Neyman and E.~S. Pearson, ``On the problem of the most efficient tests of
  statistical hypotheses,'' \emph{Philosophical Trans. Royal Soc. A}, vol. 231,
  pp. 289--337, 1933.

\bibitem{nagaoka2001}
H. Nagaoka, ``Strong converse theorems in quantum information theory,''
in \emph{Proc. ERATO Conference on Quantum Information Science (EQIS) 2001}, p.~33, Tokyo, Japan, Sep.~2001.

\bibitem{cover06-a}
T.~M. Cover and J.~A. Thomas, \emph{Elements of Information Theory},
  2nd~ed.\hskip 1em plus 0.5em minus 0.4em\relax New Jersey: Wiley, 2006.

\bibitem{verdu96-01a}
S.~Verd\'{u}, ``The exponential distribution in information theory,''
  \emph{{Probl. Inf. Transm.}}, vol.~32, no.~1, pp. 86--95, 1996.

\bibitem{polyanskiy11-08a}
Y.~Polyanskiy and S.~Verd\'{u}, ``Scalar coherent fading channel: dispersion
  analysis,'' in \emph{Proc. IEEE Int. Symp. Inf. Theory (ISIT)}, Saint
  Petersburg, Russia, Aug. 2011, pp. 2959--2963.

\bibitem{shannon57}
C.~E. Shannon, ``Certain results in the coding theory for noisy channels,''
  \emph{Inf. Contr.}, vol.~1, pp. 6--25, 1957.

\bibitem{wang2009-07a}
L.~Wang, R.~Colbeck, and R.~Renner, ``Simple channel coding bounds,'' in
  \emph{Proc. IEEE Int. Symp. Inf. Theory (ISIT)}, Seoul, Korea, Jul. 2009.

\bibitem{polyanskiy10-09a}
Y.~Polyanskiy and S.~Verd\'{u}, ``Arimoto channel coding converse and
  {R}\'{e}nyi divergence,'' in \emph{Proc. 48th Allerton Conf. Commun., Contr.,
  Comp.}, Monticello, IL, USA, Sep. 2010.

\bibitem{polyanskiy13}
Y.~Polyanskiy, ``Saddle point in the minimax converse for channel coding,''
  \emph{{IEEE} Trans. Inf. Theory}, vol.~59, no.~5, pp. 2576--2595, May 2013.

\bibitem{molavianJazi15-12a}
E.~Molavian{J}azi and J.~N. Laneman, ``A second-order achievable rate region
  for {G}aussian multi-access channels via a central limit theorem for
  functions,'' \emph{{IEEE} Trans. Inf. Theory}, vol.~61, no.~12, pp.
  6719--6733, 2015.

\bibitem{hoydis15-12}
J.~Hoydis, R.~Couillet, and P.~Piantanida, ``The second-order coding rate of
  the {MIMO} quasi-static {R}ayleigh fading channel,'' \emph{{IEEE} Trans. Inf.
  Theory}, vol.~61, no.~12, pp. 6591--6622, Dec 2015.

\bibitem{shannon49}
C.~E. Shannon, ``Communication in the presense of noise,'' \emph{Proc. IRE},
  vol.~37, pp. 10--21, 1949.

\bibitem{kennedy69}
R.~S. Kennedy, \emph{Fading Dispersive Communication Channels}.\hskip 1em plus
  0.5em minus 0.4em\relax New York, NY, USA: Willey, 1969.

\bibitem{polyanskiy11-08b}
Y.~Polyanskiy, H.~V. Poor, and S.~Verd\'{u}, ``Minimum energy to send $k$ bits
  through the {G}aussian channel with and without feedback,'' \emph{{IEEE}
  Trans. Inf. Theory}, vol.~57, no.~8, pp. 4880--4902, Aug. 2011.

\bibitem{yang2016-to-appear}
W.~Yang, G.~Durisi, and Y.~Polyanskiy, ``Minimum energy to send $k$ bits over
  multiple-antenna fading channels,'' \emph{{IEEE} Trans. Inf. Theory},
  vol.~62, no.~12, pp. 6831--6853, Dec. 2016.

\bibitem{collins2016-07a}
A.~Collins and Y.~Polyanskiy, ``Dispersion of the coherent {MIMO} block-fading
  channel,'' in \emph{Proc. IEEE Int. Symp. Inf. Theory (ISIT)}, Barcelona,
  Spain, Jul. 2016.

\bibitem{anantharam1996-01a}
V.~Anantharam and S.~Verd\'{u}, ``Bits through queues,'' \emph{{IEEE} Trans.
  Inf. Theory}, vol.~42, no.~1, pp. 4--18, Jan. 1996.

\bibitem{martinez2011-06a}
A.~Martinez, ``Communication by energy modulation: The additive exponential
  noise channel,'' \emph{{IEEE} Trans. Inf. Theory}, vol.~57, no.~6, pp.
  3333--3351, Jun. 2011.

\bibitem{riedl2011-10a}
T.~J. Riedl, T.~P. Coleman, and A.~C. Singer, ``Finite block-length achievable
  rates for queuing timing channels,'' in \emph{Proc. IEEE Inf. Theory Workshop
  (ITW)}, Paraty, Oct. 2011, pp. 200--204.

\bibitem{feller70a}
W.~Feller, \emph{An Introduction to Probability Theory and Its
  Applications}.\hskip 1em plus 0.5em minus 0.4em\relax New York, NY, USA: John
  Wiley \& Sons, 1970, vol.~1.

\bibitem{telatar99-11a}
{\.I}.~E. Telatar, ``Capacity of multi-antenna {Gaussian} channels,''
  \emph{Eur. Trans. Telecommun.}, vol.~10, pp. 585--595, Nov. 1999.

\bibitem{collins14-07}
A.~Collins and Y.~Polyanskiy, ``Orthogonal designs optimize achievable
  dispersion for coherent {MISO} channels,'' in \emph{Proc. IEEE Int. Symp.
  Inf. Theory (ISIT)}, Honolulu, HI, USA, Jul. 2014.


\bibitem{yang15-08}
W.~Yang, ``Fading channels: Capacity and channel coding rate in the
  finite-blocklength regime,'' Ph.D. dissertation, Department of Signals and Systems, Chalmers University of
  Technology, Gothenburg, Sweden, Aug. 2015.

\bibitem{polyanskiy13-07a}
\BIBentryALTinterwordspacing
Y.~Polyanskiy, ``Finite blocklength methods in channel coding,'' in
  \emph{{Proc. IEEE Int. Symp. Inf. Theory (ISIT)}}, Istanbul, Turkey, Jul.
  2013, tutorial. [Online]. Available:
  \url{http://people.lids.mit.edu/yp/homepage/data/ISIT13\_tutorial.pdf}
\BIBentrySTDinterwordspacing

\bibitem{verdu98}
S.~Verd\'{u}, \emph{Multiuser Detection}.\hskip 1em plus 0.5em minus
  0.4em\relax Cambridge, UK: Cambridge University Press, 1998.

\bibitem{renyi1961}
A.~R\'{e}nyi, ``On measures of entropy and information,'' \emph{Proc. 4th
  Berkeley Symp. Math. Statist. and Prob.}, vol.~1, pp. 547--561, 1961.

\bibitem{dembo98}
A.~Dembo and O.~Zeitouni, \emph{Large deviations techniques and
  applications}.\hskip 1em plus 0.5em minus 0.4em\relax New York: Springer
  Verlag, 1998.

\bibitem{iri2015-06a}
N.~Iri and O.~Kosut, ``Third-order coding rate for universal compression of
  {M}arkov sources,'' in \emph{Proc. IEEE Int. Symp. Inf. Theory (ISIT)}, Hong
  Kong, China, Jun. 2015, pp. 1996--2000.

\bibitem{haroutunian68-04}
E. A. Haroutunian, ``Bounds for the exponent of the probability of error for a semicontinuous memoryless
channel,'' \emph{Probl. Peredachi Inf.}, vol.~4,
no.~4, pp.~37–48, 1968.



\bibitem{wendel1948-11a}
J.~G. Wendel, ``Note on the {G}amma function,'' \emph{Amer. Math. Monthly},
  vol.~55, no.~9, pp. 563--564, Nov. 1948.

\bibitem{rudin87a}
W.~Rudin, \emph{Real and Complex Analysis}, 3rd~ed.\hskip 1em plus 0.5em minus
  0.4em\relax New York, NY, U.S.A.: McGraw-Hill, 1987.

\bibitem{polyanskiy2012-10a}
Y.~Polyanskiy, ``$\ell_p$-norms of codewords from capacity- and
  dispersion-achieveing {G}aussian codes,'' in \emph{Proc. 50th Allerton Conf.
  Commun., Contr., Comp.}, Monticello, IL, USA, Oct. 2012.

\bibitem{raginsky13}
M.~Raginsky and I.~Sason, ``Concentration of measure inequalities in
  information theory, communications and coding,'' in \emph{Foundations and
  Trends in Communications and Information Theory}.\hskip 1em plus 0.5em minus
  0.4em\relax now Publishers, 2013, vol.~10, no. 1--2, pp. 1--246.

\end{thebibliography}
\end{document}